\documentclass[preprint]{aastex}

\usepackage{epsfig,amsmath}
\usepackage{natbib}
\usepackage[dvipdfmx, 3Dmenu, 3Dtoolbar]{media9}
\usepackage{graphicx}
\usepackage{rotating, morefloats}
\usepackage{lscape}
\newcommand{\mopex}{{\sc mopex }}
\newcommand{\apex}{{\sc apex }}
\newcommand{\spitzer}{{\it Spitzer }}

\shorttitle{}
\shortauthors{}

\begin{document}

%\title{The Carnegie Hubble Program: The Cepheid population of the SMC and the effect of metallicity on mid--infrared Cepheid colors.}
\title{The Carnegie Hubble Program: The Distance and Structure of the SMC as Revealed by Mid--infrared Observations of Cepheids}

\author{\bf Victoria~Scowcroft, Wendy~L.~Freedman, Barry~F.~Madore, Andy~Monson, S.~E.~Persson, Jeff~Rich, Mark~Seibert}
\affil{Observatories of the Carnegie Institution of Washington \\ 813
Santa Barbara St., Pasadena, CA~91101}
\author{\bf \&}
\author{\bf Jane~R.~Rigby}
\affil{Observational Cosmology Lab, NASA Goddard Space Flight Center, Greenbelt MD 20771}
\email{vs@obs.carnegiescience.edu, wendy@obs.carnegiescience.edu, 
barry@obs.carnegiescience.edu, amonson@obs.carnegiescience.edu,
persson@obs.carnegiescience.edu, jrich@obs.carnegiescience.edu,
mseibert@obs.carnegiescience.edu, jane.r.rigby@nasa.gov}

%\affil{}

%\author{name\altaffilmark{4,5}}
%\affil{}
%\email{}

%and

%\author{name\altaffilmark{5}}
%\affil{}

%% Abstract goes here
%\slugcomment{Draft last edited \today}

\begin{abstract}
Using \textit{Spitzer} observations of classical Cepheids  we have measured the true average distance modulus of the SMC to be $18.96 \pm 0.01_{stat} \pm 0.03_{sys}$~mag (corresponding to $62 \pm 0.3$~kpc), which is $0.48 \pm 0.01$~mag more distant than the LMC. This is in agreement with previous results from Cepheid observations, as well as with measurements from other indicators such as RR Lyrae stars and the tip of the red giant branch. 

Utilizing the properties of the mid--infrared Leavitt Law we measured precise distances to individual Cepheids in the SMC, and have confirmed that the galaxy is tilted and elongated such that its eastern side is up to 20~kpc closer than its western side. This is in agreement with the results from red clump stars and dynamical simulations of the Magellanic Clouds and Stream. 
\end{abstract}


\section{Introduction}
\label{sec:intro}

The Small Magellanic Cloud (SMC) is a highly disrupted satellite of
the Milky Way. It is estimated to be approximately 60 kpc from the
Sun. Because of its proximity and its complex three--dimensional form,
the precise distance to the SMC is ultimately entangled in the
simultaneous determination of this galaxy's line-of-sight geometry. The irregular nature of this late--type dwarf
galaxy brings into question what defines the ``center" of such a system
in any case. In their study of a selection of eclipsing binary systems in the SMC \citet{2014ApJ...780...59G} compiled the available literature values of the distance of the SMC. What they found was unsurprising --- the distances had a large spread as the no two groups were legitimately measuring the same value. Even when using near--infrared observations of Cepheids where the effects of extinction and temperature are greatly reduced  \citep[e.g.][]{2013ApJ...764...84I}, the true distance of the SMC is unclear. Perhaps the best that can be done is to map the 3D
spatial distribution of any given population and let the question of
some abstract ``mean" distance remain moot. 

%That is the main purpose of
%this paper.

In a seminal paper, entitled ``Aspects of the Structure of the Small
Magellanic Cloud", \citet{1981A&A....96..158F} made the first
quantitative and largely convincing case for measurable differences in
distance back-to-front, as well as across the line of sight, to the
SMC.  Magnitude differences in the supergiant population of stars over
the face of the SMC indicated that the southern end of the bar was
systematically more distant than the central and northern parts of
the galaxy. Informed arguments were also made against the suggestion
that these effects might be purely the result of varying amounts of
extinction. Soon thereafter, \citet[hereafter CC86][]{1986MNRAS.218..223C} and then \citet{1987ApJ...321..162W} also probed the geometry of the
SMC, this time using Classical Cepheids whose period--luminosity
relation can provide individually higher--precision distances than the
general supergiant populations used by \citeauthor{1981A&A....96..158F}  The results
however were qualitatively the same: the south--west side of the SMC
was seen to be further away than main body or the north--east
quadrant. CC86 dealt with reddening using multi--color (BVI)
photometry; \citeauthor{1987ApJ...321..162W} used infrared (JHK) photometry to minimize
the effects of extinction. The total (peak-to-peak)
effect quoted by CC86 amounted to 0.6 mag from side-to-side, and was
in excess of 0.3~mag back-to-front at any given point in the main body
of the SMC. \citeauthor{1987ApJ...321..162W} found somewhat reduced, but still
significant, depths and trends, with differences between the two
investigations being ascribed to the smaller sample size in the CC86
study and their erroneous inclusion of some W~Virginis (Pop II) stars
in their Classical Cepheid (Pop~I) analysis.

Most recently depth estimates of the Magellanic Clouds have come from
\citet{2009A&A...496..399S, 2012ApJ...744..128S} and from \citet{2012AJ....144..107H} using Cepheids, RR~Lyrae variables and Red Clump
stars as observed by OGLE~II \citep{1998AcA....48..147U} and OGLE~III
\citep{2010AcA....60..165S}. The sample sizes and areal coverage are now
quite impressive, as are the newly employed techniques for visualizing
these new datasets.

In moving from the optical to the mid--infrared (mid--IR) the intrinsic
dispersion in the Cepheid period--luminosity relation (the Leavitt Law;
hereafter LL) monotonically decreases. Indeed, the intrinsic
dispersion of the $3.6~\mu$m LL, as determined for the LMC and Milky Way
(MW), is $\pm$0.10~mag \citep{2011ApJ...743...76S, 2012ApJ...759..146M}.
Scatter in excess of this will be attributed hereafter to geometry of
the SMC.

The study presented here is being undertaken as part of the Carnegie
Hubble Program (CHP); a full description of which is given in Freedman
et al. (2011). Briefly, the CHP aims to measure the local Hubble
constant ($H_0$) to a systematic accuracy of 2\% using mid--infrared data
from the Warm Spitzer mission, and in the future, JWST. The SMC is
important to the CHP effort as it is a low metallicity system which
can be compared to the Milky Way and LMC in search of possible
metallicity effects on the Cepheids. This will be discussed in
Scowcroft et al. (2015). Here we focus on the structure of the SMC as
revealed by mid--infrared observations of its longest-period Cepheids.

\section{Observations}
\label{sec:observations}
\subsection{Target Selection}
\label{sec:target_selection}

The SMC has a large sample of well--studied Cepheids.  For the present survey we selected 92 Harvard variable Cepheids spread over the SMC. These objects have all been extensively studied over the years; archival optical and/or near--IR data is available for all of them and the vast majority were measured by OGLE \citep{1992AcA....42..253U, 1997AcA....47..319U, 2010AcA....60...17S}. The chosen Cepheids are free from crowding and have periods greater than six days to match with the period distribution of the calibrating samples from the LMC and MW.

%The OGLE studies \citep{1992AcA....42..253U, 1997AcA....47..319U, 2010AcA....60...17S}, while having a primary goal of observing microlensing events, have produced large Cepheid catalogs for the LMC and SMC in the $V$ and $I$ bands.

\begin{figure}
\begin{center}
\includegraphics[width=165mm]{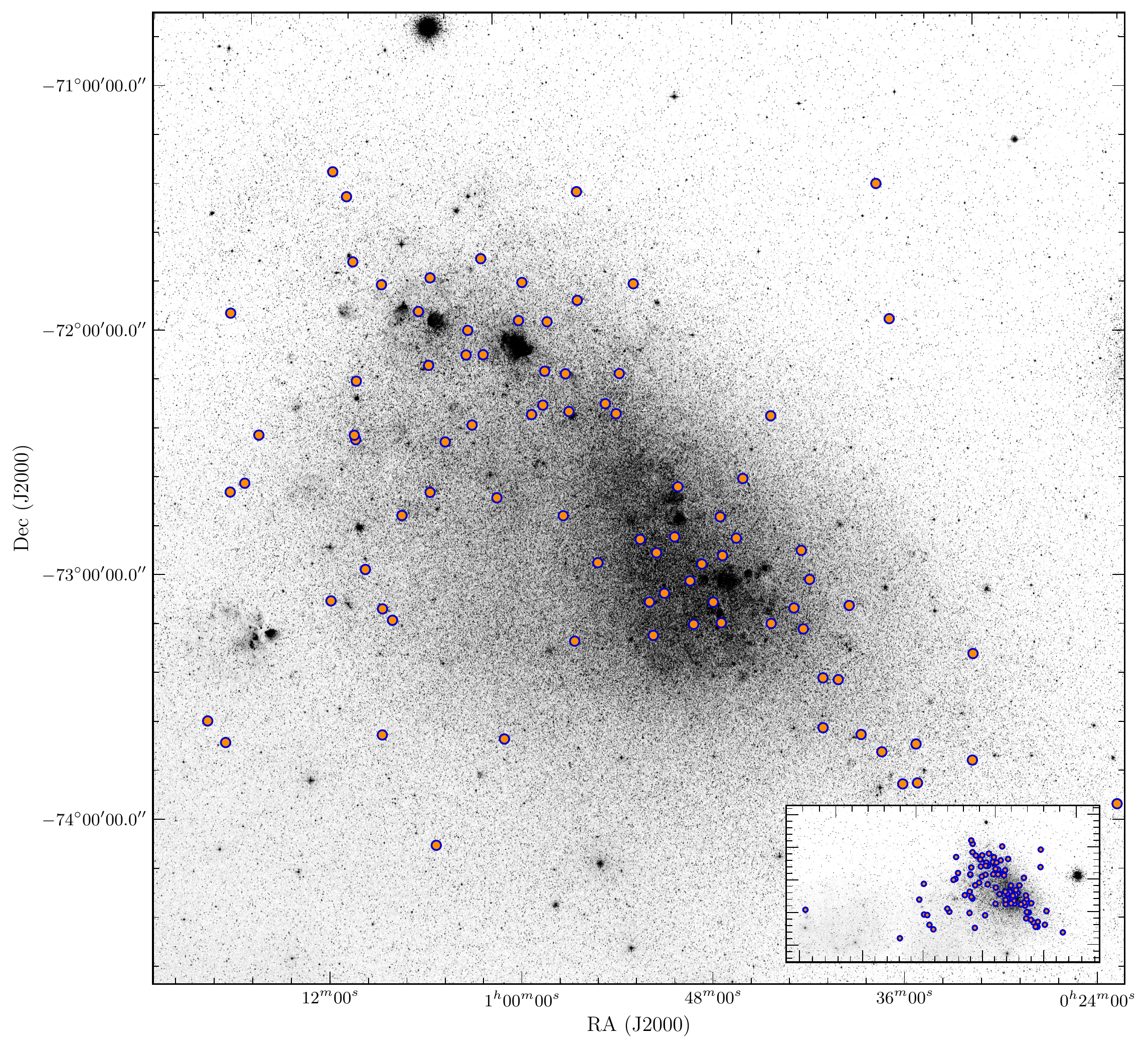}
\caption{DSS2 image of the SMC with the locations of the CHP Cepheids indicated by circles. The main image shows the body of the SMC. The inset image covers $10^{\circ}$ in RA and $5^{\circ}$ in Dec, showing all the Cepheids in the sample.}
\label{fig:smc_image}
 \end{center}
\end{figure}

The spatial distribution of the selected Cepheids is given in Figure~\ref{fig:smc_image}. The main figure shows the body of the SMC. The inset shows a larger region covering $10^{\circ} \times 5^{\circ}$ to display our full sample of Cepheids. Although some Cepheids are several degrees from the center they are confirmed members of the SMC.

\subsection{Warm Spitzer Data}
\label{sec:warm_spitzer}

Ninety fundamental mode SMC type I (and two type II) Cepheids were observed with \spitzer between August 2010 and January 2012 (Program ID 70010, P.I. Madore). Each light curve has twelve epochs, each spaced by approximately $P/12$ days, where $P$ is the Cepheid's pulsation period. This methodology is similar to the CHP LMC program \citep[][hereafter S11]{2011ApJ...743...76S}, but with twelve epochs as opposed the the LMC's twenty-four. We made the decision to reduce the number of observations for the SMC as we already know Cepheid light and color curves extremely well from the LMC and MW studies \citep[][hereafter M12]{2012ApJ...759..146M}. By using equally spaced phase--points for these well defined systems we can obtain highly accurate mean magnitudes and colors without overburdening the \spitzer schedule. Details of how the phasing of observations affects the uncertainty of mean magnitudes are given in the appendix of S11. 

The data were reduced identically to the LMC data in S11; the full details can be found in that paper. In summary, the photometry was performed using the Point Response Function (PRF) model provided by the Spitzer Science Center (SSC), using the \mopex and \apex software \citep{2005PASP..117.1113M}. Our photometry is calibrated to the standard system set by \citet{2005PASP..117..978R}. The photometry of the individual data points is given in Table~\ref{tab:photometric_data}\footnote{The full data set is available with the online version of the paper.}

An example IRAC light curve is shown in Figure~\ref{fig:light_curves}; the full sample can be found in Figure~\ref{fig:light_curves_appendix} in the Appendix. The light curves were fit using GLOESS, a Gaussian local estimation algorithm which has been used throughout the CHP (e.g. see S11, M12). The mean--light mid--infrared magnitudes are shown in Table~\ref{tab:midir_phot}.

Also shown in Figure~\ref{fig:light_curves} is the $[3.6]-[4.5]$ color curve. The change in color through the Cepheid's pulsation cycle is induced by the change in opacity due to dissociation and recombination of CO in the Cepheid's atmosphere. There is evidence that the MW (M12), LMC (S11) and SMC have distinct period--color relations in the mid--infrared. We will explore this in Scowcroft et al.~(2015) where we compare the mid--IR properties of the Cepheids in the three galaxies. It is important to note that this effect does not have an effect on the analysis presented in this paper. The effect is confined to the 4.5~$\mu$m band, which we do not use for distance measurements in this paper. 

\subsection{Archival Data}
\label{sec:archival}

In addition to the mid--IR data from \textit{Spitzer}, archival data for the Cepheids at all wavelengths from $U$ to $K$ were compiled from the literature; the sources are listed in Table~\ref{tab:archival}. The archival data span several decades, long enough for the period of a Cepheid to change by a noticeable amount. To take this into account the data were phased using the period at the time of the observations. The light curves for all bands were fit using GLOESS. 

\section{Results}
\label{sec:results}

\begin{figure}
\begin{center}
\includegraphics[width=180mm]{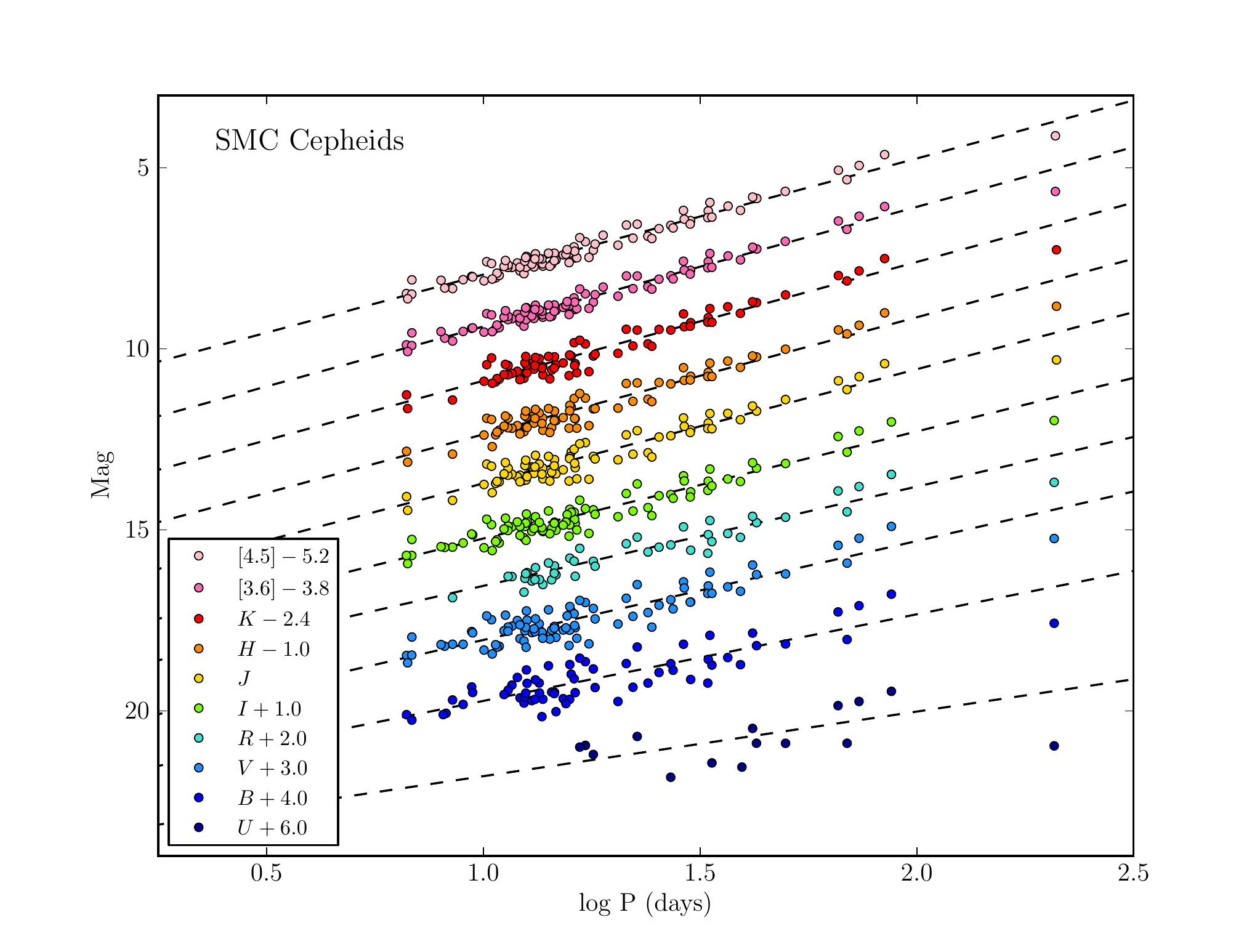}
\caption{Leavitt laws for the 10 bands from $U$ to $[4.5]$. The drop in dispersion and increase in slope in moving from the optical to the mid--infrared can clearly be seen.}
\label{fig:10bandpl}
 \end{center}
\end{figure}

Figure~\ref{fig:10bandpl} and Table~\ref{tab:plfits} give the Leavitt laws (LL) for the 10 bands from U to [4.5].
%We constructed LLs for the 10 bands from $U$ to $[4.5]$, these are shown in Table~\ref{tab:plfits} and Figure~\ref{fig:10bandpl}. 
The LLs take the form
\begin{equation}
m_{i}= a_{i} (\log P - 1.0) + b_{i} \text{,}
\end{equation}
\label{eqn:plform}
where $m_{i}$ is the apparent magnitude of the Cepheid in band $i$. For the nine longest wavelengths the relations were fit by fixing the slopes to the literature values for LMC Cepheids --- \citet{2007A&A...476...73F} for $B$ to $I$, \citet{2004AJ....128.2239P} for $J$ to $K$, and \citet{2012ApJ...759..146M} for the mid--IR\footnote{The optical and near--IR LLs were transformed to use $a_{i}(\log P - 1.0)$ instead of $a_{i}\log P$ to be consistent with the mid--IR fits.}. The zero--points of the relations were derived using unweighted least squares fits. For the $U$ band there is no published relation, so the slope and zero--point were both fit. 

The decision to use the LMC LLs as fiducial was made in Monson et al. (2012), where we compared the slopes of the LMC and MW midÑIR PL relations and found them to be the same. We therefore came to the conclusion that the entirety of the CHP project we would adopt the LMC derived relations. The LMC was chosen as it was derived from more Cepheids, hence has an smaller uncertainty.

It is possible that the slight changes in slope between the LMC and SMC LL that are sometimes reported \citep[for example, see table 4 of][]{2010ApJ...715..277B} are due to the increased dispersion in the SMC PL from distance effects. This would be exacerbated at shorter wavelengths where the effect of extinction is larger and is harder to disentangle from the structural properties of the galaxy.

The distance modulus from each band was derived using the same literature fits and assuming an LMC distance modulus of $18.48 \pm 
0.01$~mag \citep{2011AJ....142..192F, 2012ApJ...759..146M}.
The fit parameters and distance moduli are shown in Table~\ref{tab:plfits}. There is 
no $U$ band distance modulus as there was no canonical relation with which to compare our fit. We have included our relation for 
completeness. 

The increasing slope and decreasing dispersion of the relation as a function of wavelength can clearly be seen, confirming our expectations regarding LLs in the mid--IR. This was our motivation for moving to the mid--infrared --- at $3.6~\mu$m the dispersion in the LL is so low that we can obtain distance moduli for individual Cepheids accurate to 0.10~mag, or 4.7\% in distance. However, as Table~\ref{tab:plfits} shows, the $[3.6]$ LL has a dispersion of 0.16~mag, as opposed to the 0.10~mag dispersion seen in the LMC (S11). This is due to the spread in distances of SMC Cepheids, as discussed in Section~\ref{sec:smc_structure}.

\begin{figure}
\begin{center}
\includegraphics[width=180mm]{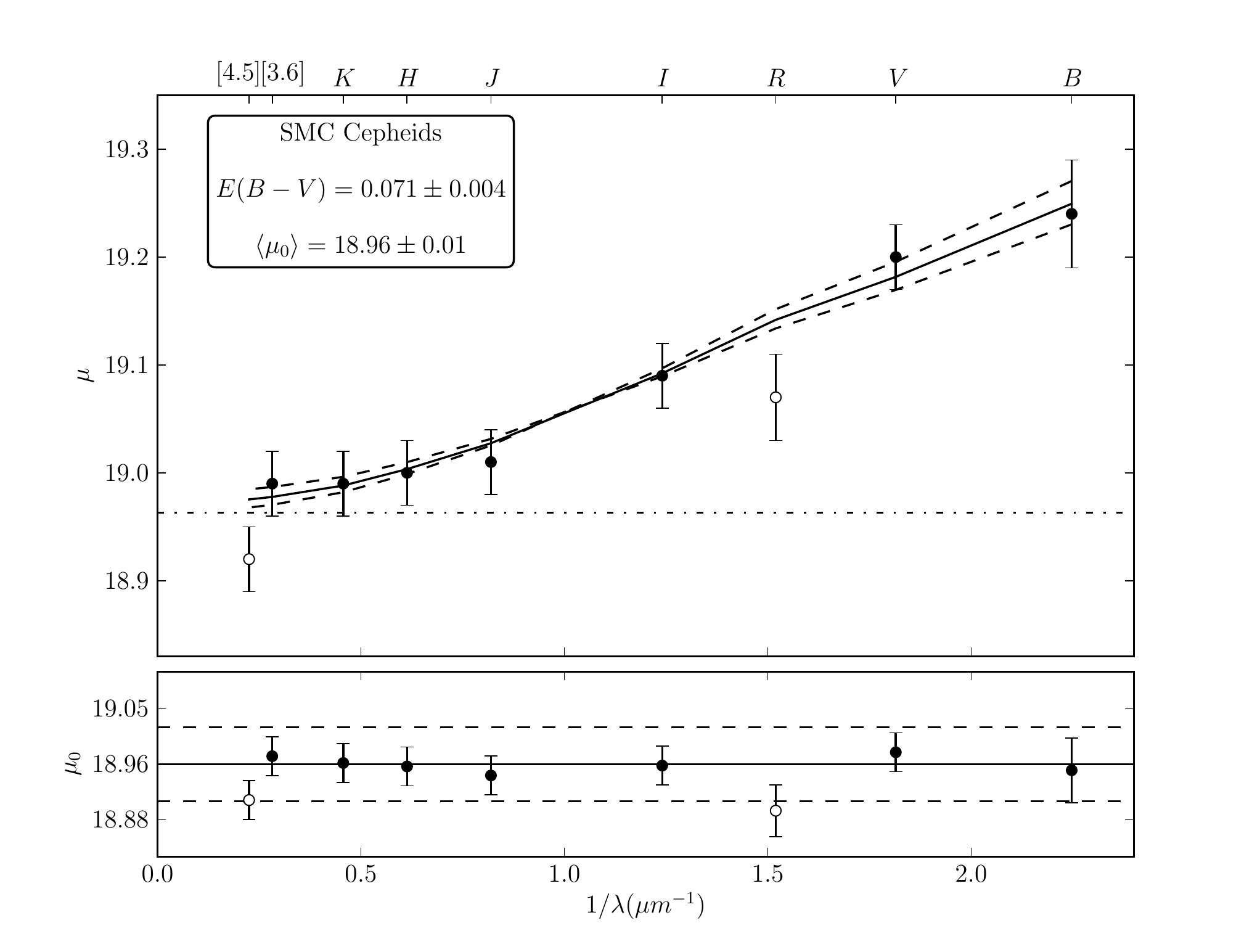}
\caption{Reddening fit to the distance moduli from Table~\ref{tab:plfits}. We assume $\mu_{LMC} = 18.48$~mag, $R_{V} = 3.1$, and fit the optical and near--IR \citep{1989ApJ...345..245C} and mid--IR reddening laws \citep{2005ApJ...619..931I} simultaneously.  
In the top panel the solid line shows the best fit reddening law. The dashed lines show how the fit changes if $E(B-V)$ is changed by $2\sigma$. The dashed-dot line denotes  $\mu_{0}$. The bottom panel shows each $\mu$ value corrected for extinction. The horizontal lines show $\mu_{0}\pm 2\sigma$. 
The open points ($R$ and $[4.5]$) were not used to derive $E(B-V)$ or $\mu_{0}$ but are included on the plot for completeness. The phenomenon of the $[4.5]$ distance modulus falling significantly below the curve (i.e. $[4.5]$ appears too bright) has been documented previously \citep{2013ApJ...773..106S} and is due to the ionization and recombination of CO in the Cepheid's atmosphere affecting the [4.5] magnitude. The $R$ distance modulus was removed from the fit as this band has the least data archival data available and does not uniformly fill the width of the Leavitt Law distribution. The non-uniformity of the $R$--band sample is demonstrated in Figure~\ref{fig:reddening_fit_rcut}.}
\label{fig:reddening_fit}
\end{center}
\end{figure}

The nine bands for which distance moduli are available can be combined to produce an estimate of the mean $E(B-V)$ and mean reddening--corrected distance modulus of the galaxy. We assume the ratio of total to selective absorption, $R_{V} = 3.1$, and fit three reddening laws simultaneously to the data presented in Table~\ref{tab:plfits}. For the optical and near--infrared bands we use the appropriate relations from \citet{1989ApJ...345..245C} and for the mid--infrared we use the relation from \citet{2005ApJ...619..931I}. The relations are fit by minimizing the dispersion of the distance moduli around the values predicted by the reddening law. The resulting fit is shown in Figure~\ref{fig:reddening_fit}.

As can be seen in Figure~\ref{fig:reddening_fit}, the $R$ and $[4.5]$ band data points lie furthest from the reddening prediction. The deviation of the $4.5~\mu$m point is expected; temperature changes through the Cepheid's pulsation cycle cause CO in the Cepheid's atmosphere to dissociate and recombine, which changes the opacity at $4.5~\mu$m. This has been noted in our previous works \citep[S11, M12,][]{2013ApJ...773..106S}. We believe that the magnitude and direction of this effect ($\Delta\mu_{CO}$) is dependent on metallicity, and will discuss this in Scowcroft et al.~(2015).

\begin{figure}
\begin{center}
\includegraphics[width=180mm]{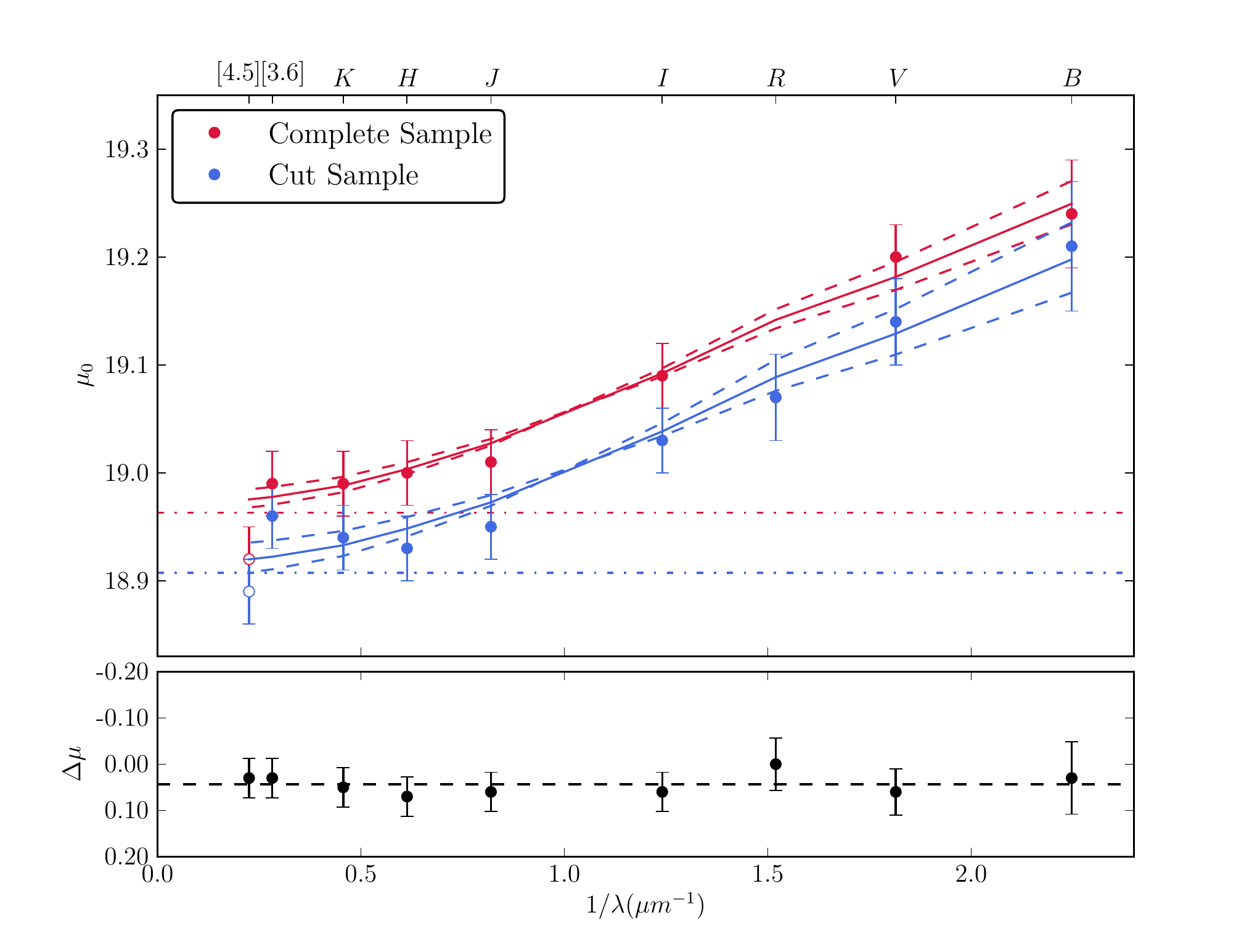}
\caption{Reddening fit to the distance moduli from Table~\ref{tab:plfits_rcuts}. The red points show the distance moduli from the complete samples, the blue points show the distance moduli derived only using Cepheids that were observed in all nine bands. The $R$ band point is now in line with the rest of the data, but they have all been systematically shifted to smaller distances. This is because the cut sample does not uniformly sample the Cepheid instability strip nor the range in distances. 
The bottom panel shows the offset at each waveband. The mean offset induced by this effect is $\Delta\mu=\mu_{com} - \mu_{min} = 0.04 \pm 0.02$~mag.}
\label{fig:reddening_fit_rcut}
\end{center}
\end{figure}

The $R$ band distance modulus also lies far from the prediction. The $R$ band has the least archival data available with 44 Cepheids compared to the full sample of 90 stars. We tested whether this could be the cause of the discrepant distance modulus by fitting each LL zero--point in the same manner as before, but this time only using the 44 stars that have $R$ band data. The results are given in Table~\ref{tab:plfits_rcuts}. The $R$ band sample does not uniformly sample the Cepheid instability strip and is slightly biased towards the bright edge inducing a bias of $0.04 \pm 0.02$~mag towards lower distance moduli when the smaller sample is used. This is demonstrated in Figure~\ref{fig:reddening_fit_rcut}.

%The $R$ band distance modulus also lies far from the prediction. The $R$ band has the least archival data available with 44 Cepheids compared to the full sample of 90 stars. We tested whether this could be the cause of the discrepant distance modulus by fitting each LL zero--point in the same manner as before, but this time only using the 44 stars that have $R$ band data. The results are given in Table~\ref{tab:plfits_rcuts}. The $R$ band sample does not uniformly sample the Cepheid instability strip and is slightly biased towards the bright edge inducing a bias of $0.04 \pm 0.02$~mag towards lower distance moduli when the smaller sample is used. This is demonstrated in Figure~\ref{fig:reddening_fit_rcut}.

%The fitting process was tested using all nine bands and removing either or both of the $R$ or $[4.5]$. In each case the resulting distance modulus changed by less than 0.01 mag and $E(B-V)$ by the same amount. 

We decided to remove the $R$ and $[4.5]$ bands from the fit in our final result. Our estimate for the mean reddening--corrected distance modulus of the SMC is $\mu_{0} = 18.96 \pm 0.01$~mag, with a color excess $E(B-V) = 0.071 \pm 0.004$~mag, assuming an LMC distance modulus of 18.48~mag.

\section{Discussion}
\label{sec:discussion}
\subsection{Mean Distance of the SMC}
\label{sec:smc_distance}
As we discussed in our previous work on IC~1613 \citep{2013ApJ...773..106S}, comparing distance measurements from different techniques can prove problematic. 
Both systematic and statistical uncertainties in both the relative distance measurements (e.g. to the LMC) and their assumed zero--points contribute to the total uncertainty in the actual (physical) distance to the SMC. Differential measurements, on the other hand, can clearly be more easily compared with the results of others.

The mean reddening--corrected distance modulus derived in the previous section ($\mu_{0} = 18.96 \pm 0.01$~mag) assumes an LMC distance modulus of $18.48 \pm 0.01_{stat} \pm 0.03_{stat}$~mag. The differential distance modulus is $\Delta\mu = \mu_{SMC} - \mu_{LMC} = 0.48 \pm 0.01$~mag. 
 
Table 11 of \citet{2014ApJ...780...59G} compares their eclipsing binary differential distance measurement with several other techniques. The expectation value of the gaussian mixture distribution of these results is $0.47 \pm 0.02$~mag, which agrees with our result of $0.48 \pm 0.01$~mag.

%The mean reddening--corrected distance modulus derived in the previous section ($\mu_{0} = 18.96 \pm 0.01$~mag) is consistent with many recent measurements of the SMC with different techniques. The result leads to a distance between the MCs of $\Delta\mu = \mu_{SMC} - \mu_{LMC} = 0.48 \pm 0.02$~mag. 

%Different techniques have different systematic uncertainties, and may have been calibrated to different zero--points, for example different fiducial distances to the LMC. One way to take this into account is to make a differential comparison; in this case between the LMC and SMC distances. By using a differential measurement, the effect of different zero--points disappears and we can make a meaningful comparison between distance indicators. 

%In Figure~\ref{fig:smc_distance_comparison} we plot those measurements (thin, black lines) along with our own estimate (thicker red line) and the gaussian mixture distribution (defined as the frequentist sum of the probability density functions). Each of the measurements is represented by a unit--area gaussian, with the uncertainty ($\sigma$) related to the full--width at half--maximum of the distribution by $FWHM = 2\sqrt{2 \ln 2} \sigma$. The gaussian mixture distribution shows that there is still some dispersion in the measurements, but that they do appear to be approaching a consensus. The expectation value of the mixture distribution is $0.47 \pm 0.02$~mag, which agrees with our result of $0.48 \pm 0.02$~mag.

\subsection{Structure of the SMC}
\label{sec:smc_structure}
When considering the distance to the SMC as defined by a given tracer population, it is not only the average distance, but also its dispersion, that bears study. The SMC is known to have a large line--of--sight depth; \citet{2012ApJ...744..128S} found that the SMC is elongated from the NE -- SW, and has a tidal radius of 7 -- 12~kpc. Most recently, \citet{2013ApJ...779..145N} examined SMC red clump stars in several regions and found line--of--sight depths of 23~kpc in the eastern regions.
%When considering the distance of the SMC we must not only think about the average distance, but the dispersion of the distances of the individual points around the average. 

The major advantage of studying Cepheids in the mid--infrared is that we are in a wavelength regime where the width of the instability strip is intrinsically small. This translates directly into the small width of the mid--IR LL. In our work on the MW and LMC we demonstrated that the intrinsic width of the $3.6~\mu$m LL is 0.10~mag. We purposely designed our SMC observations to achieve mean magnitudes with equally small photometric uncertainties so that any additional dispersion we observe in the LL would be due entirely to depth effects. The dispersion we see in the SMC $3.6~\mu$m LL is 0.16~mag, thus we conclude that the depth effects are producing a scatter of $\pm 0.125$~mag. This would correspond to an additional 3.5~kpc.

\begin{figure}
\begin{center}
\includegraphics[width=180mm]{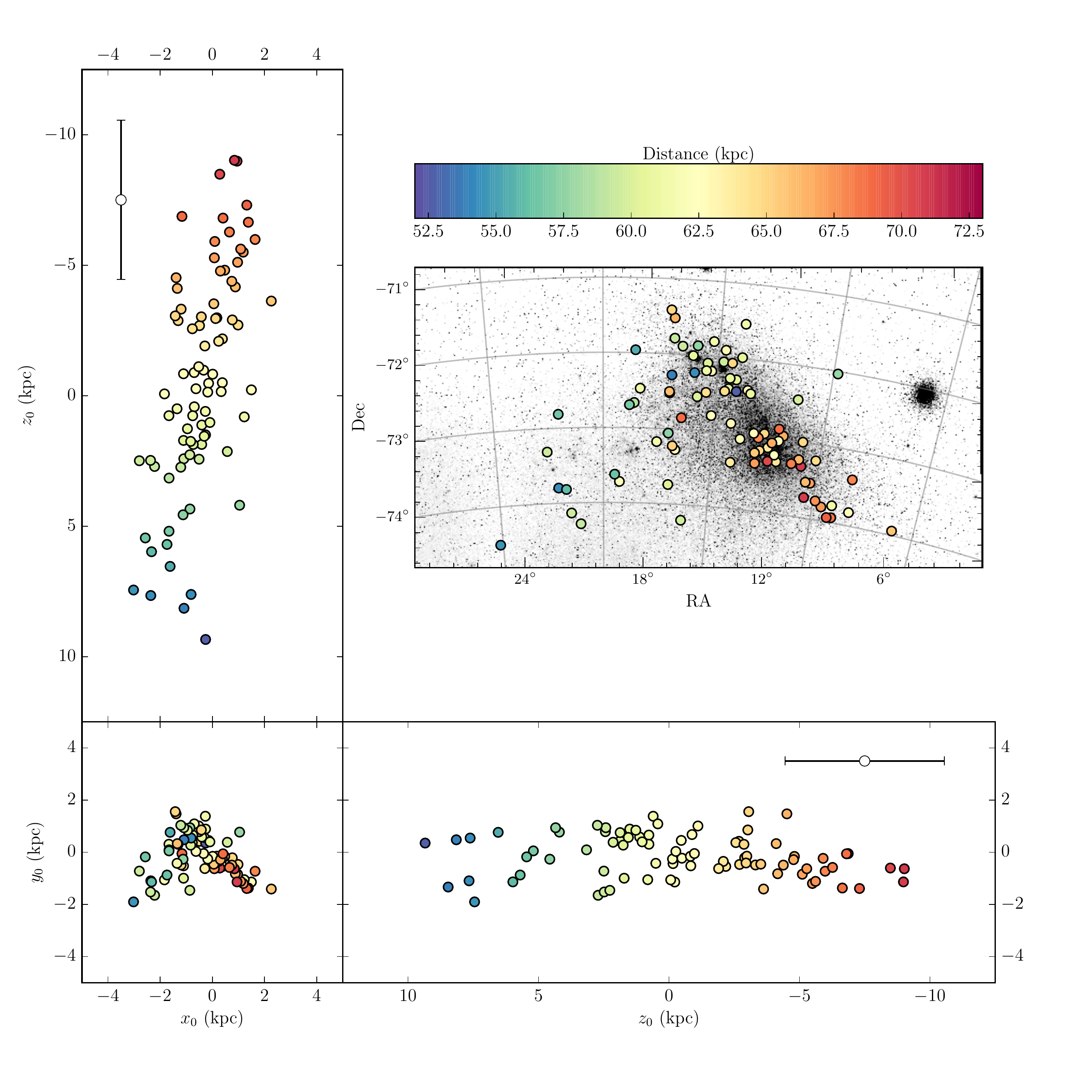}
\caption{Three--dimensional structure of the SMC traced by Cepheids. The coordinates are in units of kpc from the center of the SMC ($x_0, y_0, z_0$ corresponds to $\alpha=00^{h}52^{m}44.8^{s}$, $\delta=-72^{\circ}49'43''$, $R_{SMC} = 61.94$~kpc), following the transformation defined by \citet{2001ApJ...548..712W}. The white points at $z_{0}=-7.5$~kpc show the typical uncertainty on the individual distances. The color-coding of the points represents the distance of each Cepheid from Earth. Although the SMC appears to have an irregular shape on the sky, as evidenced by the bottom left and top right plots, it is clear that the Cepheids actually trace an 'elongated sausage' shape, with a slight tilt to the west. The irregular on--sky shape is due to projection effects. The distances of the individual Cepheids have been dereddened as described in Figures~\ref{fig:ind_reds} and ~\ref{fig:reddening_map}. }
%Tilt =  -0.132244665189 degrees / kpc
\label{fig:3d_smc}
\end{center}
\end{figure}

In Figure~\ref{fig:3d_smc} we break down the three-dimensional structure of the SMC as traced by the Cepheids.  The distances were calculated for each individual Cepheid using the LMC $3.6~\mu$m LL. The individual Cepheids were dereddened as described later in this section. To properly examine the three-dimensional structure, we first transform the coordinate system from $\alpha$, $\delta$, $\mu$ to the cartesian system $x_0, y_0, z_0$ described in the appendix of \citet{2001ApJ...548..712W}. The center of the SMC is defined as $\alpha_{0} = 00^{h}52^{m}44.8^{s}, \delta_{0} = -72^{\circ}49^{m}43^{s}$\footnote{Center of SMC as defined by NED: \url{http://ned.ipac.caltech.edu/ngi/?objectname=smc}}, $R_{SMC} = 61.94$~kpc.

The three cartesian plots show the SMC split into the $x_0, y_0$ and $z_0$ components. The axes are to scale, such that the galaxy truly is almost 5 times more extended in the $z_0$ direction than the $x_0$ and $y_0$ directions. The central plot shows the positions and distances of the Cepheids as seen on--sky. In each panel the Cepheids are color--coded according to distance. From this figure we can see that there is a significant line--of--sight elongation along the E--W axis, with the E region being $\sim20$~kpc closer than the W region. 
%While there is only a slight trend seen in the distance vs. $\Delta \delta$ plot, it is clear that there is a significant line--of--sight elongation in the distance vs. $\Delta \alpha$ plane. From this figure we can see that there is a significant line--of--sight elongation along the E--W axis, with the E region being $\sim20$~kpc closer than the W region. The dashed line shows the fitted $\Delta \alpha$ distance gradient of $-7.56$~kpc~degree$^{-1}$.
 
Although we are in the low--extinction mid--IR region, it is imperative that we confirm the spread in distance is not due to differential reddening. The $A_{V}$ values for each star were calculated in the same way as in Figure~\ref{fig:reddening_fit}, this time fitting every star individually and only using OGLE $V$ and $I$ plus \textit{Spitzer} $3.6~\mu$m observations. This ensures that the individual reddenings are on a consistent scale. 

\begin{figure}
\begin{center}
\includegraphics[width=180mm]{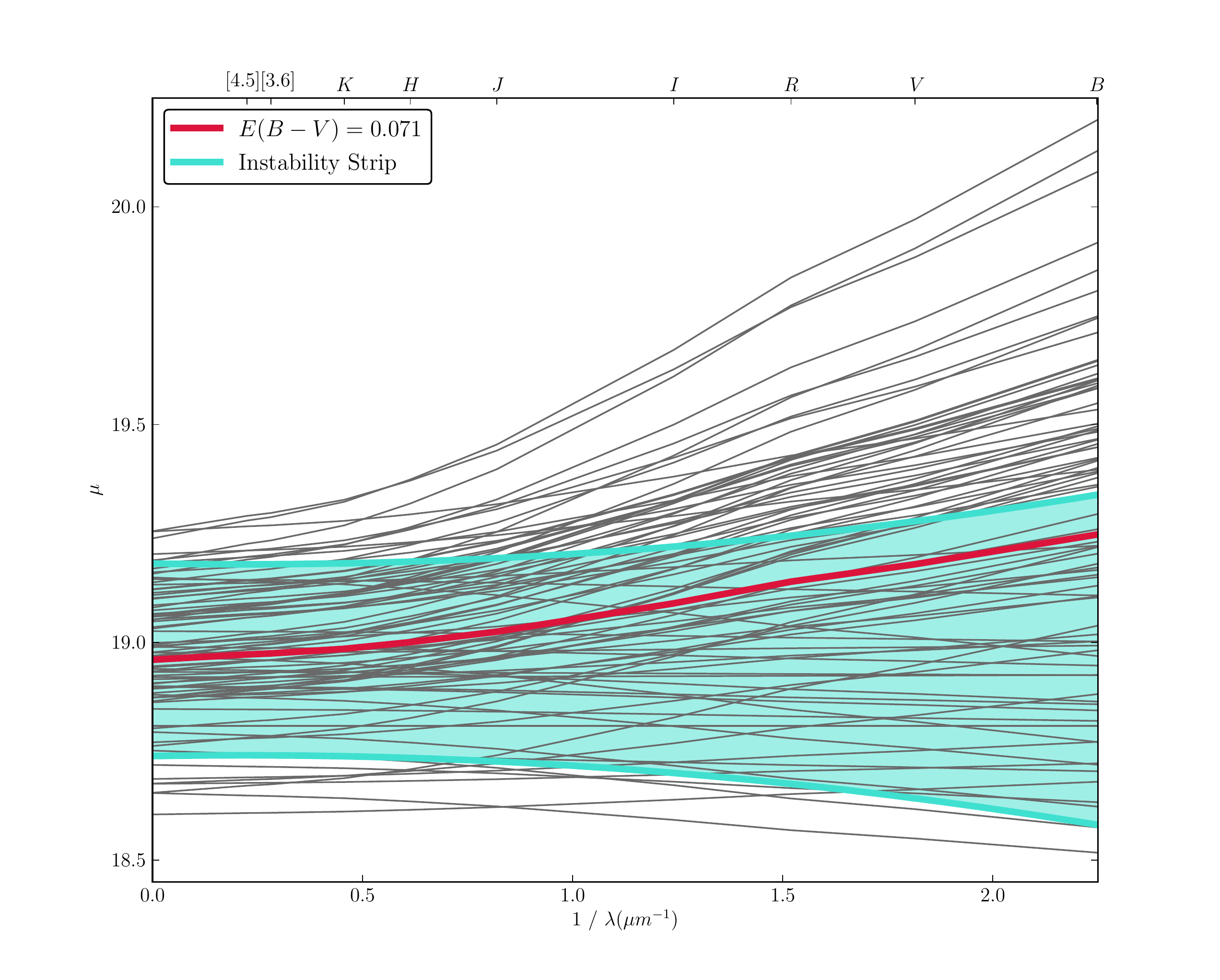}
\caption{Pseudo--reddening fits for individual Cepheids with OGLE $V, I$ and \textit{Spitzer} [3.6] data available (gray lines). The dispersion introduced by the intrinsic width of the Cepheid Instability Strip (the mean Leavitt law $\pm 2 \sigma_{LMC}$) is shown by the blue shaded region. The red line denotes the mean distance modulus with the mean extinction of $E(B-V) = 0.071$~mag applied. The grey lines have a significantly larger dispersion at $1 / \lambda = 0.0$ than is predicted from the width of the instability strip. At  At this wavelength the effect of extinction should be minimal. This demonstrates once again that the large dispersion in the SMC LLs and derived distances must come from the depth of the SMC.}
%Tilt =  -0.132244665189 degrees / kpc
\label{fig:ind_reds}
\end{center}
\end{figure}

The reddening fits to each star are shown in Figure~\ref{fig:ind_reds}. In this Figure we also show the dispersion that the intrinsic width of the Cepheid instability strip (IS) will cause at different wavelengths. As is expected, the dispersion of the IS increases with decreasing wavelength (increasing $1\lambda$). When Cepheids are considered individually rather than as an ensemble population we must also consider the effect that the with of the IS will have on the Cepheid's magnitude. When we derive the reddening of a population in this manner the IS effects are averaged over, but for individual stars where the reddening is comparable to the dispersion induced by the IS width we can only derive a pseudo--reddening.

If the large range of measured distance moduli in the SMC were due to reddening then we would see an extreme range of extinctions in Figure~\ref{fig:ind_reds}. However, at $1/\lambda=0.0$, where extinction is minimized we see a large range of distance moduli -- larger than can be explained by the width of the IS. If the width of the IS were the explanation then 87 of the 90 Cepheids (97\% of the sample) would lie within the blue shaded region at $1/\lambda=0.0$. Over twice the expected number lie outside the shaded region. This shows that neither the IS width nor large reddenings of individual stars can explain the large range of derived Cepheid distances.

\begin{figure}
\begin{center}
\includegraphics[width=180mm]{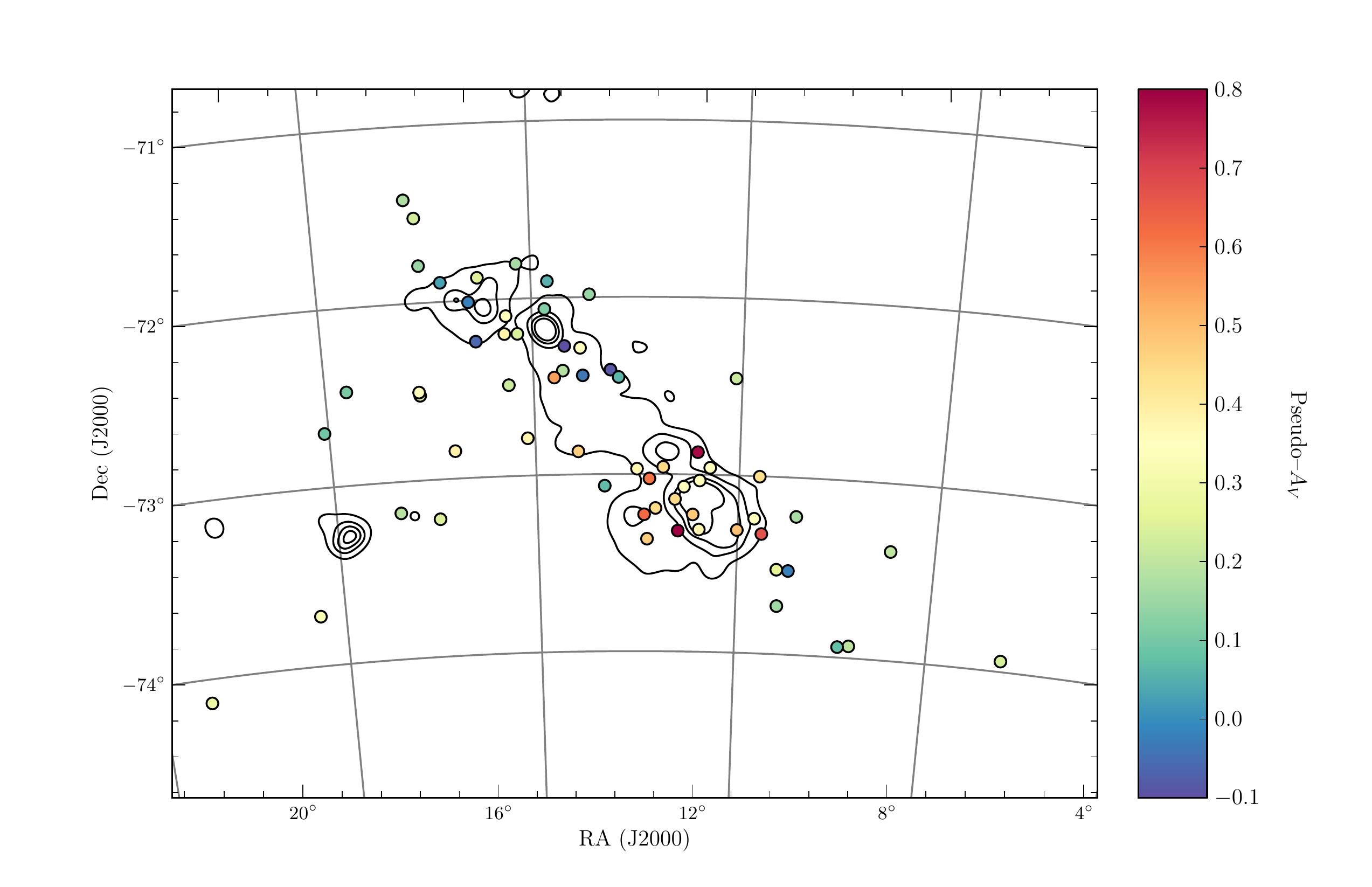}
\caption{Reddening map of the SMC. $A_{V}$ values were determined using the OGLE $V$ and $I$ and \textit{Spitzer} $3.6~\mu$m mean magnitudes. The contours show the $100~\mu$m IRAS image of the SMC, representing cool dust. The most extincted regions line up with the highest density of dust. Note that although the SW has the highest $A_{V}$ and highest density of dust, the extinction here is over an order of magnitude too low to explain the difference in distance.}
%Tilt =  -0.132244665189 degrees / kpc
\label{fig:reddening_map}
\end{center}
\end{figure}

In Figure~\ref{fig:reddening_map} we show $A_{V}$ for the Cepheids as a function of position. The color--coding represents $A_{V}$; the contours show the IRAS $100~\mu$m image of the galaxy, corresponding to cool dust\footnote{$100~\mu$m image from \url{http://dirty.as.arizona.edu/$\sim$kgordon/research/mc/smc\_iras.html}}. The $A_{V}$ values range between 0.1 and 0.8~mag, with the highest extinction regions lining up with the highest density of cool dust. However, these levels of extinction cannot explain the large spread in derived distances of the Cepheids. An $A_{V}$ of 0.8 corresponds to an $A_{3.6}$ of 0.07. It is possible that derived distances of the most extincted Cepheids in the SMC are slightly affected by extinction, but this effect is not sufficient to explain the full range. To account for the full 20~kpc range an $A_{V}$ of over 10~mag is required. 

\begin{figure} 
 \begin{center} 
\includemedia[
	width=\textwidth,
	height=0.5\textwidth,
	activate=onclick,
	3Dtoolbar,
	3Dviews=f9a.vws
] {\includegraphics[width=0.8\textwidth]{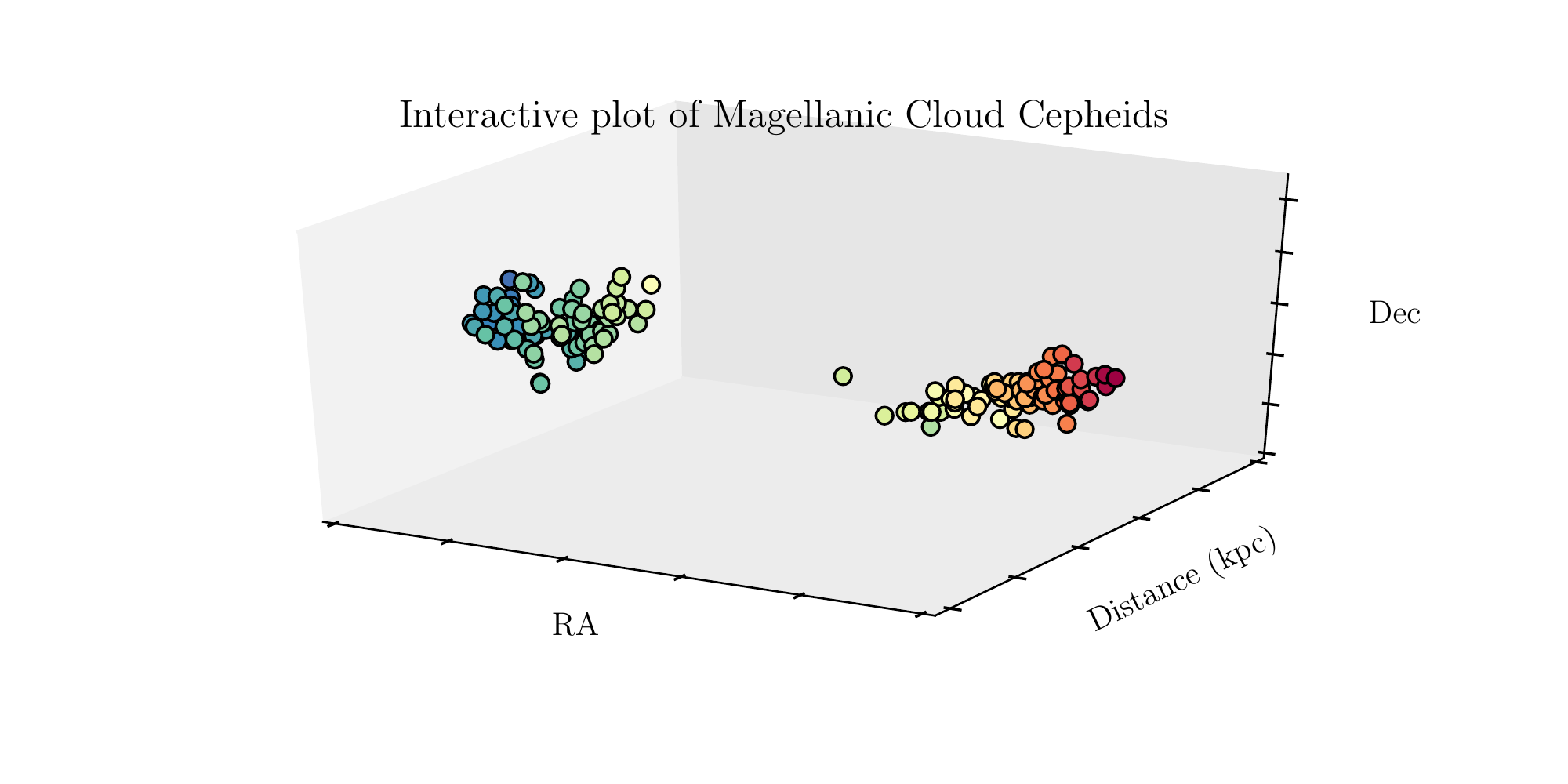}} {f9a.u3d} 
\mbox{\bf (a)} 
%\caption{Interactive 3D plot of the Magellanic Cloud Cepheids}
%\label{fig:interactive_plot}
 \includegraphics[width=180mm]{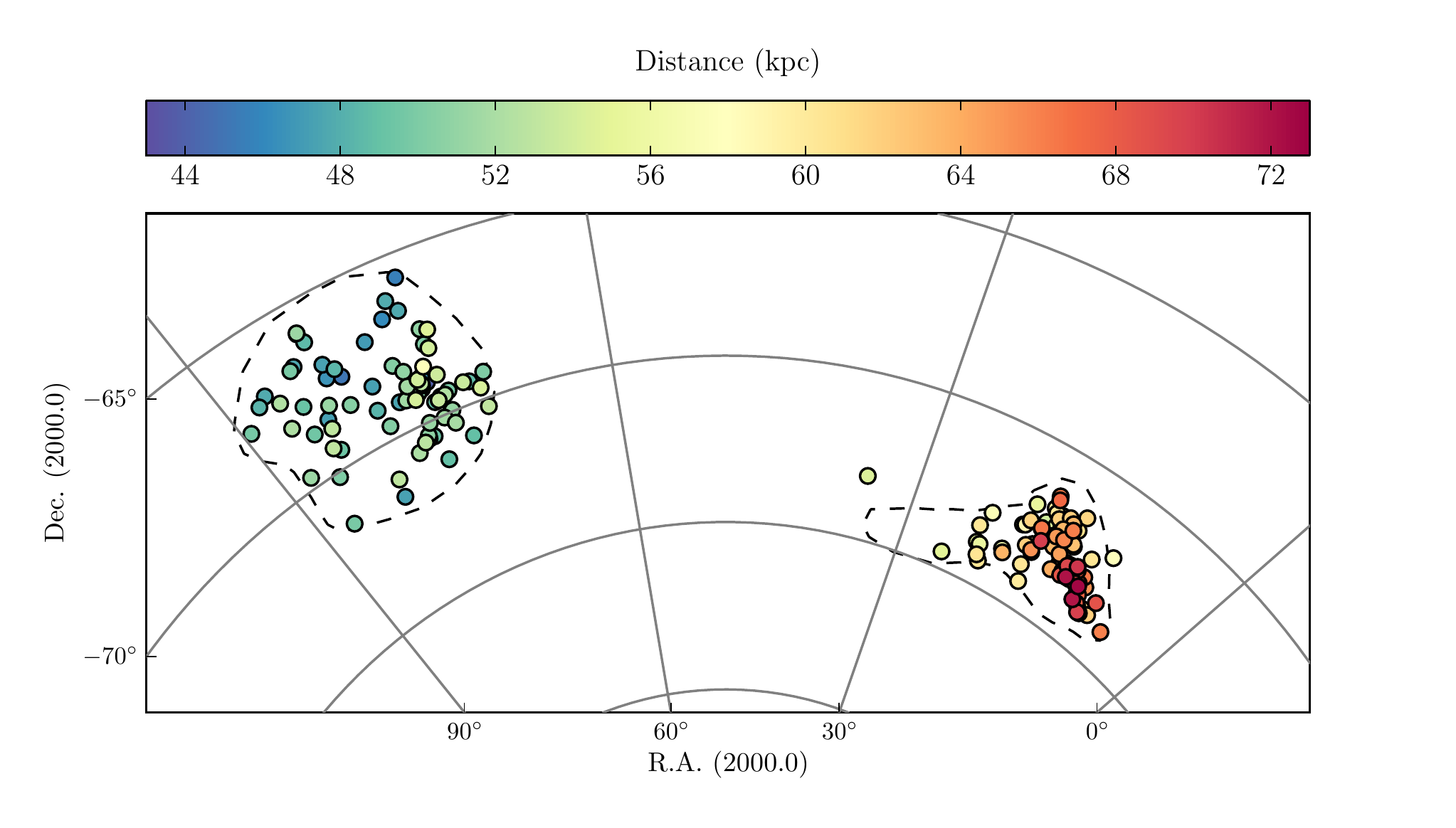} 
\mbox{\bf (b)}

% \caption{On--sky projection of Magellanic cloud Cepheids}
\end{center} 
\caption{(a) Interactive 3D plot of the Magellanic Cloud Cepheids (b) On--sky projection of Magellanic cloud Cepheids.  An interactive version Figure 9a is available in the online journal.}\label{fig:3d_structure}
\end{figure}

In Figure~\ref{fig:3d_structure} we explore the structure of the whole Magellanic Clouds system as seen through Cepheids. Figure~\ref{fig:3d_structure}a is an interactive 3D model of the Clouds\footnote{To access the interactive figure you must use Adobe Reader version 9 or newer. Click the figure to activate it.}; Figure~\ref{fig:3d_structure}b shows the 2D on--sky projection. The color--coding shows the distances of each Cepheid as derived from the LMC LL. The contours in Fig.~\ref{fig:3d_structure}b show the approximate outlines of the galaxies. 

In the initial view of Fig.~\ref{fig:3d_structure}a we show the RA (red axis) and Dec (blue axis) projection of the Cepheids in the LMC and SMC. Changing the view to RA and Distance (green axis) shows that the Cepheids in the SMC are being pulled towards the LMC, and that the closest SMC Cepheids are in fact at the same distance as the furthest LMC Cepheids. This is again made clear in the Dec -- Distance view, where the two samples become almost indistinguishable. By rotating the axes you can see that the SMC is extremely elongated compared to the LMC, and is clearly trailing off towards its neighbor.

The closest of the SMC Cepheids appear to be trailing off towards the Magellanic Bridge, connecting the LMC and SMC, suggesting that the elongated shape of the young population may be evidence of interaction between the Clouds.

Unlike previous works (for example,  \citet{2004ApJ...604..176S}, \citet{2012ApJ...744..128S}, \citet{2012AJ....144..107H}) we do not believe that the standard parameters of the position angle and inclination angle of a plane are appropriate to describe this galaxy. As can be seen in Figure~\ref{fig:3d_structure} and ~\ref{fig:3d_structure}a, the SMC is clearly a very disturbed galaxy and is not well described by a plane. In this case, the structure traced by the Cepheids trace a cylindrical shape, which we are viewing from one end. Previous measurements of the inclination and position angles of the SMC have showed significant spread, and have been highly dependent on which tracer is being used. \citet{2009IAUS..256...81V} makes the point that differences may also be arising in these results due to the differing spatial coverage of the tracers employed in the calculations. We believe that a detailed study of the SMC utilizing high precision distance measurements of Cepheids {\it and} RR Lyrae over the whole galaxy will help lay some of these issues to rest.

%% Removing the comparison with the Diaz simulation. Its not a fair comparison as the Cepheid observations were pointed. Or is it? They were only pointed in RA, Dec, not in distance? Need to fit the slope in the simulation and compare it to the slope we get in the observations. 

\subsubsection{Comparison with other recent structural studies}
\label{sec:structure_comparison}

Recently, \citet{2012AJ....144..107H} have studied the structure of the SMC using Cepheids and RR Lyrae in the optical bands. They performed a similar analysis to the one presented here, measuring the distances to the individual stars from the Cepheid LL and the RR Lyrae metallicity--luminosity relation. Their results for the geometry of the Cepheid population are very similar to the structure found here. They find that the majority of Cepheids have distances between 55 and 75~kpc and that the sample is distributed such that the more distant Cepheids lie in the south--western region. Interestingly, they do not see the same distribution for the old stellar population, i.e. the RR Lyrae stars, which are much more uniformly distributed. 

\citet{2012ApJ...744..128S} also performed a detailed analysis of the SMC structure using RR Lyrae and red clump (RC) stars. They found that these stars form a uniform spheroid with a line--of--sight depth of 14~kpc. They suggest that the distribution could be explained by the SMC experiencing a merger with another dwarf galaxy around 4--5~Gyr ago which took around 2--3~Gyr to complete. This resulted in a population of stars over 2~Gyr old forming the spheroidal structure. 

Although we do not study the old population in this work, our results do add weight to this theory. The star formation event triggered by the merger produced the shape seen in the old population. The young population would have been triggered by a different mechanism, resulting in the less spheroidal, more elongated volume. The extended shape of the SMC is predicted by theoretical models of the interactions of the Magellanic Clouds. In Section~\ref{sec:theory_comparison}, we compare our observations with such works. 

\subsubsection{Comparision with theoretical models}
\label{sec:theory_comparison}

In this section we consider how our observations match up with recent theoretical works. This will provide insight into the mechanism that produced the SMC ``wing" --- the region of the SMC that is being drawn off towards the LMC.

In Figure~\ref{fig:diaz_comparison} we compare the Cepheid positions with the SMC--spheroid model from \citet{2012ApJ...750...36D}. The contours represent the density of points in the simulation and the colored circles represent the SMC Cepheids. Although we did not cover as large an area with the CHP observations, it is clear that the Cepheids follow the distance gradient found in the simulation with the most distant Cepheids appearing in the most westerly regions. The distance gradient is a feature that is readily reproduced in Magellanic Cloud simulations.

To better understand the formation mechanism of the Magellanic Bridge and the SMC wing we look to \citet{2012MNRAS.421.2109B}. In this paper they consider two models of the Magellanic Cloud system: Model 1 where the clouds pass close to each other, and Model 2 where the SMC collides directly with the LMC and passes through it. In both cases the Clouds are on their first passing of the Milky Way. We compare our results to their Model 2 (henceforth known as the collision model) as it reproduces the column density of the Magellanic Bridge well, while Model 1 does not. It also reproduces the LMC's observed offset bar and one--armed spiral structure. 

One feature of the collision model that we would expect to see is tails of stellar material pulled out from each galaxy, left behind after the interaction. We see precisely this evidence in the interactive version of Figure~\ref{fig:3d_structure}. By rotating the figure it becomes clear that the regions of each galaxy that are closest to each other on the sky are also closest in distance. This is made clearest by changing to the Dec--Distance projection where the two galaxies become indistinguishable. The observations show that it is not just the SMC that has a wing; the LMC has a tail of material that leads off towards the SMC.

The collision model also predicts star formation along the Magellanic Bridge. Young stars (which are too young to have migrated from the SMC) have been observed in the Bridge, adding weight to this theory \citep{1998AJ....115..154D}. We intend to test this further by looking for Cepheids in the Bridge. This will not only help confirm the collision hypothesis but will also give us precise distances to points along the Bridge to help constrain the past trajectories of the galaxies.

The present models (\citeauthor{2012MNRAS.421.2109B}, \citeauthor{2012ApJ...750...36D} and others) cannot precisely reproduce all of the features of the Magellanic system. For instance, the collision model results in the Clouds being separated by 10~kpc which is much closer than the $\sim$20~kpc we observe. This has implications for the timing of the proposed collision, which can be tested by searching for the oldest stars in the Magellanic Bridge. \citet{2007ApJ...658..345H} found stars as old as 300~Myr in the Bridge; assuming these formed in situ this places an upper limit on the time of collision. Most recently, \citet{2014ApJ...795..108S} made the first detections of Cepheids in the Magellanic Bridge, as part of OGLE~IV. They found four short period classical Cepheids in the bridge, with tentative ages of 100--250~Myr. This is consistent with an interaction of the Magellanic Clouds in approximately the last 200~Myr. 

We have now reached a stage where we can measure precise distances to Cepheids on a star--by--star basis and use these exquisite three--dimensional maps to inform the simulations. Cepheids are no longer just to be used as aggregate distance indicators; they can be used to delve deep into the structure of galaxies and help us understand their dynamical histories.

\section{Conclusions}
\label{sec:conclusions}

Using \textit{Spitzer} observations of classical Cepheids  we have measured the SMC's average distance modulus to be $18.96 \pm 0.01$~mag, $0.48 \pm 0.01$~mag greater than the LMC. This is in agreement with previous results from Cepheid observations, as well as with measurements from other indicators such as RR Lyrae stars and the tip of the red giant branch. 

Using the mid--infrared has enabled us to derive some of the most precise distances to date to Cepheids within the SMC. Utilizing the small dispersion in the mid--IR Leavitt Law, we have examined the three-dimensional structure of the young stars in the Magellanic Clouds, confirming that the Cepheid distribution does not just have a large line--of--sight depth, but is elongated from the north--east to the south--west, such that the south-western side is up to 20~kpc more distant than the north--east. This is consistent with previous observational work, such as the studies by \citet{2012AJ....144..107H} and \citet{2012ApJ...744..128S}.

We also compare our work to the dynamical simulations of \citet{2012MNRAS.421.2109B}. The elongation of the young Cepheid population, compared to the spheroidal shape of the old population seen by  \citet{2012AJ....144..107H} and \citet{2012ApJ...744..128S} adds weight to the theory that the LMC and SMC underwent a direct collision in the recent past which triggered star formation in both galaxies and the Magellanic Bridge.

%The study presented here highlights the power of infra--red observations of variable stars. An essential follow--up to this work would be an infra--red study of the RR Lyrae stars in the Magellanic Clouds and Magellanic Bridge. In a wavelength regime essentially free from reddening and metallicity effects we can reveal the fine structure of the Magellanic Clouds. Although RR Lyrae in the SMC are beyond the capabilities of \textit{Spitzer}, they are ripe for study with the \textit{James Webb Space Telescope} (JWST). In the next decade, not only will variable star studies with \textit{JWST} and \textit{Spitzer} enable precision cosmology on the largest scale, but will provide us with the most precise studies of our nearest neighbors.

\section{Acknowledgements}
\label{sec:acknowledgements}
\begin{figure}
\begin{center}
\includegraphics[width=180mm]{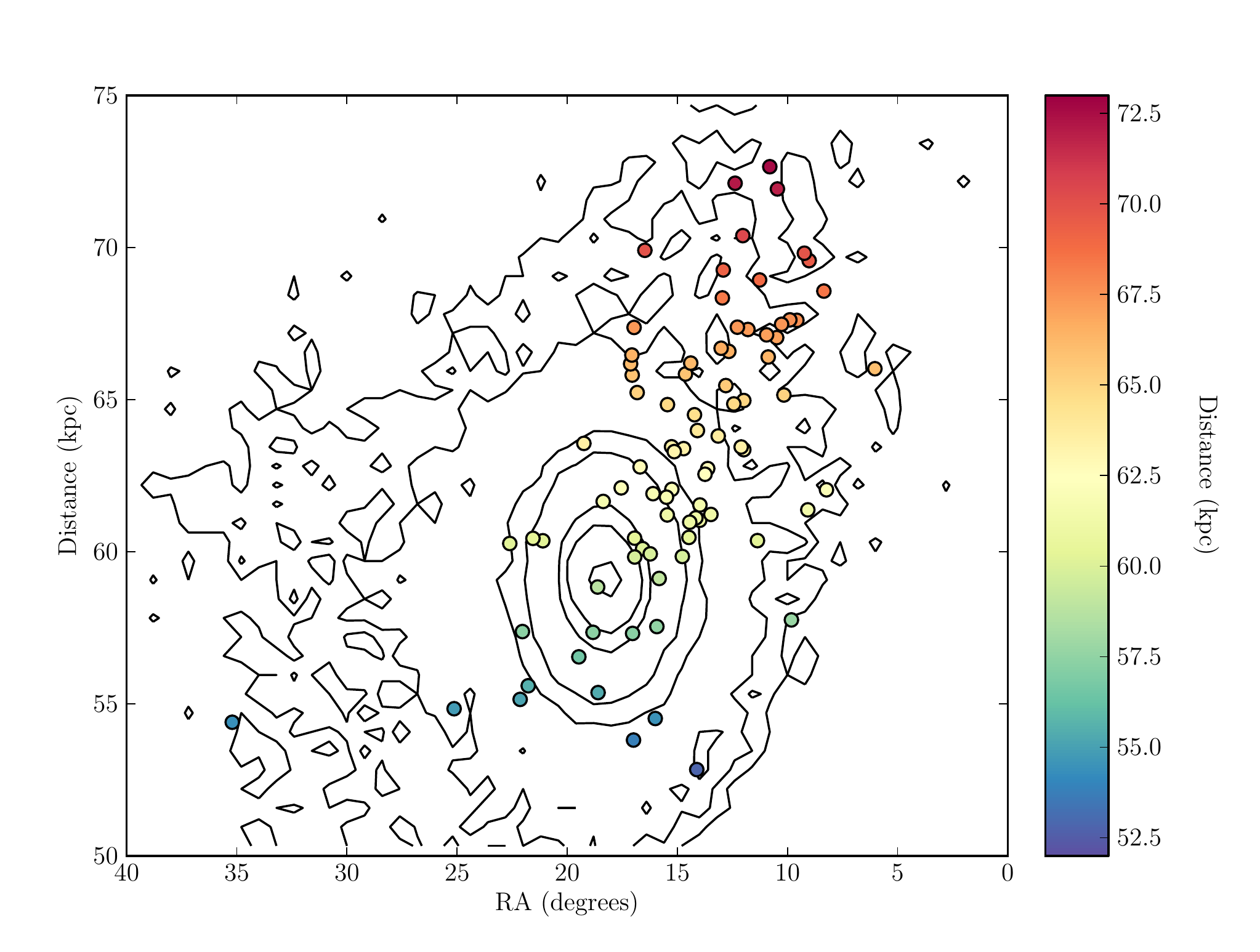}
\caption{Comparison of Cepheid distances with the SMC spheroid simulations from \citet{2012ApJ...750...36D}. The contours show the number of simulated objects in each bin. The colored points show the positions and distance of the Cepheids. Although the observed Cepheids cover a much smaller area than the simulation, there is good agreement with the distance gradient in the central region of the galaxy.}
\label{fig:diaz_comparison}
\end{center}
\end{figure}

We thank David Nidever for providing the simulation data used in Figure~\ref{fig:diaz_comparison}, and Gurtina Besla for fruitful discussions regarding the comparison of our work with theoretical models.
This work is based on observations made with the Spitzer Space Telescope, which is operated by the Jet Propulsion Laboratory, California Institute of Technology under a contract with NASA. Support for this work was provided by NASA through an award issued by JPL/Caltech. This research has made use of the NASA/IPAC Extragalactic Database (NED) which is operated by the Jet Propulsion Laboratory, California Institute of Technology, under contract with the National Aeronautics and Space Administration.  We acknowledge the use of NASA's \textit{SkyView} facility (http://skyview.gsfc.nasa.gov) located at NASA Goddard Space Flight Center.

{\it Facilities:} \facility{Spitzer}.

\label{references}
\bibliography{scowcroft2014}

\begin{table}
\begin{center}
\begin{tabular}{ l c c c c c} \hline
Cepheid ID & MHJD & [3.6] & $\sigma_[3.6]$ &  [4.5] & $\sigma_[4.5]$  \\ \hline \hline
HV00817 & 55434.902282 & 12.064 & 0.024 & ... & ... \\
HV00817 & 55436.242535 & 12.010 & 0.014 & ... & ... \\
HV00817 & 55437.715401 & 12.001 & 0.032 & ... & ...\\
HV00817 & 55439.868676 & 12.004 & 0.034 & ... & ...\\
HV00817 & 55441.242102 & 12.019 & 0.019 & ... & ... \\
HV00817 & 55443.185945 & 12.062 & 0.027 & ... & ... \\
HV00817 & 55444.358354 & 12.117 & 0.018 & ... & ... \\
HV00817 & 55446.331698 & 12.194 & 0.053 & ... & ...  \\
HV00817 & 55447.183273 & 12.208 & 0.022 & ... & ... \\
HV00817 & 55448.942353 & 12.230 & 0.022 & ... & ... \\
HV00817 & 55450.645939 & 12.196 & 0.019 & ... & ... \\
HV00817 & 55451.857901 & 12.078 & 0.023 & ... & ... \\
HV00817 & 55434.901493 & ... & ... & 12.008 & 0.026 \\
HV00817 & 55436.241747 & ... & ... & 11.981 & 0.029 \\
HV00817 & 55437.714611 & ... & ... & 11.990 & 0.033 \\
HV00817 & 55439.867883 & ... & ... & 11.968 & 0.037 \\
HV00817 & 55441.241310 & ... & ... & 11.995 & 0.028 \\
HV00817 & 55443.185157 & ... & ... & 12.051 & 0.036 \\
HV00817 & 55444.357563 & ... & ... & 12.104 & 0.025\\
HV00817 & 55446.330908 & ... & ... & 12.168 & 0.020 \\
HV00817 & 55447.182480 & ... & ... & 12.165 & 0.012 \\
HV00817 & 55448.941561 & ... & ... & 12.226 & 0.034 \\
HV00817 & 55450.645146 & ... & ... & 12.150 & 0.057 \\
HV00817 & 55451.857108 & ... & ... & 12.032 & 0.024 \\
\hline
\end{tabular}
\caption{Individual photometric data points for the SMC Cepheids. The full table is available with the online version of the paper.}
\label{tab:photometric_data}
\end{center}
\end{table}

\begin{deluxetable}{l c c c c c}
\tablecolumns{6}
\tablewidth{0pc}
\tablecaption{Mean--light magnitudes for SMC Cepheids. \label{tab:midir_phot}
}
\tablehead{
\colhead{Cepheid} & \colhead{$\log P$} & \colhead{[3.6]\tablenotemark{a}} &
\colhead{$\sigma_{[3.6]}$\tablenotemark{b}} & \colhead{[4.5]\tablenotemark{a}} &
\colhead{$\sigma_{[4.5]}$\tablenotemark{b}} \\
 & \colhead{(days)} & \colhead {(mag)} & & \colhead {(mag)} & }
\startdata
HV00817 & 1.2766 & 12.094 & 0.007 & 12.063 & 0.007 \\
HV00824 & 1.8191 & 10.270 & 0.006 & 10.267 & 0.006 \\
HV00829 & 1.9258 & 9.871 & 0.009 & 9.836 & 0.009 \\
HV00834 & 1.8670 & 10.139 & 0.006 & 10.137 & 0.004 \\
HV00836 & 0.9733 & 13.234 & 0.022 & 13.199 & 0.022 \\
HV00837 & 1.6309 & 11.042 & 0.011 & 11.047 & 0.011 \\
HV00840 & 1.5189 & 11.392 & 0.014 & 11.391 & 0.014 \\
HV00843 & 1.1677 & 12.738 & 0.012 & 12.745 & 0.011 \\
HV00847 & 1.4326 & 11.782 & 0.013 & 11.791 & 0.013 \\
HV00854 & 1.2031 & 12.518 & 0.011 & 12.493 & 0.011 \\
HV00855 & 1.5179 & 11.560 & 0.012 & 11.576 & 0.011 \\
HV00856 & 1.0848 & 13.062 & 0.011 & 13.052 & 0.011 \\
HV00857 & 1.0786 & 12.847 & 0.028 & 12.829 & 0.031 \\
HV00863 & 1.4617 & 11.381 & 0.013 & 11.378 & 0.014 \\
HV00865 & 1.5228 & 11.167 & 0.014 & 11.158 & 0.014 \\
HV01326 & 1.1375 & 12.926 & 0.010 & 12.915 & 0.008 \\
HV01328 & 1.1997 & 12.503 & 0.005 & 12.478 & 0.005 \\
HV01333 & 1.2120 & 12.711 & 0.009 & 12.689 & 0.010 \\
HV01334 & 0.9755 & 13.221 & 0.024 & 13.214 & 0.026 \\
HV01335 & 1.1578 & 12.898 & 0.009 & 12.867 & 0.008 \\
HV01338 & 0.9292 & 13.584 & 0.026 & 13.535 & 0.023 \\
HV01342 & 1.2539 & 12.511 & 0.004 & 12.476 & 0.003 \\
HV01345 & 1.1296 & 12.841 & 0.009 & 12.818 & 0.009 \\
HV01351 & 1.1170 & 12.959 & 0.005 & 12.938 & 0.006 \\
HV01363 & 1.0286 & 13.237 & 0.018 & 13.238 & 0.023 \\
HV01365 & 1.0939 & 13.174 & 0.008 & 13.124 & 0.009 \\
HV01372 & 1.1979 & 12.852 & 0.012 & 12.818 & 0.009 \\
HV01373 & 1.1370 & 12.858 & 0.010 & 12.853 & 0.009 \\
HV01382 & 1.0367 & 13.213 & 0.028 & 13.173 & 0.026 \\
HV01400 & 0.8227 & 13.690 & 0.019 & 13.674 & 0.023 \\
HV01430 & 1.3797 & 12.085 & 0.013 & 12.090 & 0.013 \\
HV01438 & 1.1351 & 12.762 & 0.006 & 12.719 & 0.005 \\
HV01442 & 1.1844 & 12.654 & 0.007 & 12.604 & 0.008 \\
HV01451 & 1.4782 & 11.631 & 0.012 & 11.659 & 0.011 \\
HV01478 & 1.2438 & 12.684 & 0.011 & 12.675 & 0.010 \\
HV01482 & 1.1993 & 12.601 & 0.009 & 12.557 & 0.008 \\
HV01501 & 1.4379 & 11.869 & 0.014 & 11.861 & 0.013 \\
HV01522 & 1.3454 & 12.138 & 0.012 & 12.144 & 0.012 \\
HV01533 & 1.2158 & 12.689 & 0.010 & 12.685 & 0.009 \\
HV01543 & 1.3108 & 12.346 & 0.012 & 12.337 & 0.012 \\
HV01553 & 1.0987 & 12.994 & 0.008 & 12.950 & 0.006 \\
HV01560 & 1.1906 & 12.594 & 0.010 & 12.589 & 0.009 \\
HV01610 & 1.0661 & 12.969 & 0.022 & 12.952 & 0.027 \\
HV01630 & 1.0569 & 12.993 & 0.027 & 12.950 & 0.023 \\
HV01682 & 1.0845 & 12.951 & 0.007 & 12.944 & 0.007 \\
HV01689 & 0.8353 & 13.359 & 0.015 & 13.295 & 0.014 \\
HV01695 & 1.1641 & 12.588 & 0.009 & 12.560 & 0.008 \\
HV01705 & 1.0317 & 13.143 & 0.021 & 13.110 & 0.021 \\
HV01744 & 1.1012 & 12.791 & 0.008 & 12.765 & 0.007 \\
HV01783 & 0.9111 & 13.504 & 0.023 & 13.521 & 0.031 \\
HV01787 & 1.2093 & 12.393 & 0.010 & 12.386 & 0.010 \\
HV01835 & 1.2107 & 12.510 & 0.010 & 12.504 & 0.010 \\
HV01855 & 0.8350 & 13.709 & 0.024 & 13.689 & 0.027 \\
HV01873 & 1.1119 & 12.889 & 0.010 & 12.861 & 0.011 \\
HV01877 & 1.6966 & 10.831 & 0.010 & 10.852 & 0.009 \\
HV01884 & 1.2579 & 12.302 & 0.010 & 12.306 & 0.009 \\
HV01925 & 1.2355 & 12.280 & 0.005 & 12.242 & 0.004 \\
HV01950 & 0.9025 & 13.322 & 0.020 & 13.307 & 0.033 \\
HV01954 & 1.2225 & 12.147 & 0.005 & 12.130 & 0.005 \\
HV01956 & 2.3197 & 9.454 & 0.018 & 9.317 & 0.012 \\
HV01967 & 1.4636 & 11.626 & 0.012 & 11.619 & 0.013 \\
HV01996 & 1.1535 & 12.915 & 0.010 & 12.919 & 0.011 \\
HV02017 & 1.0573 & 12.905 & 0.031 & 12.884 & 0.026 \\
HV02052 & 1.0996 & 12.662 & 0.009 & 12.646 & 0.007 \\
HV02060 & 1.0079 & 12.828 & 0.015 & 12.793 & 0.018 \\
HV02063 & 1.0479 & 12.926 & 0.025 & 12.926 & 0.024 \\
HV02064 & 1.5273 & 11.548 & 0.012 & 11.562 & 0.012 \\
HV02088 & 1.1638 & 12.762 & 0.013 & 12.765 & 0.013 \\
HV02103 & 0.9535 & 13.319 & 0.021 & 13.290 & 0.026 \\
HV02189 & 1.1294 & 12.722 & 0.010 & 12.707 & 0.010 \\
HV02195 & 1.6214 & 10.994 & 0.008 & 11.009 & 0.007 \\
HV02201 & 1.0512 & 12.747 & 0.020 & 12.760 & 0.029 \\
HV02202 & 1.1203 & 12.595 & 0.009 & 12.577 & 0.009 \\
HV02205 & 1.4053 & 11.876 & 0.014 & 11.882 & 0.014 \\
HV02209 & 1.3551 & 11.788 & 0.005 & 11.757 & 0.005 \\
HV02225 & 1.1191 & 12.710 & 0.009 & 12.716 & 0.010 \\
HV02227 & 1.0957 & 12.790 & 0.009 & 12.781 & 0.010 \\
HV02229 & 1.0191 & 12.862 & 0.027 & 12.843 & 0.025 \\
HV02230 & 1.0979 & 12.670 & 0.008 & 12.679 & 0.008 \\
HV02231 & 1.5646 & 11.234 & 0.006 & 11.259 & 0.006 \\
HV10355 & 1.0017 & 13.337 & 0.020 & 13.322 & 0.017 \\
HV10366 & 1.1504 & 12.591 & 0.006 & 12.558 & 0.005 \\
HV11112 & 0.8255 & 13.875 & 0.014 & 13.817 & 0.021 \\
HV11116 & 1.0206 & 13.324 & 0.023 & 13.273 & 0.018 \\
HV11129 & 1.3888 & 12.152 & 0.013 & 12.154 & 0.012 \\
HV11157 & 1.8388 & 10.501 & 0.006 & 10.529 & 0.005 \\
HV11182 & 1.5932 & 11.343 & 0.009 & 11.376 & 0.008 \\
HV11211 & 1.3302 & 11.786 & 0.013 & 11.781 & 0.012 \\
HV12108 & 1.1935 & 12.495 & 0.007 & 12.458 & 0.007 \\
HV12951 & 1.4773 & 11.735 & 0.010 & 11.753 & 0.008 \\
\enddata
\tablenotetext{a}{Magnitudes are on the Vega system.}
\tablenotetext{b}{For Cepheids with P $\leq 12$ days the uncertainties scale with 1/$\sqrt{N}$ rather than $1/N$ as the observations of short period Cepheids were not phase locked. See Appendix of S11 for details.}
\end{deluxetable}

\begin{table}
\begin{center}
\begin{tabular}{ l r} \hline
Reference & Wavelengths \\ \hline \hline
\citet{1965MNRAS.130..333G} & $B$, $V$ \\
\citet{1975ApJS...29..219M} & $U$, $B$, $V$ \\ 
\citet{1979SAAOC...1...98M} & $U$, $B$, $V$, $I$ \\
\citet{1983AAS...52..423V} & $B$, $V$ \\
\citet{1984SAAOC...8....1C} & $B$, $V$, $R$, $I$ \\
\citet{1985ApJS...59..311F}  & $B$, $V$, $R$, $I$ \\
\citet{1986SAAOC..10...51L} & $J$, $H$, $K$ \\ 
\citet{1987ApJ...321..162W} & $J$, $H$, $K$ \\ 
\citet{1992AJ....104.1430S} & $B$ \\
\citet{1994AJ....108..932S} & $V$, $I$ \\
\citet{1998ApJS..117..135M} & $B$, $V$, $R$, $I$ \\
\citet{1999PASP..111..812B} & $V$, $R$, $I$ \\
\citet{2004AA...415..521S} & $B$, $V$, $I$, $J$, $K$ \\ 
\hline
\end{tabular}
\caption{Sources of archival SMC Cepheid data used in this work.}
\label{tab:archival}
\end{center}
\end{table}

\begin{table}
\begin{center}
\begin{tabular}{l c c c c} \hline
Wavelength & Slope (a) & Zero--point (b) & Dispersion & Distance Modulus ($\mu$)  \\ \hline \hline
$U$ & $-1.784 \pm 0.629$ & $15.800 \pm 0.390$ & 0.685 & \\
$B$ & $-2.393 \pm 0.046$ & $15.722 \pm 0.048$ & 0.384 & $19.24 \pm 0.05$ \\
$V$ & $-2.734 \pm 0.031$ & $15.040 \pm 0.034$ & 0.316 & $19.20 \pm 0.03$  \\
$R$ & $-2.742 \pm 0.060$ & $14.545 \pm 0.038$ & 0.248 & $19.07 \pm 0.04$  \\
$I$ & $-2.957 \pm 0.020$ & $14.238 \pm 0.026$ & 0.245 & $19.09 \pm 0.03$ \\
$J$ & $-3.153 \pm 0.051$ & $13.716 \pm 0.029$ & 0.252 & $19.01 \pm 0.03$  \\
$H$ & $-3.234 \pm 0.042$ & $13.362 \pm 0.027$ & 0.233 & $19.00 \pm 0.03$  \\
$K$ & $-3.281\pm 0.040$ & $13.282 \pm 0.026$ & 0.225 & $18.99 \pm 0.03$  \\
$[3.6]$ & $-3.306\pm 0.050$ & $13.186 \pm 0.017$ & 0.161 & $18.99 \pm 0.03$  \\
$[4.5]$ & $-3.207\pm 0.060$ & $13.150 \pm 0.018$ & 0.164 & $18.92 \pm 0.03$ \\
\hline
\end{tabular}
\caption{Leavitt Law fits for the ten wavelengths. The $U$ band fit is included for completeness, but no distance modulus can be calculated. Uncertainties on the distance moduli are calculated by combining the uncertainties on the reference and fitted LL zero--points in quadrature.}
\label{tab:plfits}
\end{center}
\end{table}

\begin{table}
\begin{center}
\begin{tabular}{l c c c c c} \hline
Wavelength & \multicolumn{2}{c}{Full Sample}  & \multicolumn{2}{c}{Cut Sample} & $\Delta\mu$  \\
&$\mu$ &  $\sigma$ & $\mu$ &  $\sigma$ &  \\
 \hline \hline
$B$ & $19.24 \pm 0.05$ & 0.384 & $19.21 \pm 0.06$ & 0.375 & $0.03 \pm 0.08$ \\
$V$ & $19.20 \pm 0.04$ & 0.316 & $19.14 \pm 0.04$ & 0.260 & $0.06 \pm 0.05$\\
$R$ & $19.07 \pm 0.04$ & 0.248 & $19.07 \pm 0.04$ & 0.248 & 0.0 \\
$I$ & $19.09 \pm 0.03$ & 0.245 & $19.03 \pm 0.03$ & 0.214 & $0.06 \pm 0.04$\\
$J$ & $19.01 \pm 0.03$ & 0.252 & $18.95 \pm 0.03$ & 0.210 & $0.06 \pm 0.04$\\
$H$ & $19.00 \pm 0.03$ & 0.233 & $18.93 \pm 0.03$ & 0.207 & $0.07 \pm 0.04$\\
$K$ & $18.99 \pm 0.03$ & 0.225 & $18.94 \pm 0.03$ & 0.194 & $0.05 \pm 0.04$\\
$[3.6]$ & $18.99 \pm 0.02$ & 0.161 & $18.96 \pm 0.02$ & 0.157 & $0.03 \pm 0.03$\\
$[4.5]$ & $18.92 \pm 0.02$ & 0.164 & $18.89 \pm 0.03$ & 0.164 & $0.03 \pm 0.03$\\
\hline
\end{tabular}
\caption{Changes in the derived distance moduli when only the Cepheids appearing in all nine bands are used. The $R$ band sample is the smallest, therefore $\mu$ does not change from the original result. The Cepheids in the $R$ sample do not uniformly sample the instability strip but have a slight bias towards the bright edge, leading to the systematic trend towards smaller distance moduli when using this sample. The mean $\Delta\mu$ induced by this effect is $0.04 \pm 0.02$~mag.}
\label{tab:plfits_rcuts}
\end{center}
\end{table}

\clearpage

%% Put figures here

\begin{figure}
\begin{center}
\includegraphics[width=50mm]{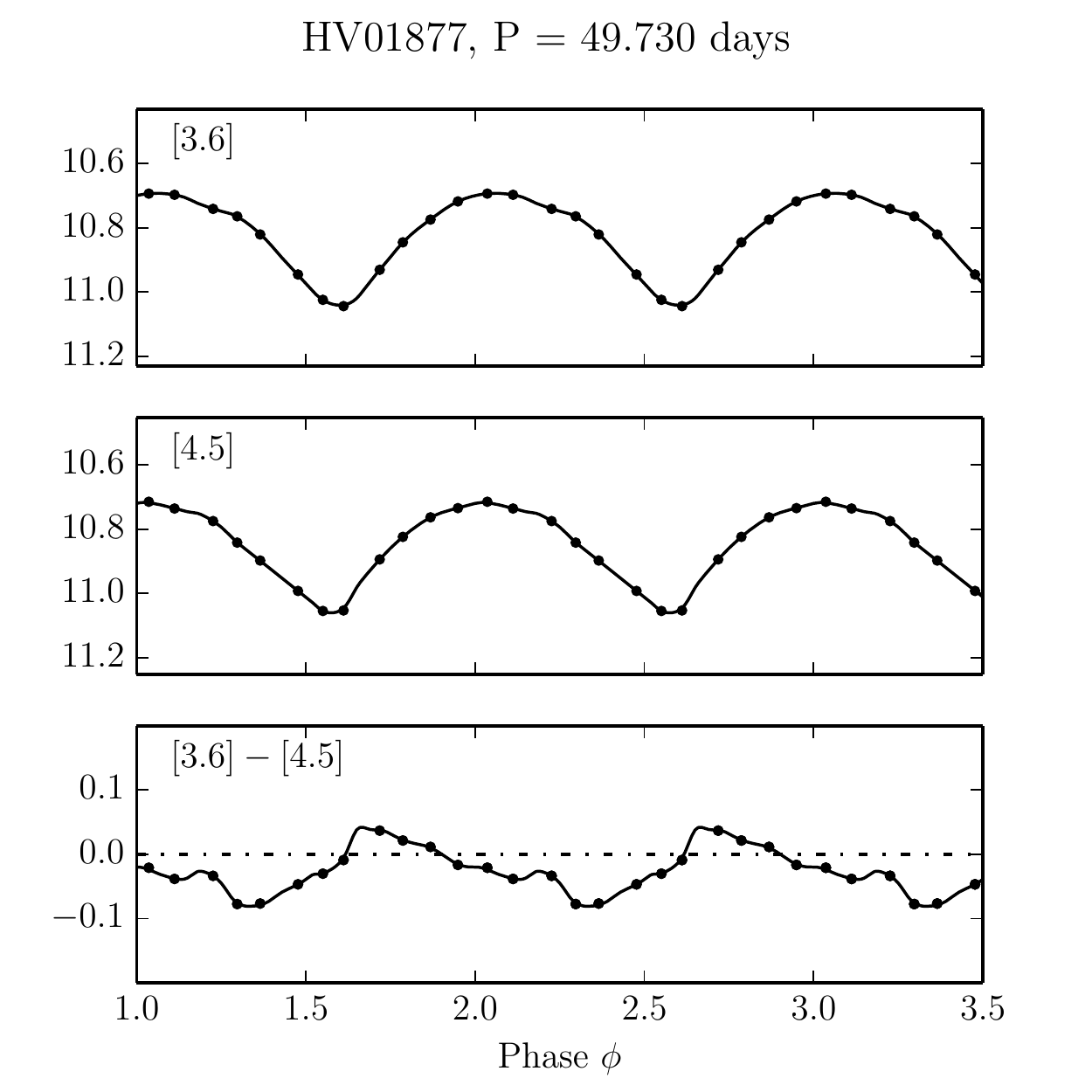} 
\caption{Example IRAC light curve. The whole sample is shown in Figure~\ref{fig:light_curves_appendix}. Point sizes are comparable to the uncertainties in the $[3.6]$ (top) and $[4.5]$ (middle) panels. The bottom panel shows the variation of the IRAC $[3.6]-[4.5]$ color with phase. More negative in color corresponds to greater absorption by CO.  The light curves were fit using GLOESS, a Gaussian local estimation algorithm which has been used throughout the CHP (e.g. see S11, M12).}
\label{fig:light_curves}
 \end{center}
\end{figure}

%\begin{figure}
%\begin{center}
%\includegraphics[width=180mm]{plots/SMC_distances-eps-converted-to.pdf}
%\caption{Comparison of LMC -- SMC distances. The dashed red line is the CHP result and the dashed vertical line is its median value. The solid black lines are the individual measurements from table 11 of \citet{2013arXiv1311.2340G}, and the thick blue line is the combined probability density distribution. The point shows the modal value of the cumulative distribution. The large error bars show the 68\% confidence limits around this value, the smaller error bars show the error on the mean. }
%\label{fig:smc_distance_comparison}
%\end{center}
%\end{figure}

%\begin{figure}
%\begin{center}
%\includegraphics[width=180mm]{plots/SMC_mixture_model.pdf}
%\caption{Mixture model of SMC distances from individual stars. The black gaussians represent individual distances derived from the [3.6] Leavitt Law, with their widths representing their uncertainties. The blue line is the mixture model of these measurements. The red line shows the distance derived from the $[3.6]$ LL. The green line shows the mean and standard deviation of the mixture model. It is clear from this plot that the SMC Cepheids can be thought of as a main body population and the wing population leading off to the LMC.}
%Tilt =  -0.132244665189 degrees / kpc
%\label{fig:smc_mixture}
%\end{center}
%\end{figure}

%\begin{figure}
%\begin{center}
%\includegraphics[width=180mm]{plots/plot_mcs.pdf}
%\includemedia[
%	width=\textwidth,
%	height=0.5\textwidth,
%	activate=pageopen,
%	3Dtoolbar,
%	3Dviews=plots/plot_mcs.vws
%] {\includegraphics[width=0.5\textwidth]{plots/plot_mcs.png}} {plots/plot_mcs.u3d}
%\caption{3D map of the MCs}
%Tilt =  -0.132244665189 degrees / kpc
%\label{fig:rotateable_smc}
%\end{center}
%\end{figure}

\begin{figure} 
 \begin{center}$ 
 \begin{array}{ccc} 
\includegraphics[width=50mm]{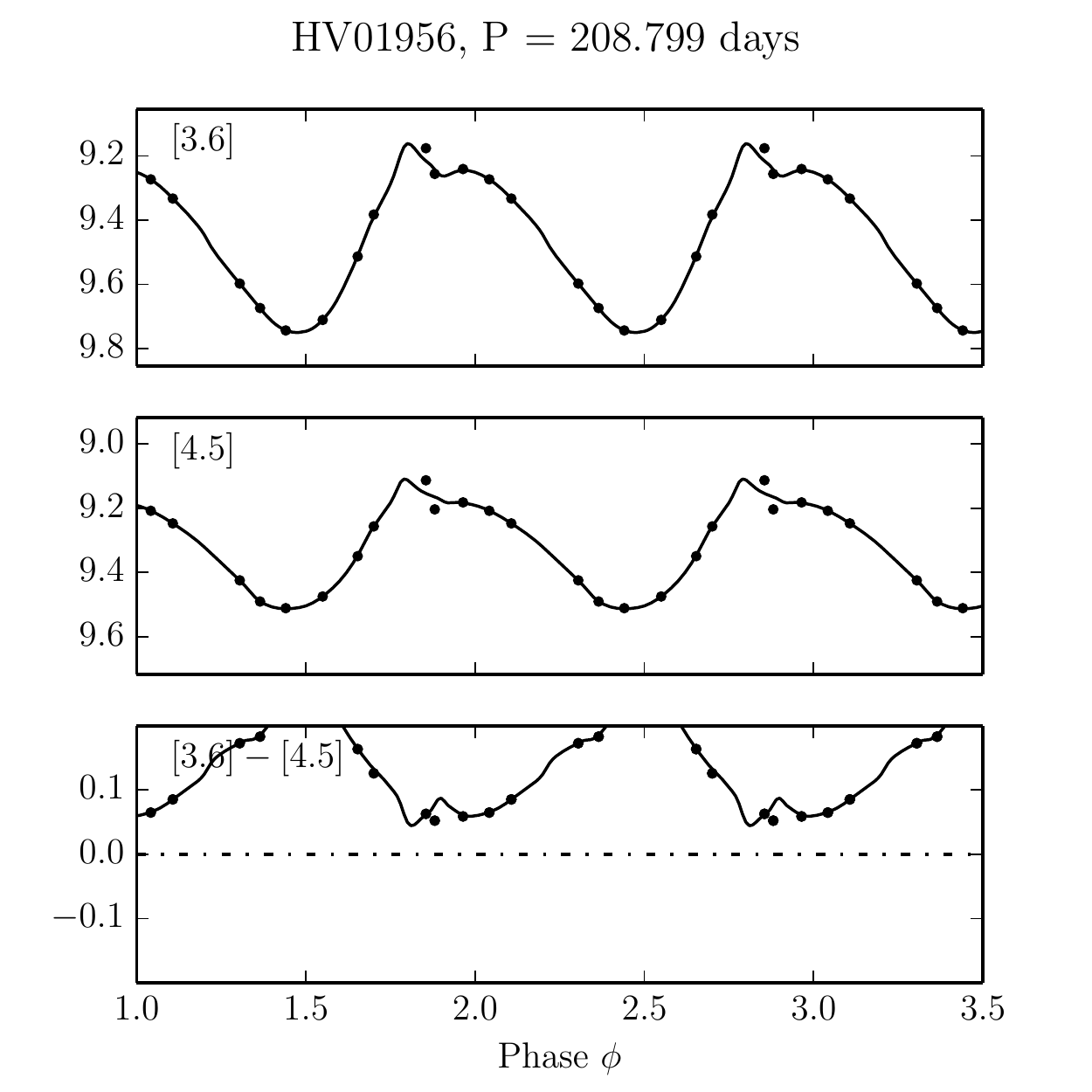} &
\includegraphics[width=50mm]{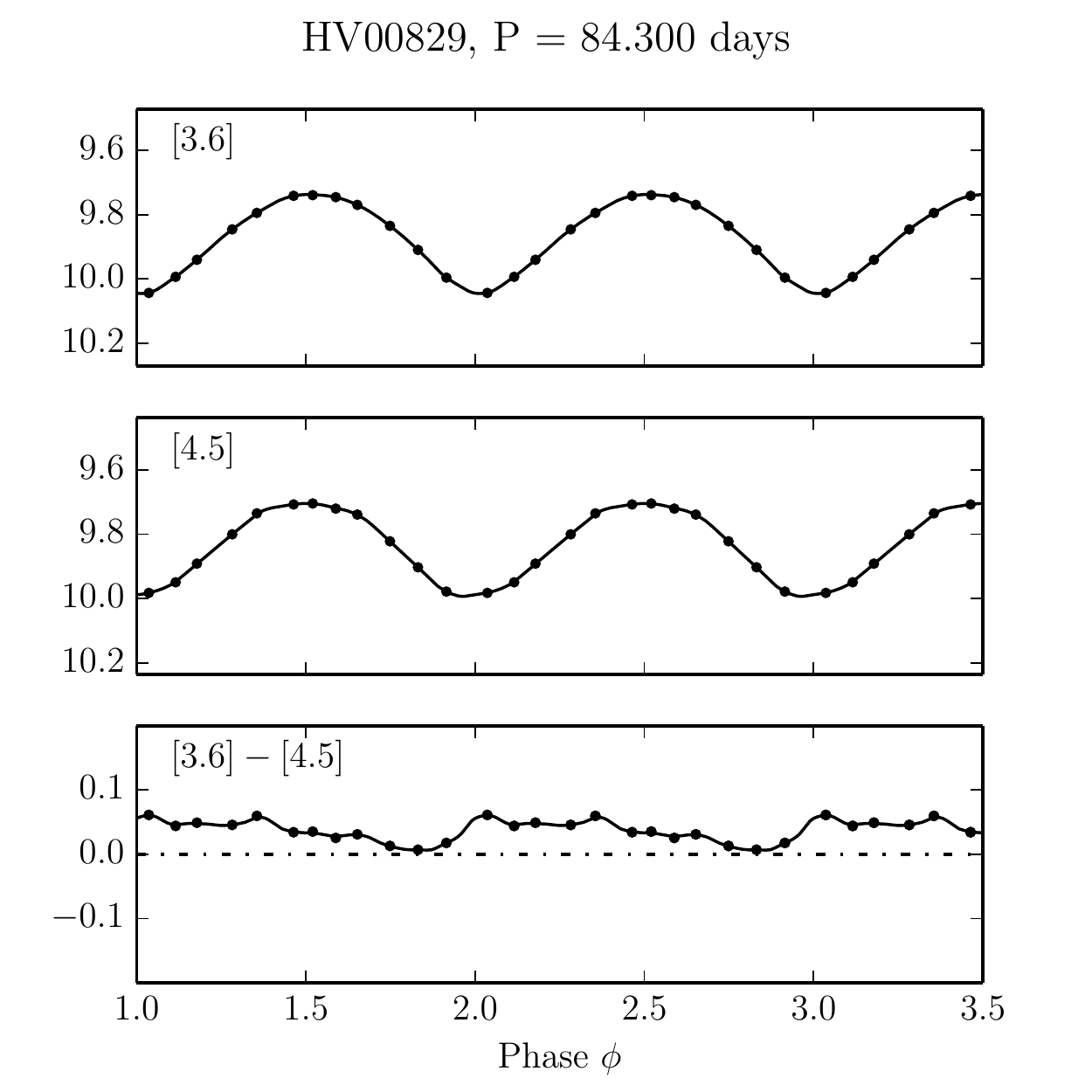} &
\includegraphics[width=50mm]{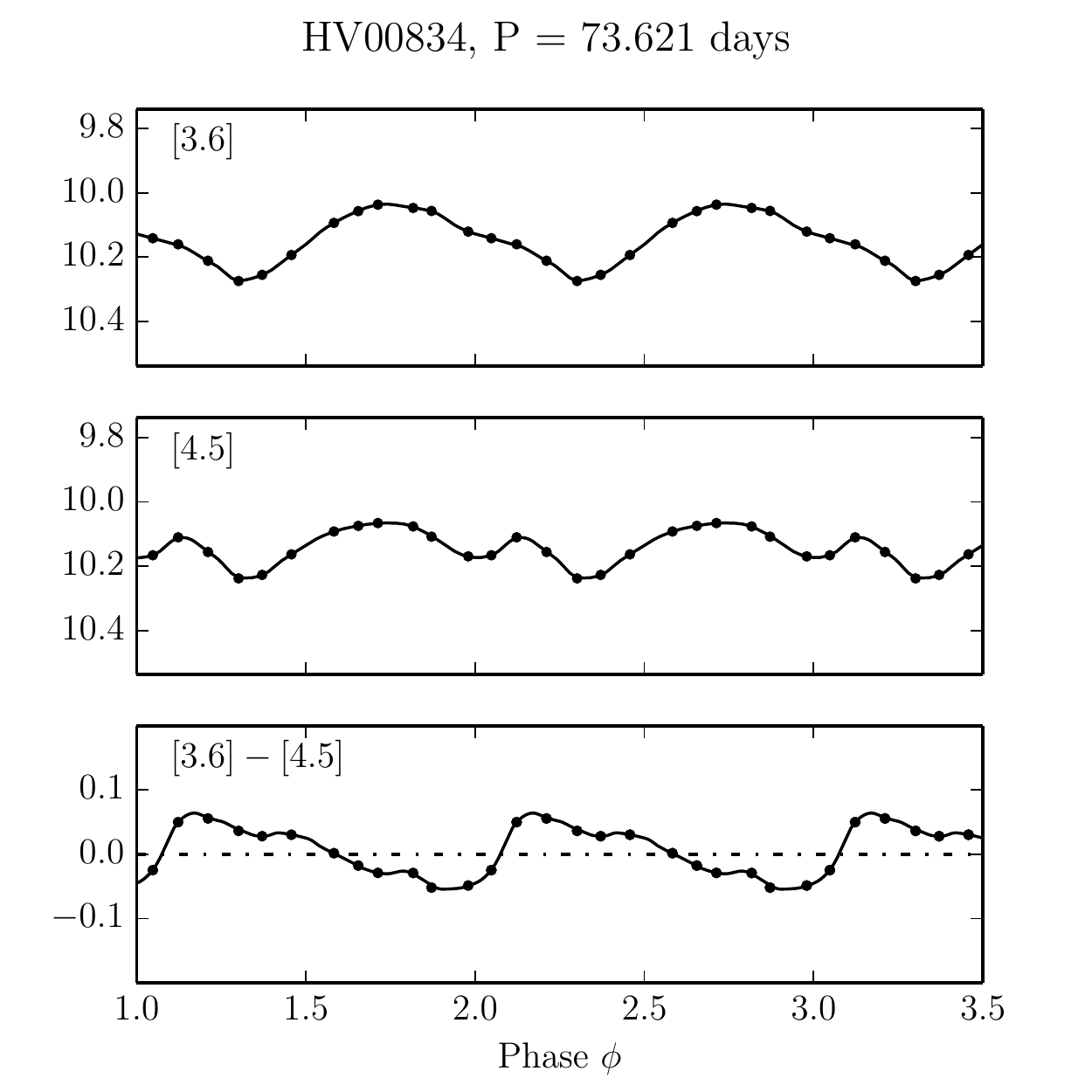} \\ 
\includegraphics[width=50mm]{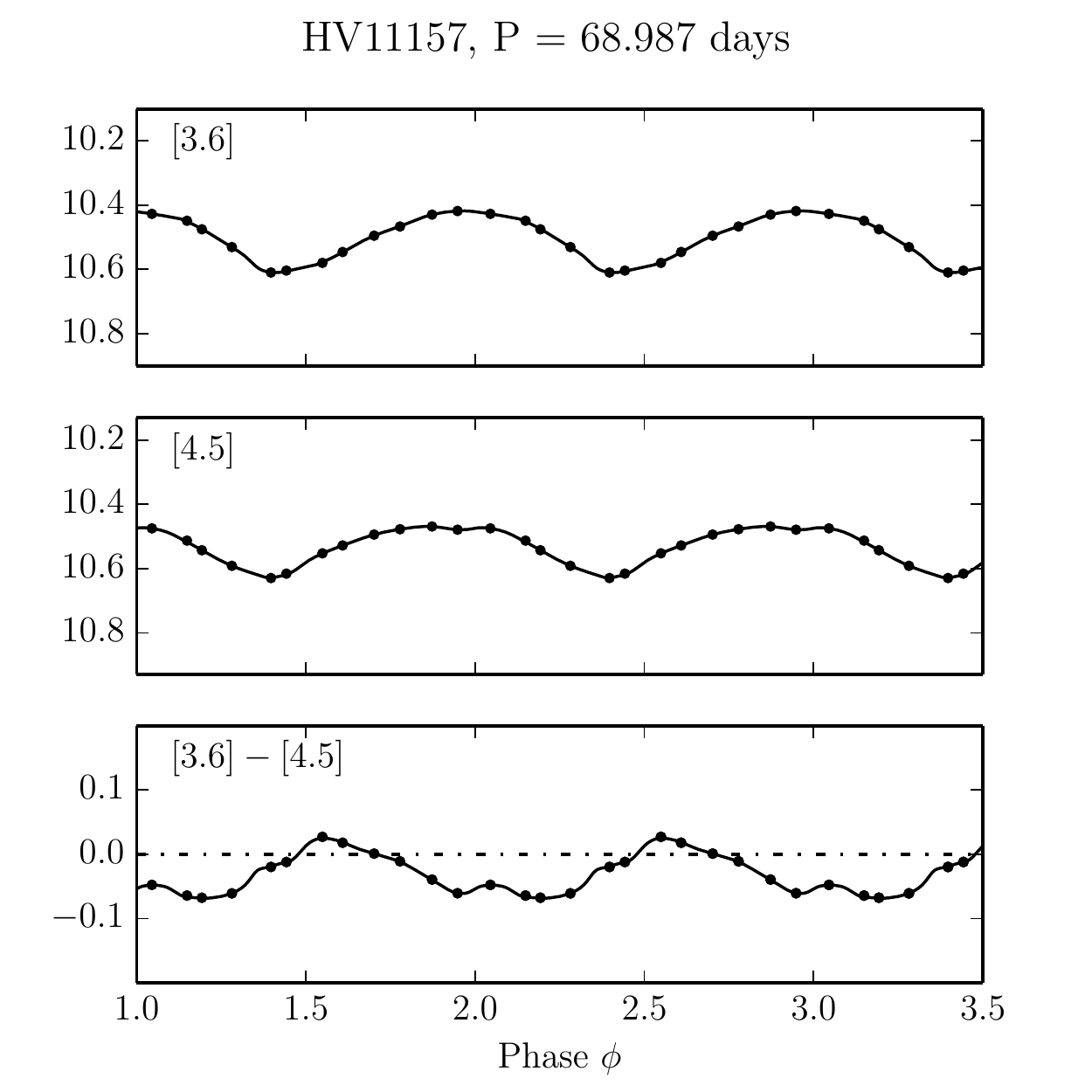} & 
\includegraphics[width=50mm]{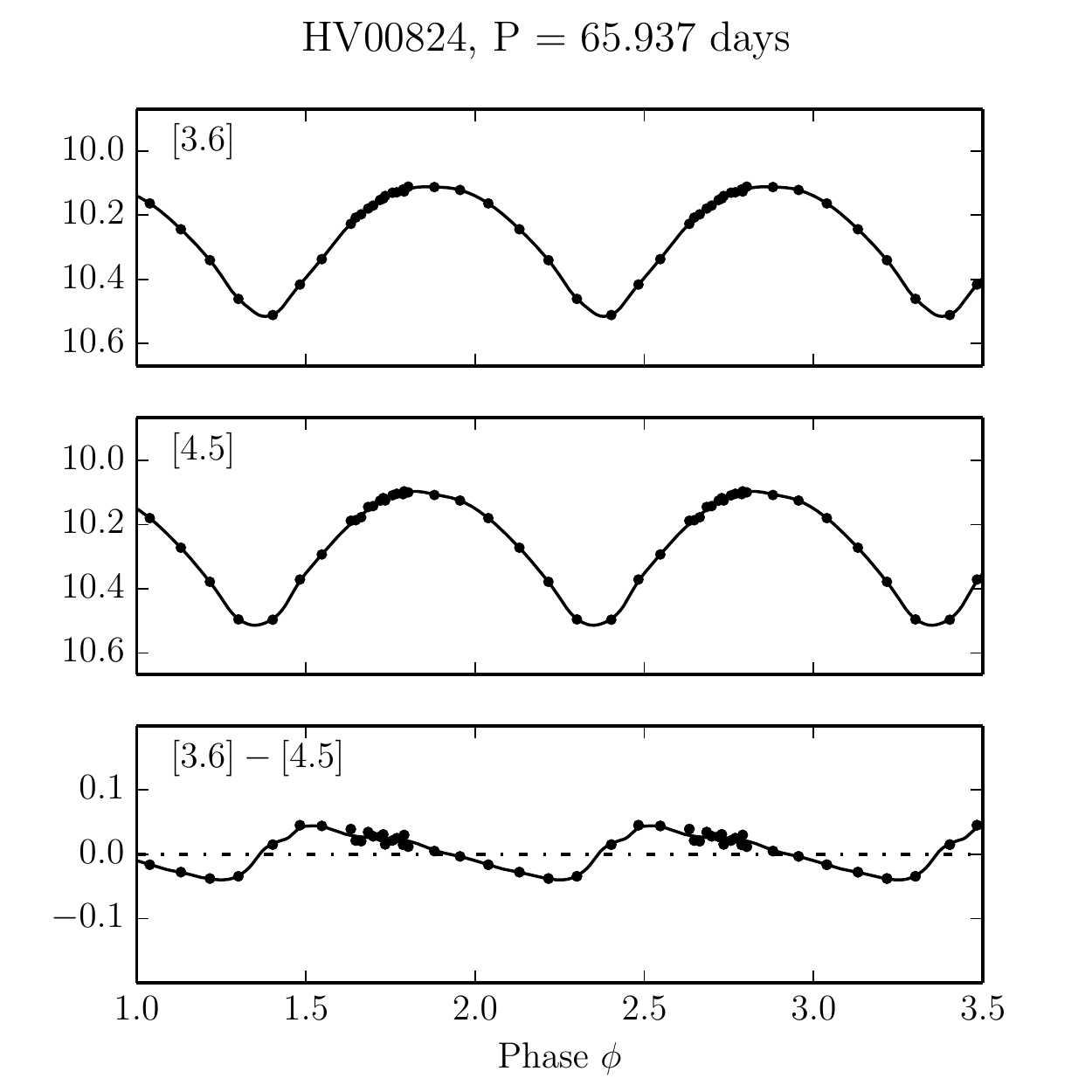} &
\includegraphics[width=50mm]{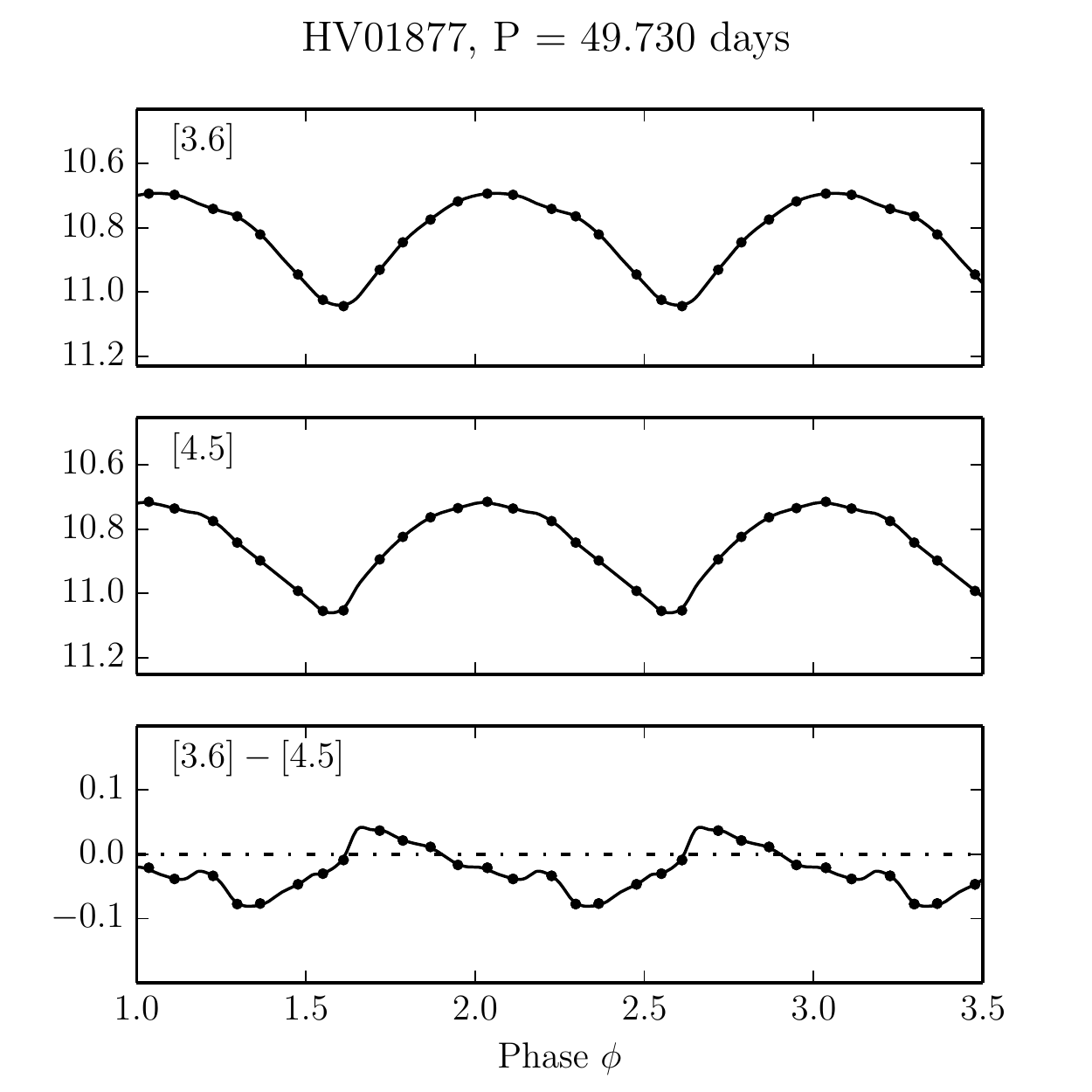} \\
\includegraphics[width=50mm]{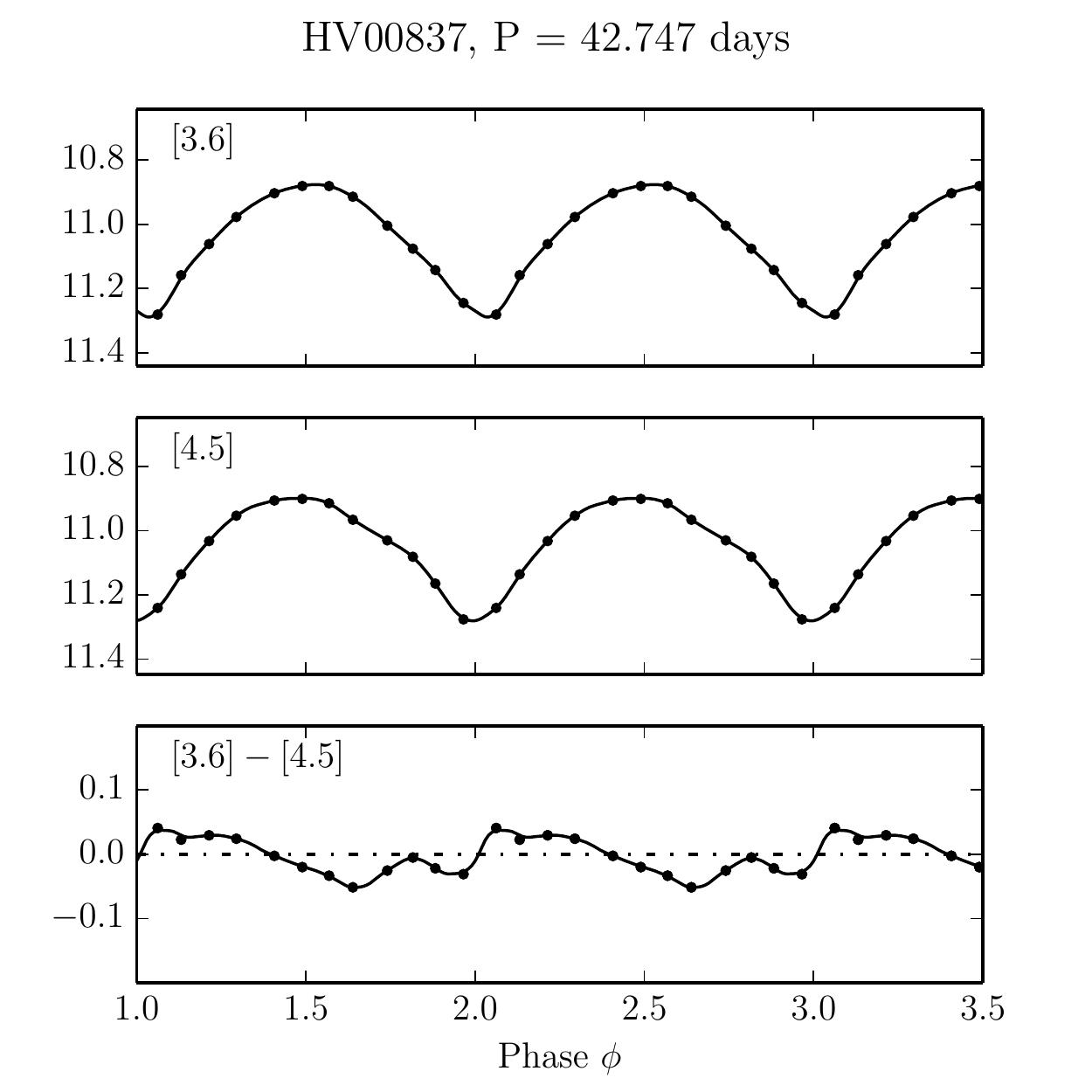} & 
\includegraphics[width=50mm]{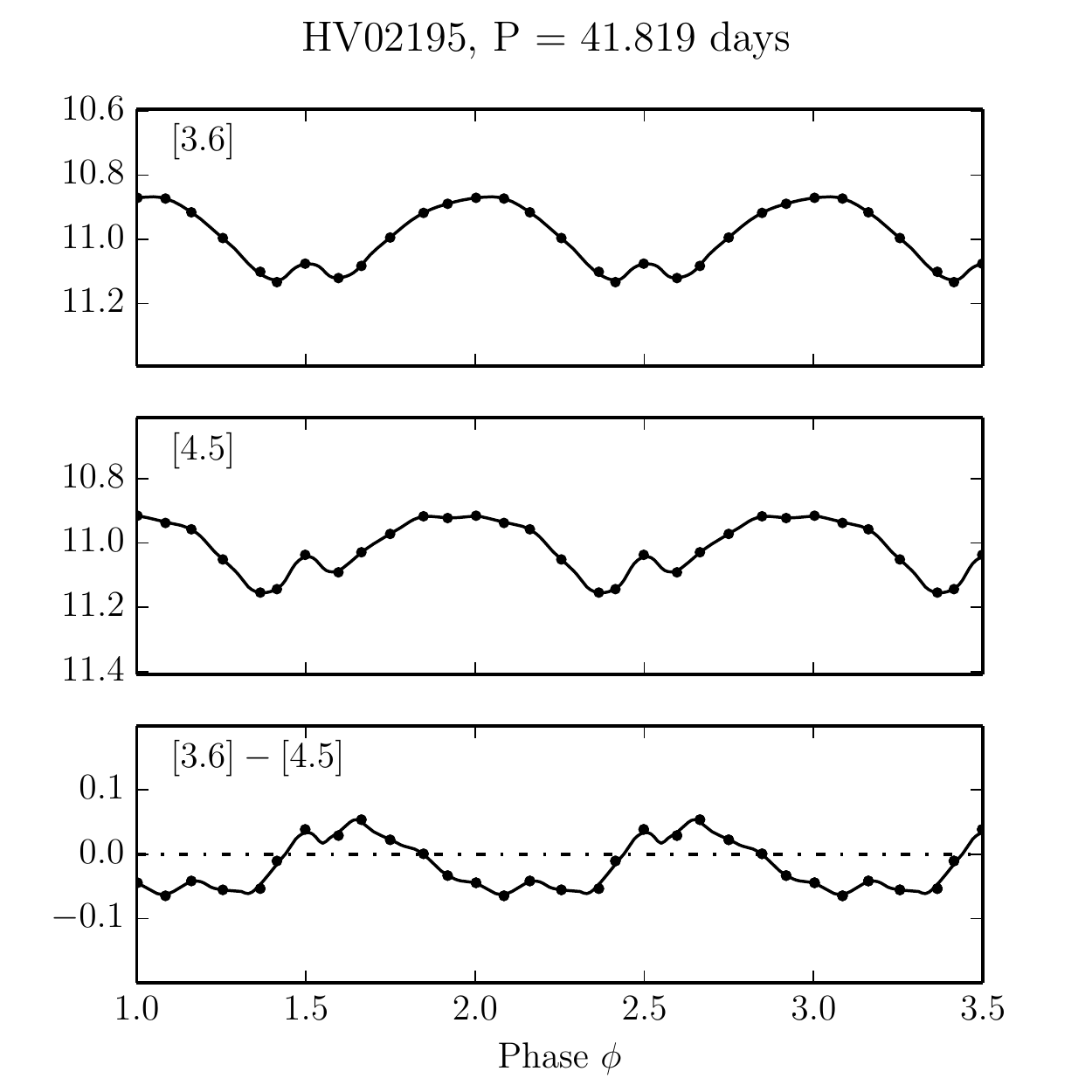} & 
\includegraphics[width=50mm]{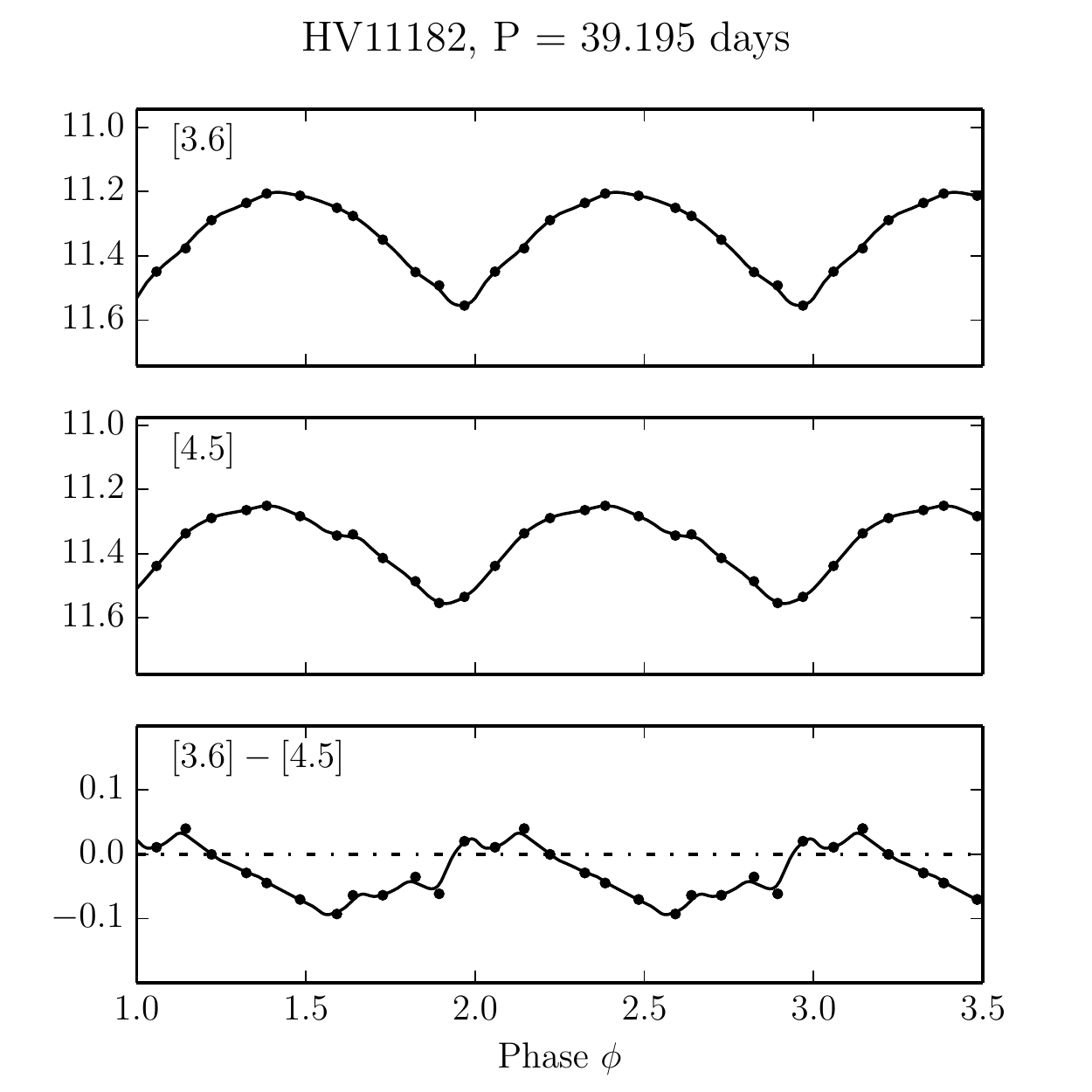} \\
\includegraphics[width=50mm]{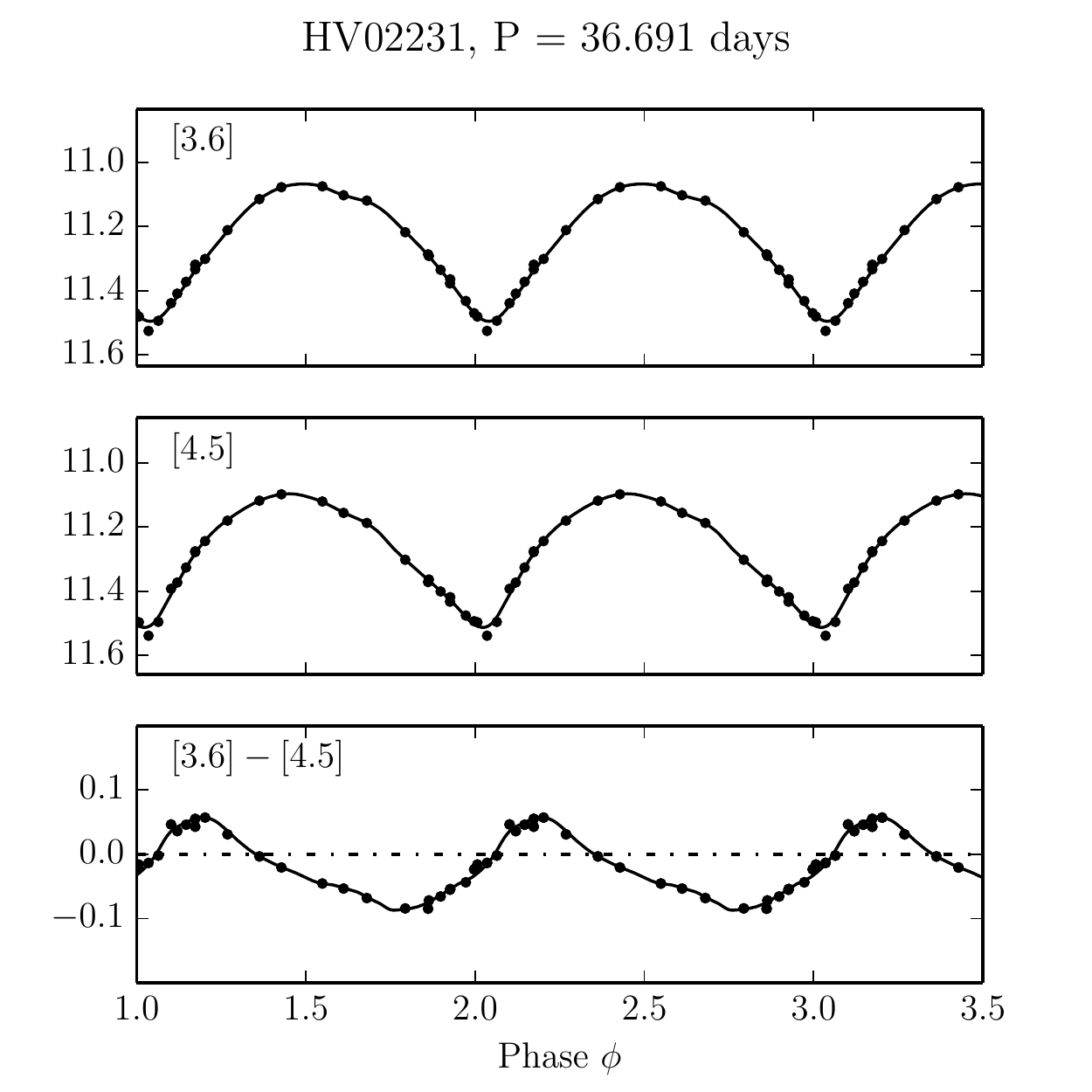} & 
\includegraphics[width=50mm]{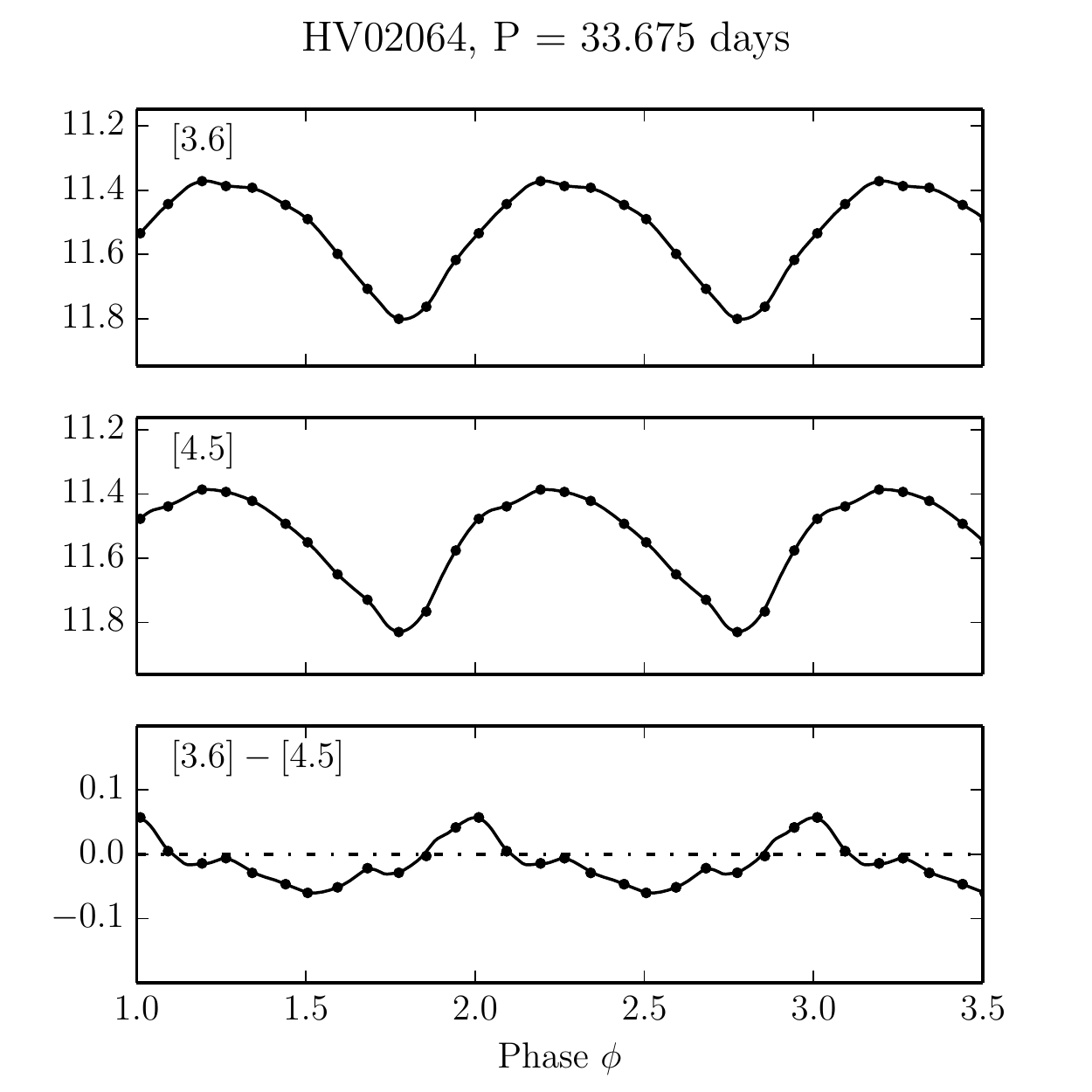} & 
\includegraphics[width=50mm]{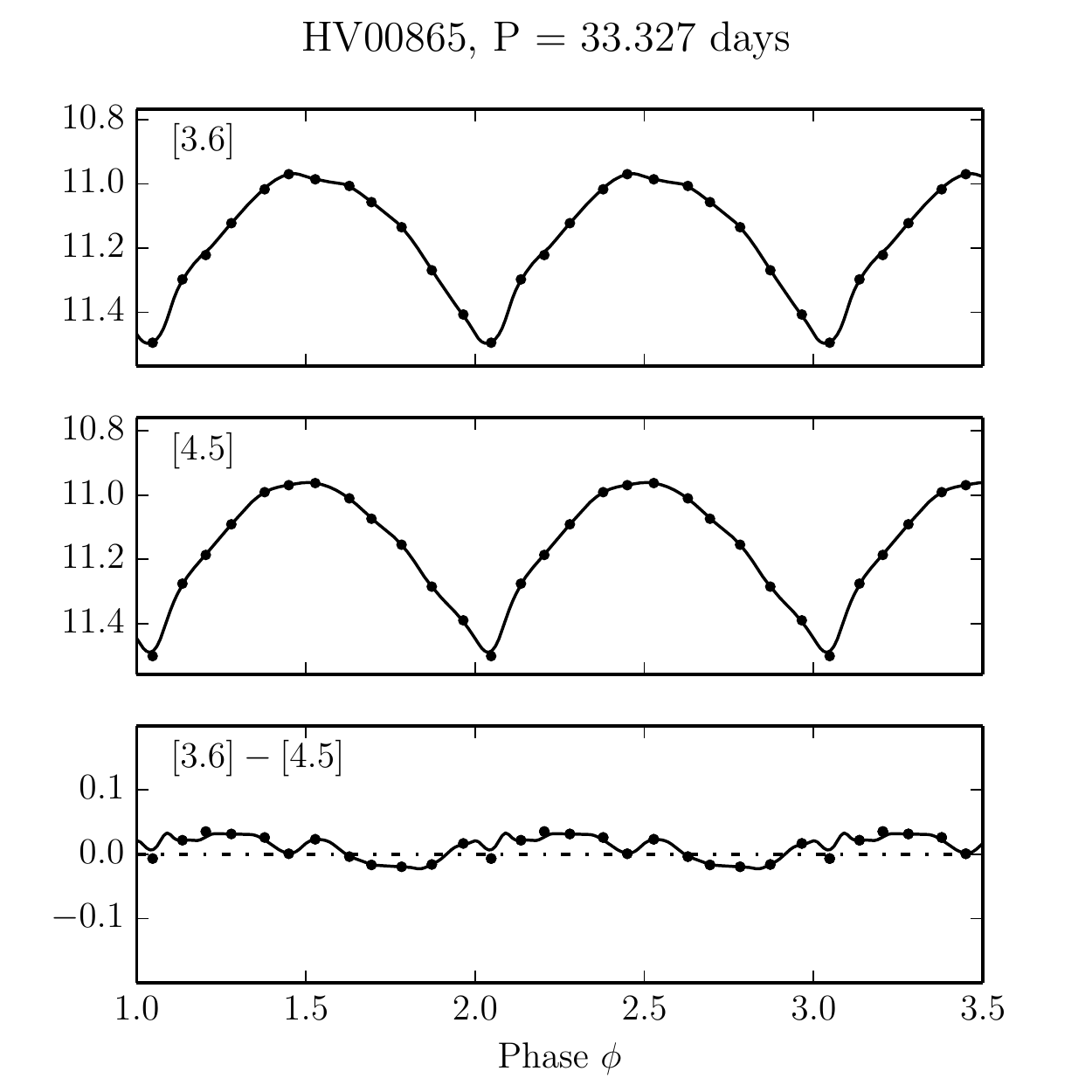} \\
\end{array}$ 
\end{center} 
\end{figure}
\begin{figure} 
 \begin{center}$ 
 \begin{array}{ccc} 
\includegraphics[width=50mm]{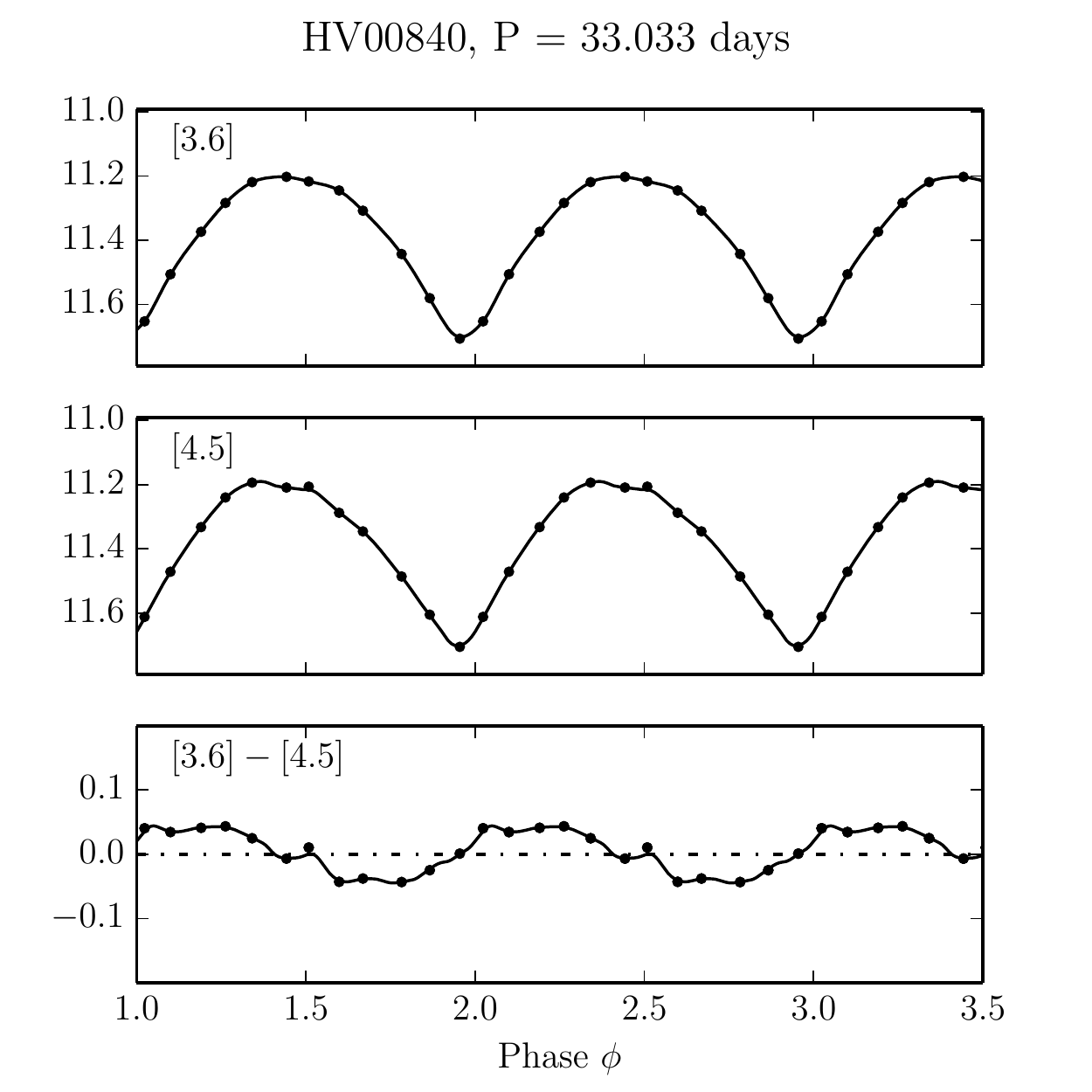} &
\includegraphics[width=50mm]{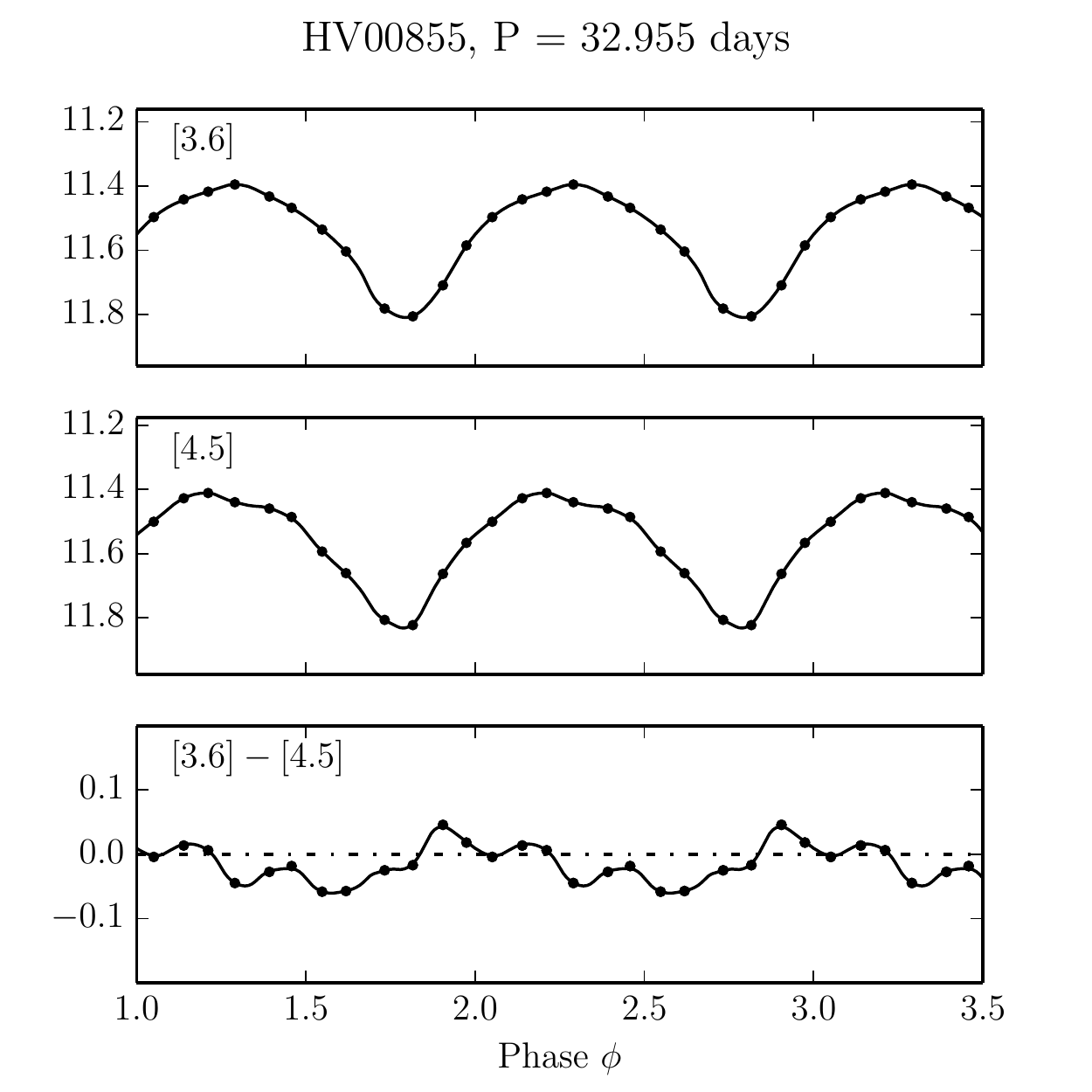} &
\includegraphics[width=50mm]{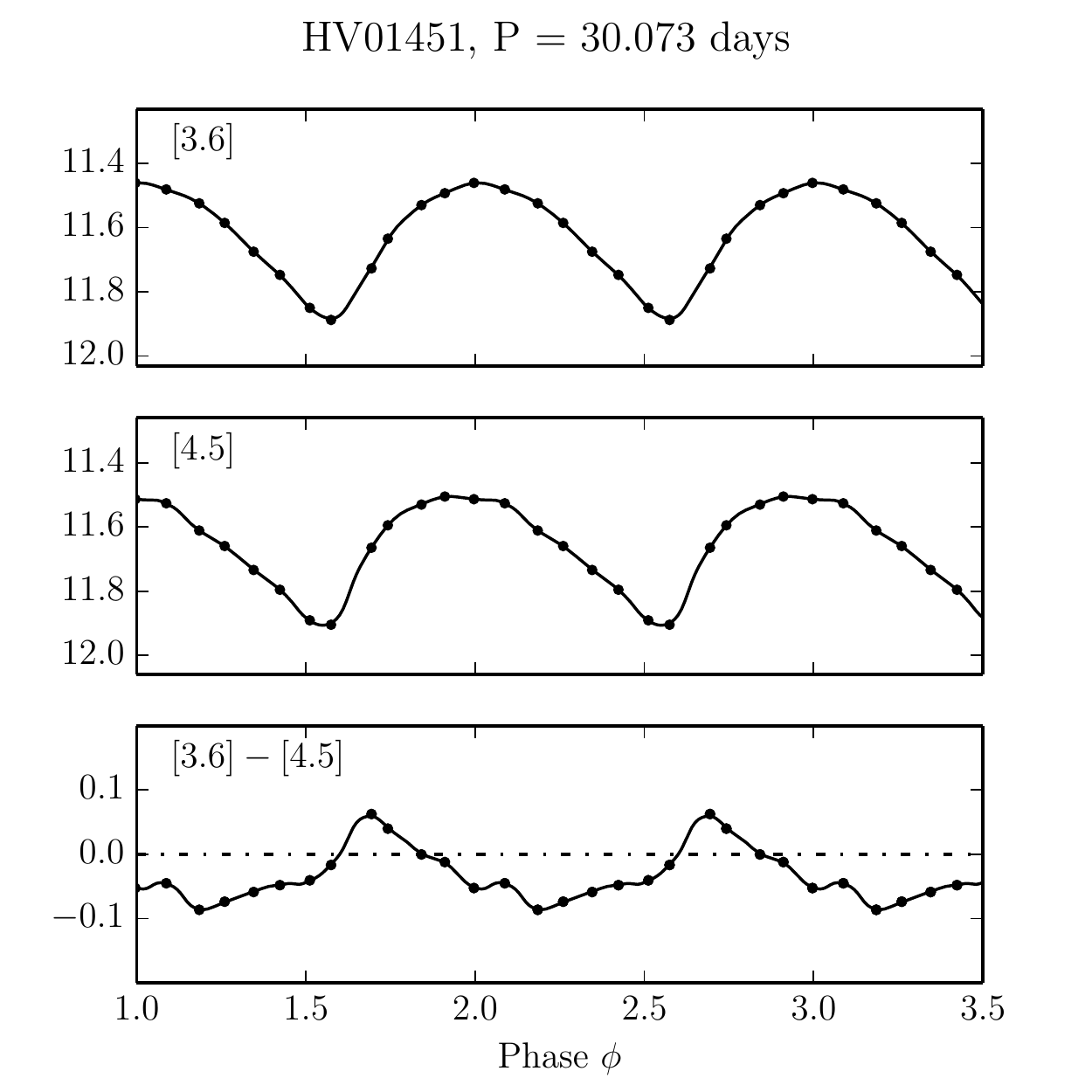} \\
\includegraphics[width=50mm]{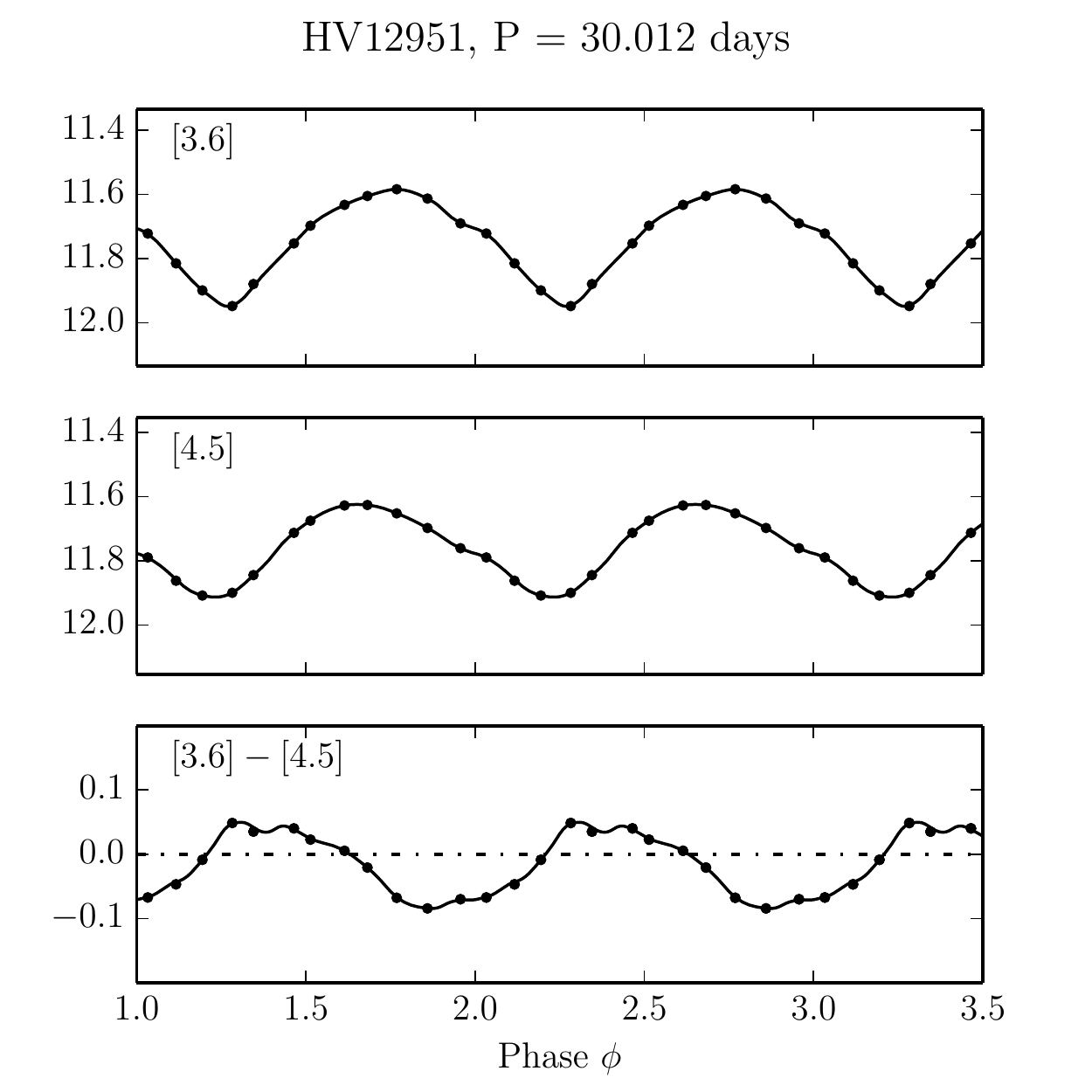} &
\includegraphics[width=50mm]{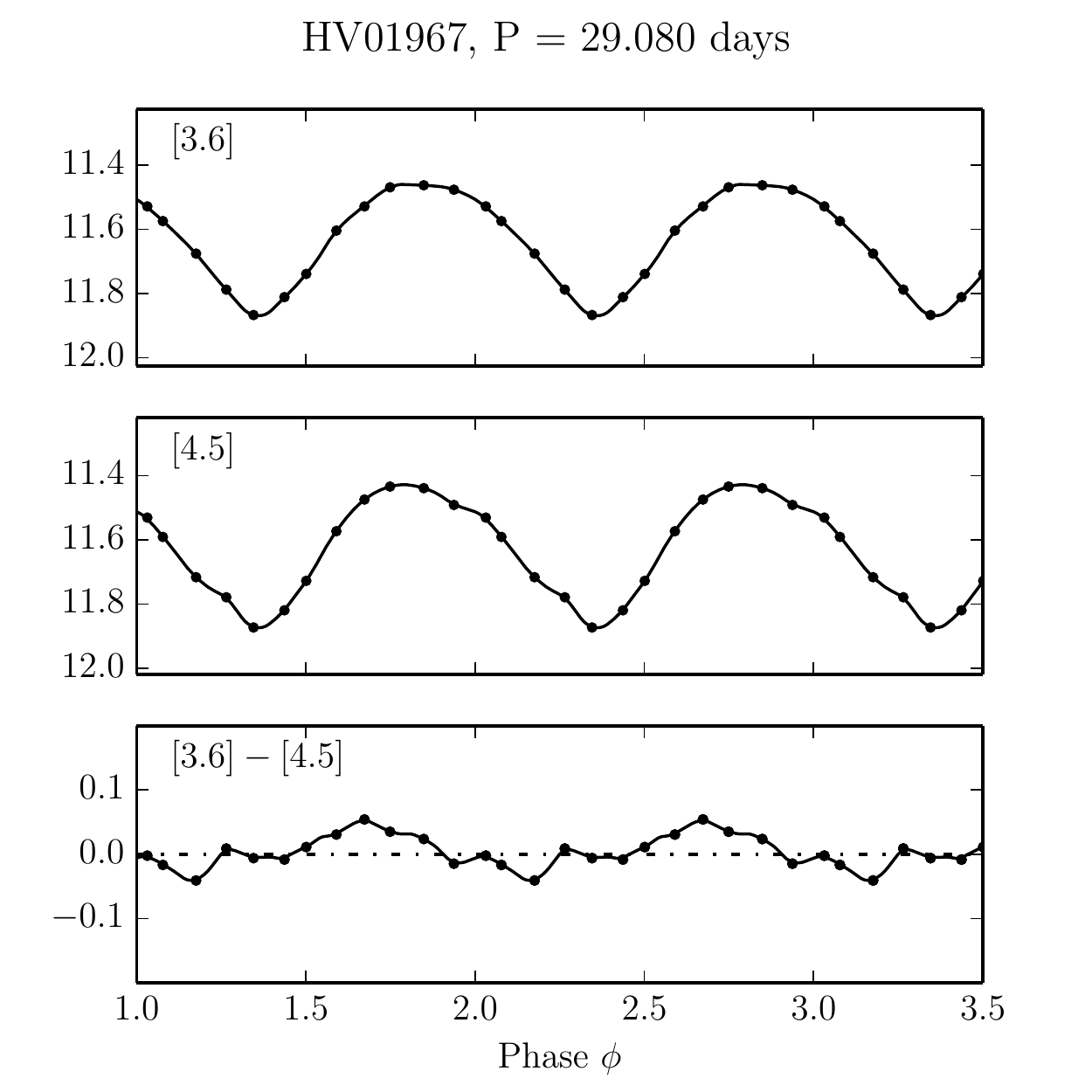} &
\includegraphics[width=50mm]{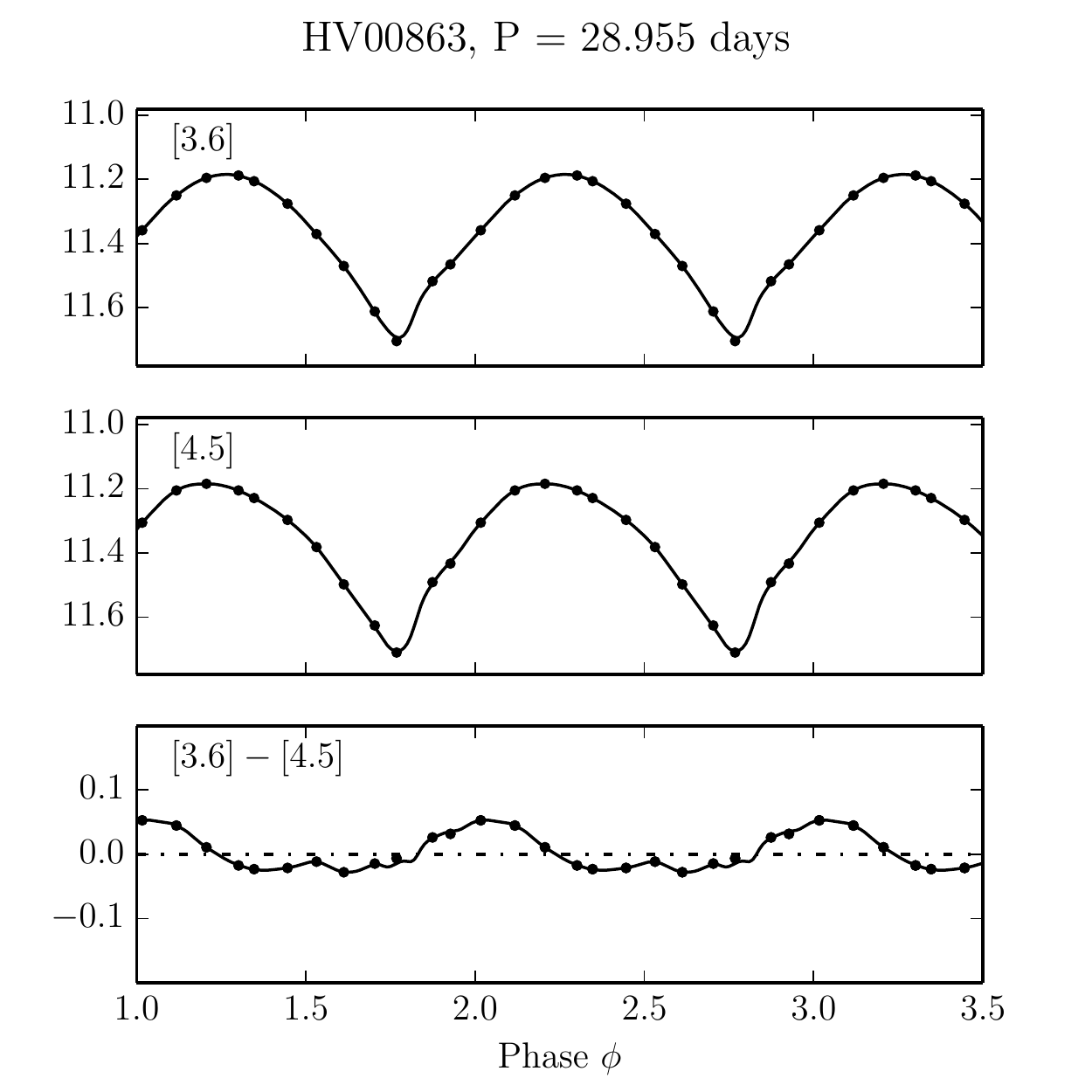} \\ 
\includegraphics[width=50mm]{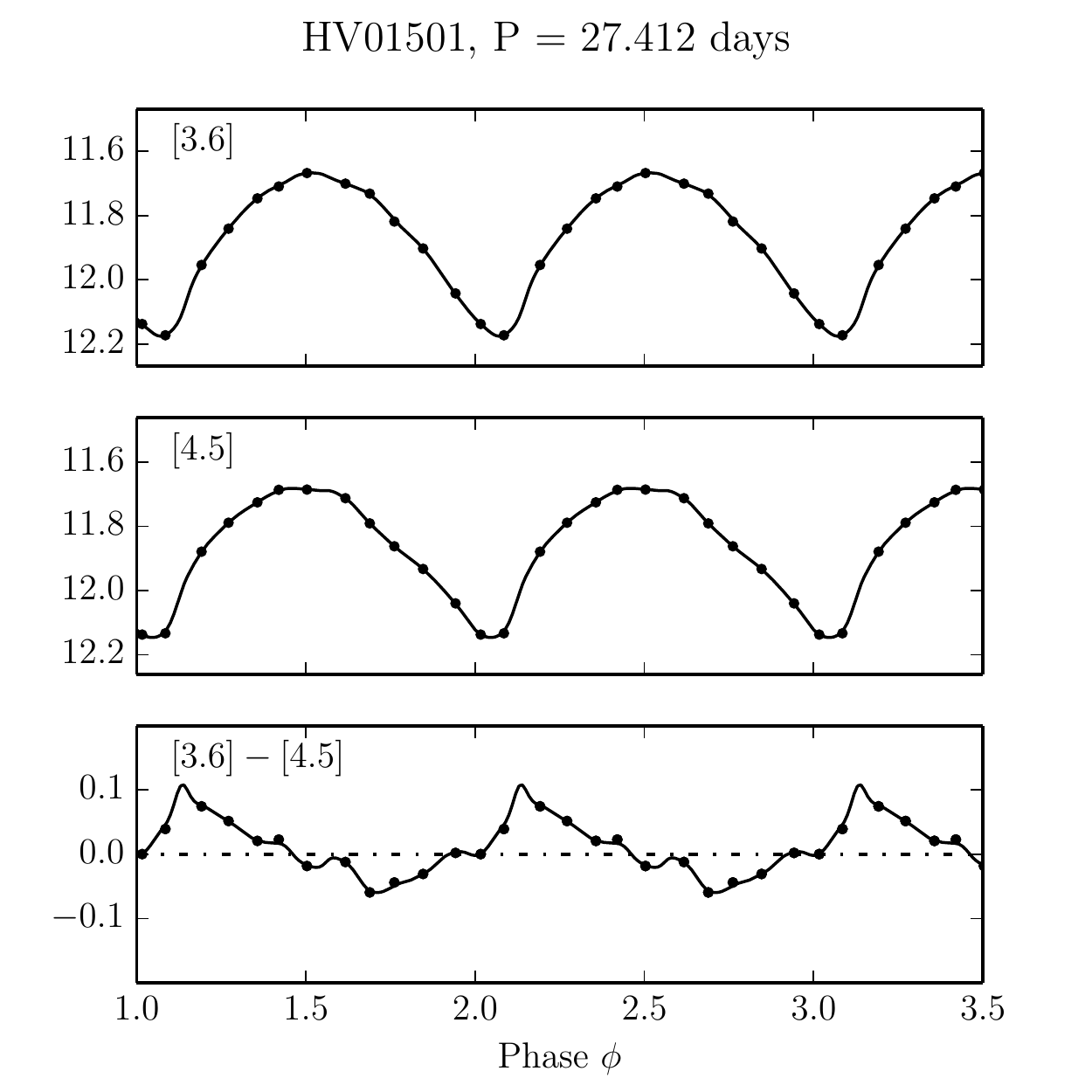} & 
\includegraphics[width=50mm]{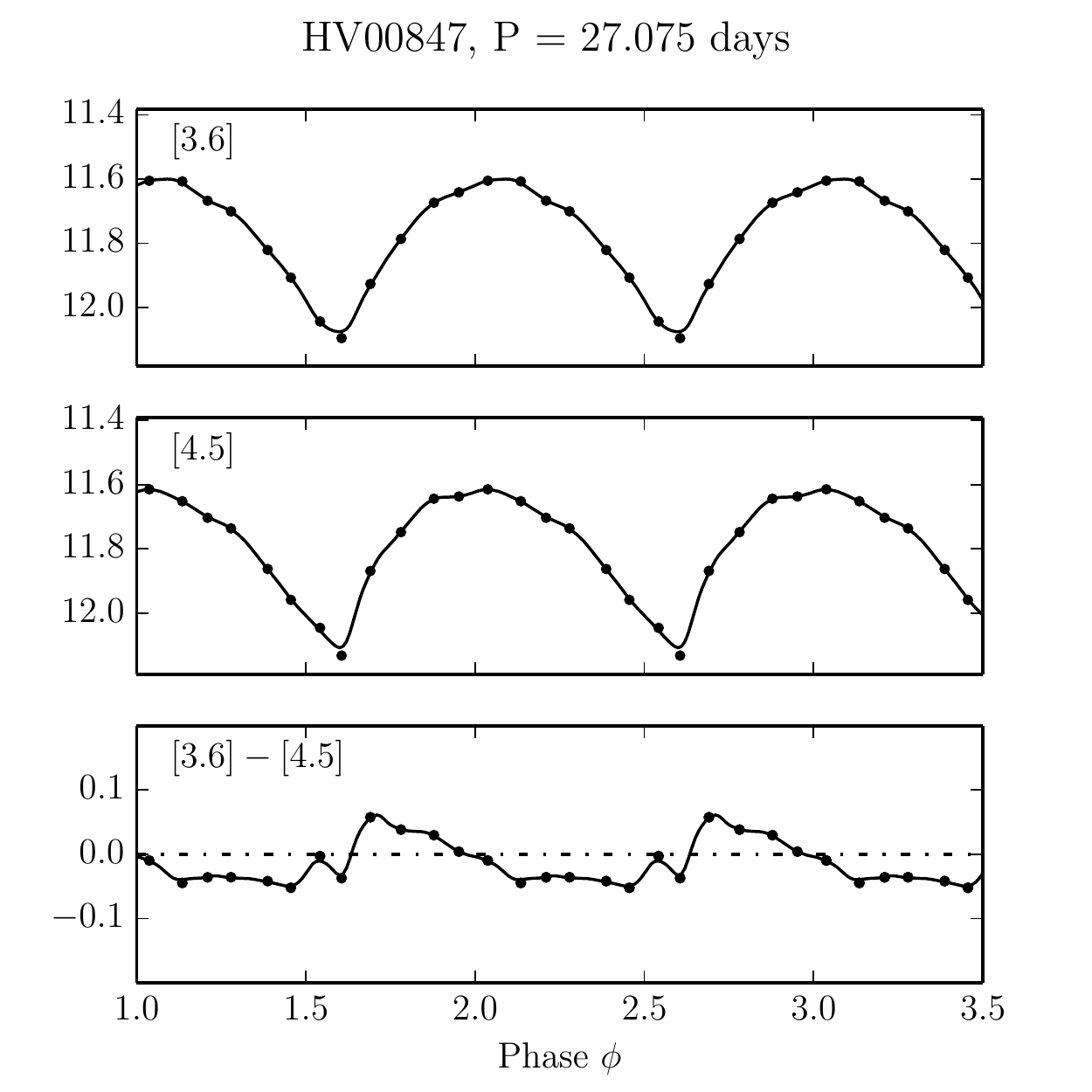} &
\includegraphics[width=50mm]{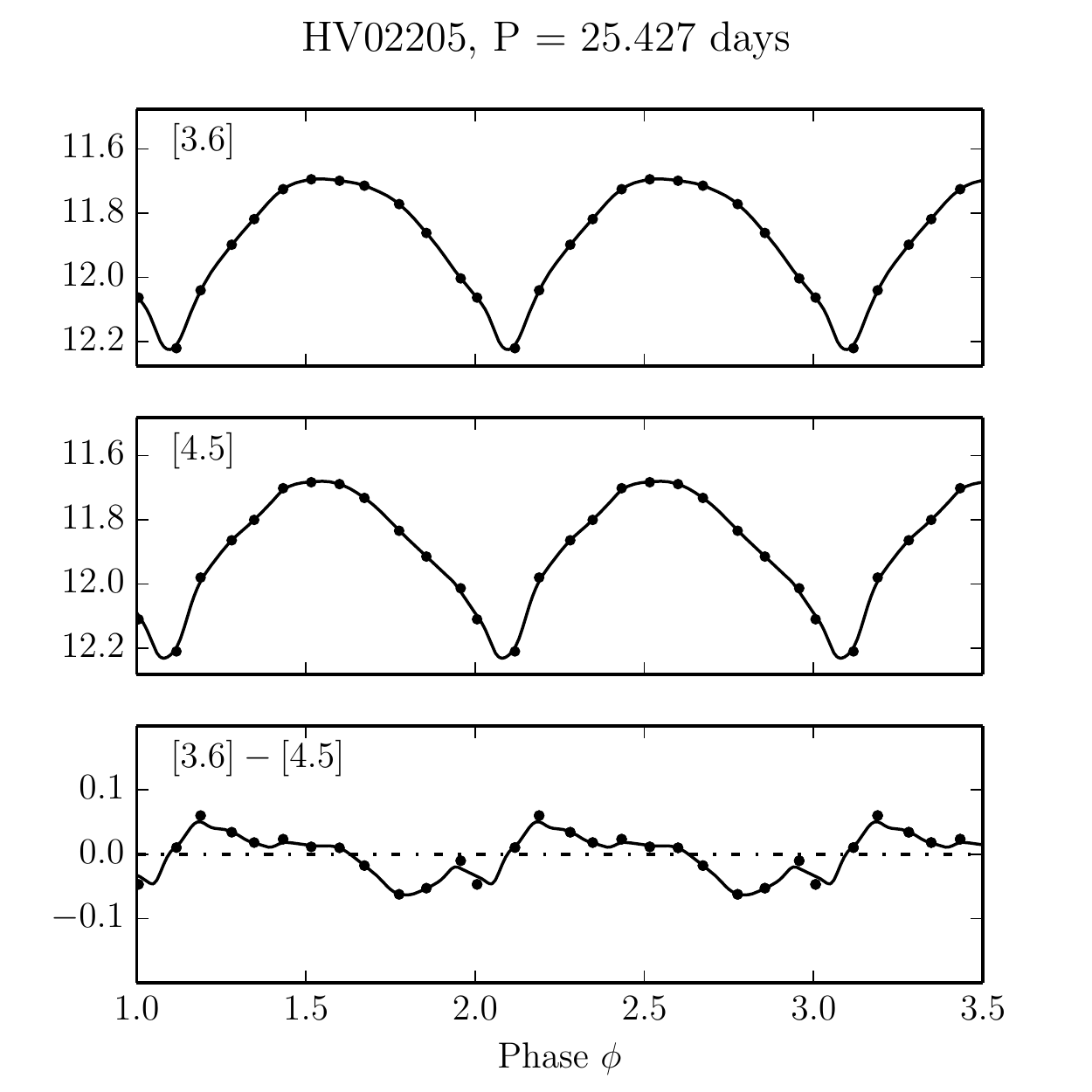} \\ 
\includegraphics[width=50mm]{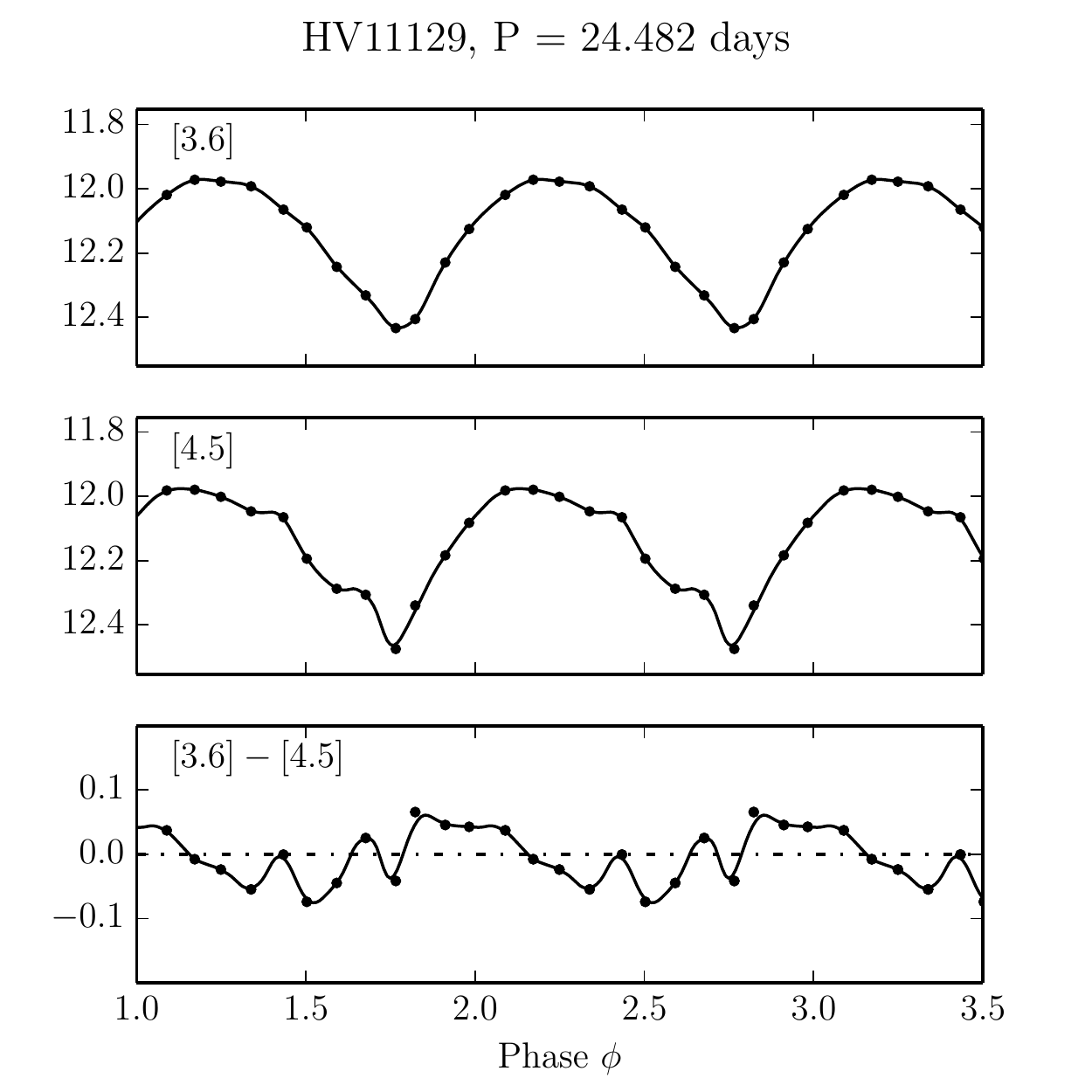} & 
\includegraphics[width=50mm]{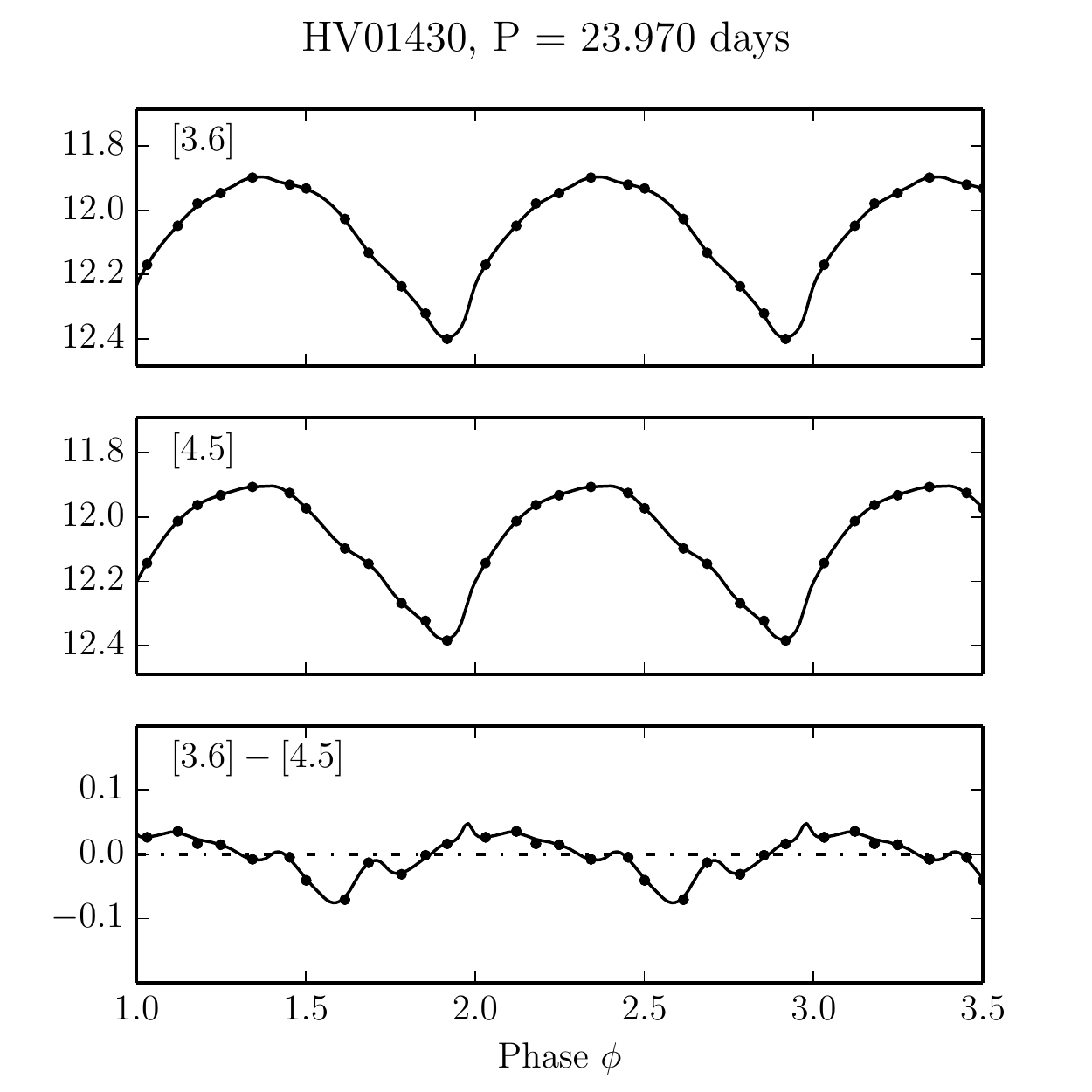} &
\includegraphics[width=50mm]{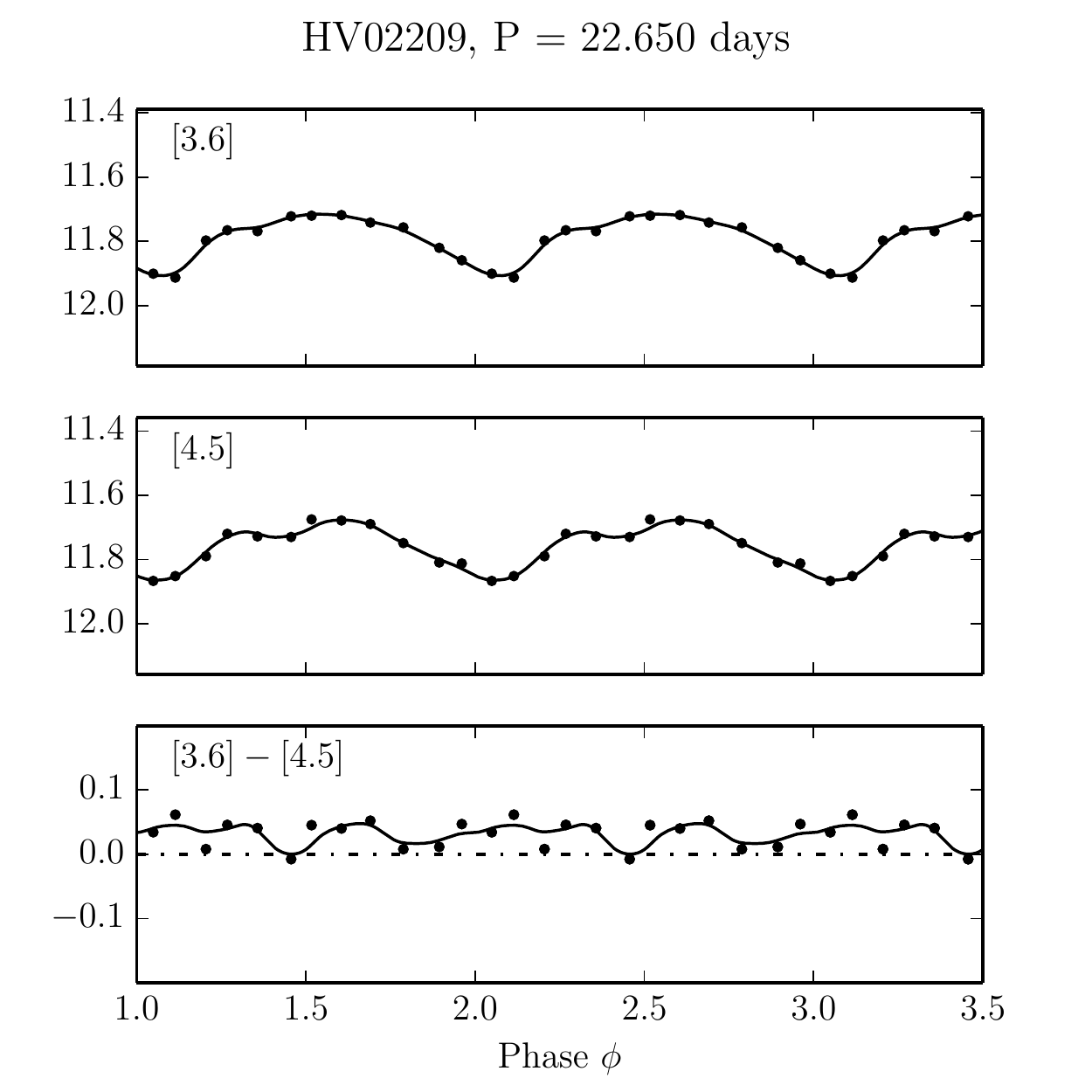} \\

\end{array}$ 
\end{center} 
\end{figure}
\begin{figure} 
 \begin{center}$ 
 \begin{array}{ccc} 
\includegraphics[width=50mm]{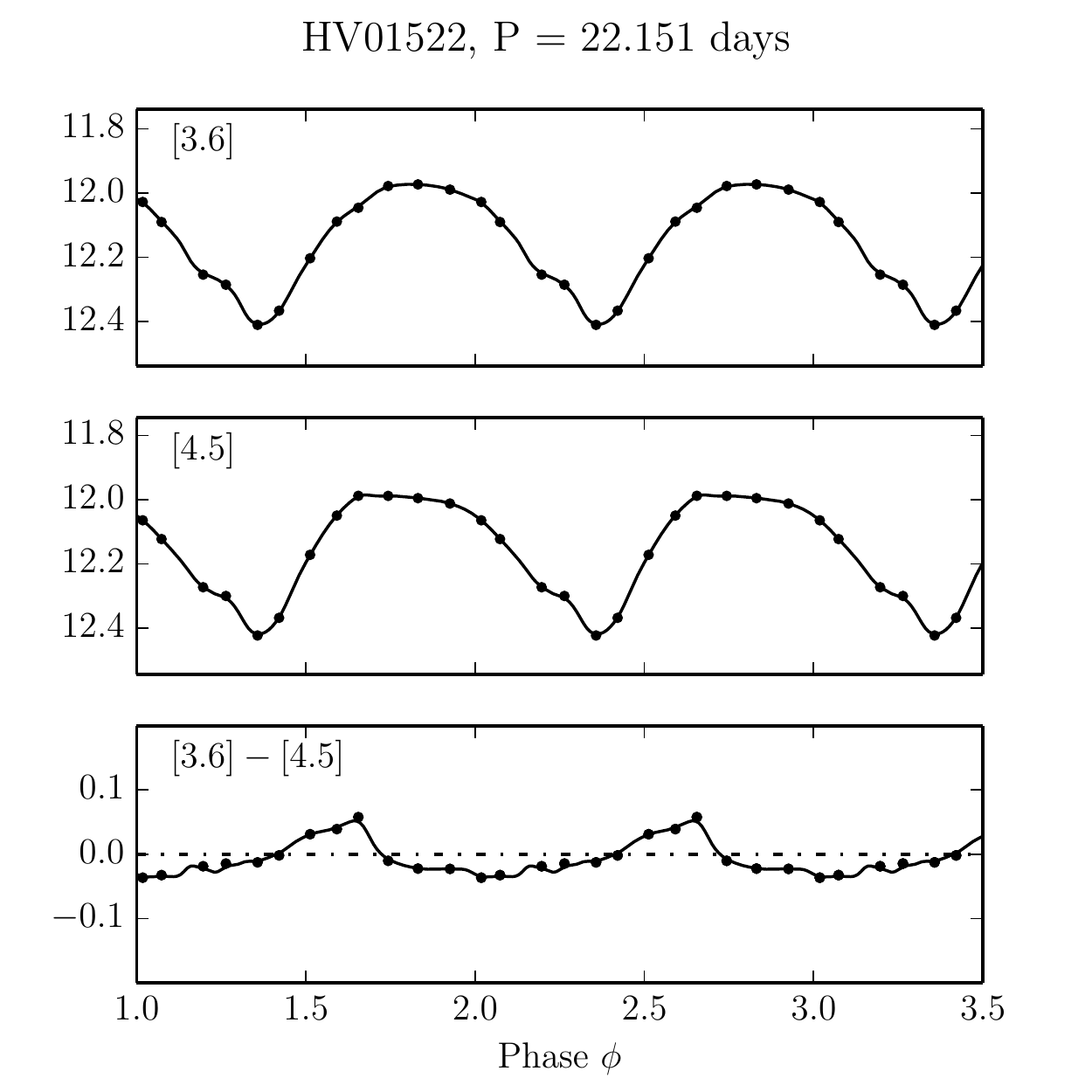} &
\includegraphics[width=50mm]{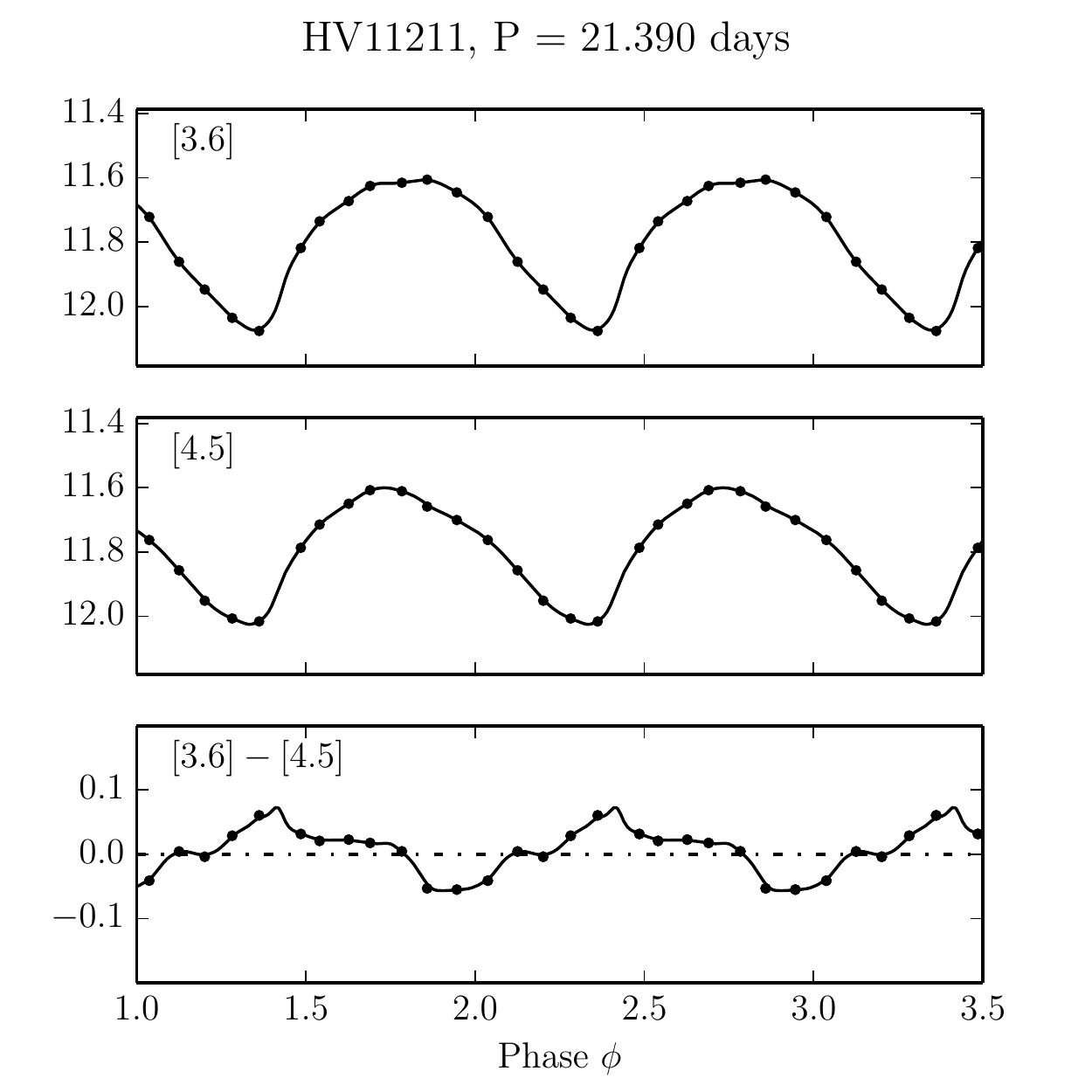} &
\includegraphics[width=50mm]{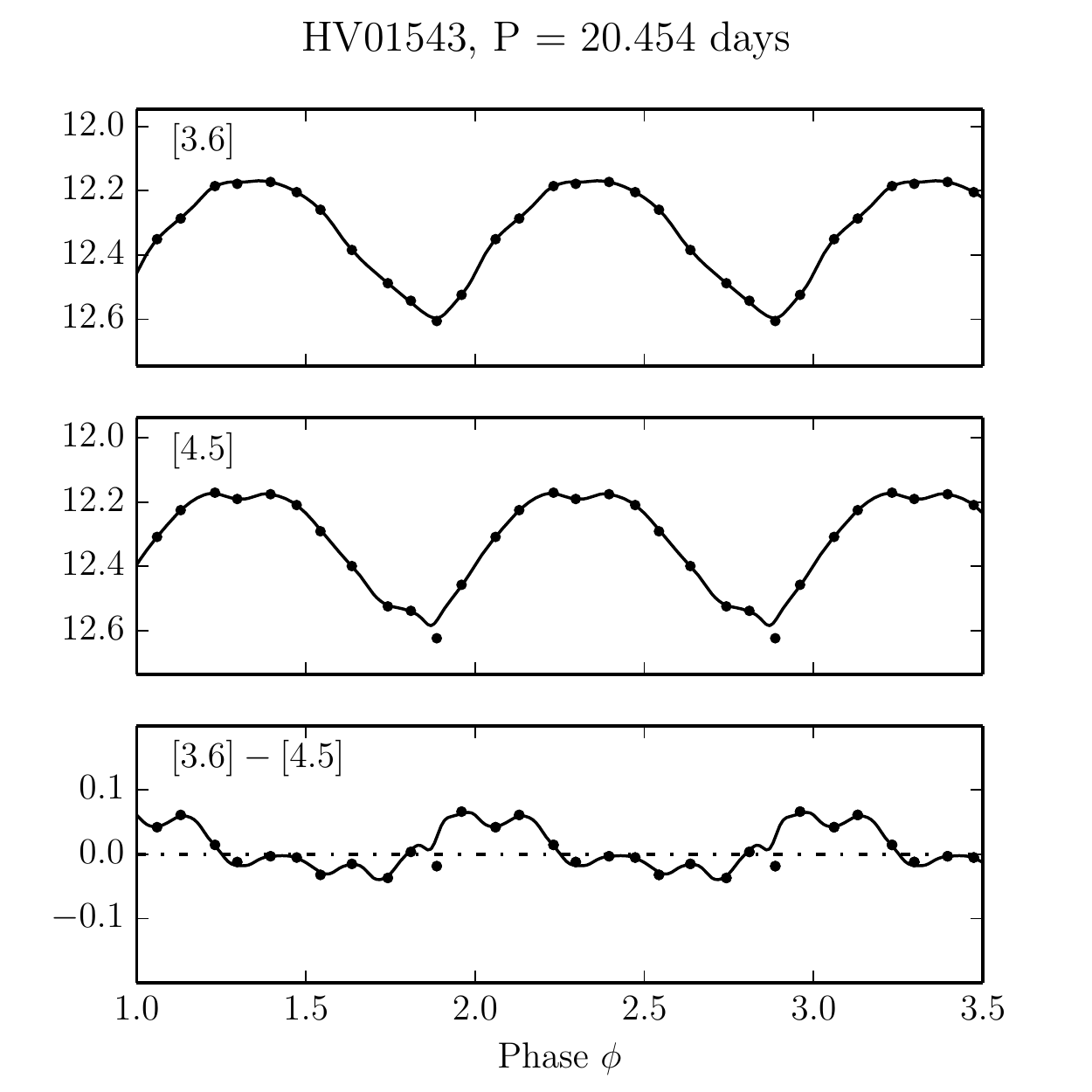} \\ 
\includegraphics[width=50mm]{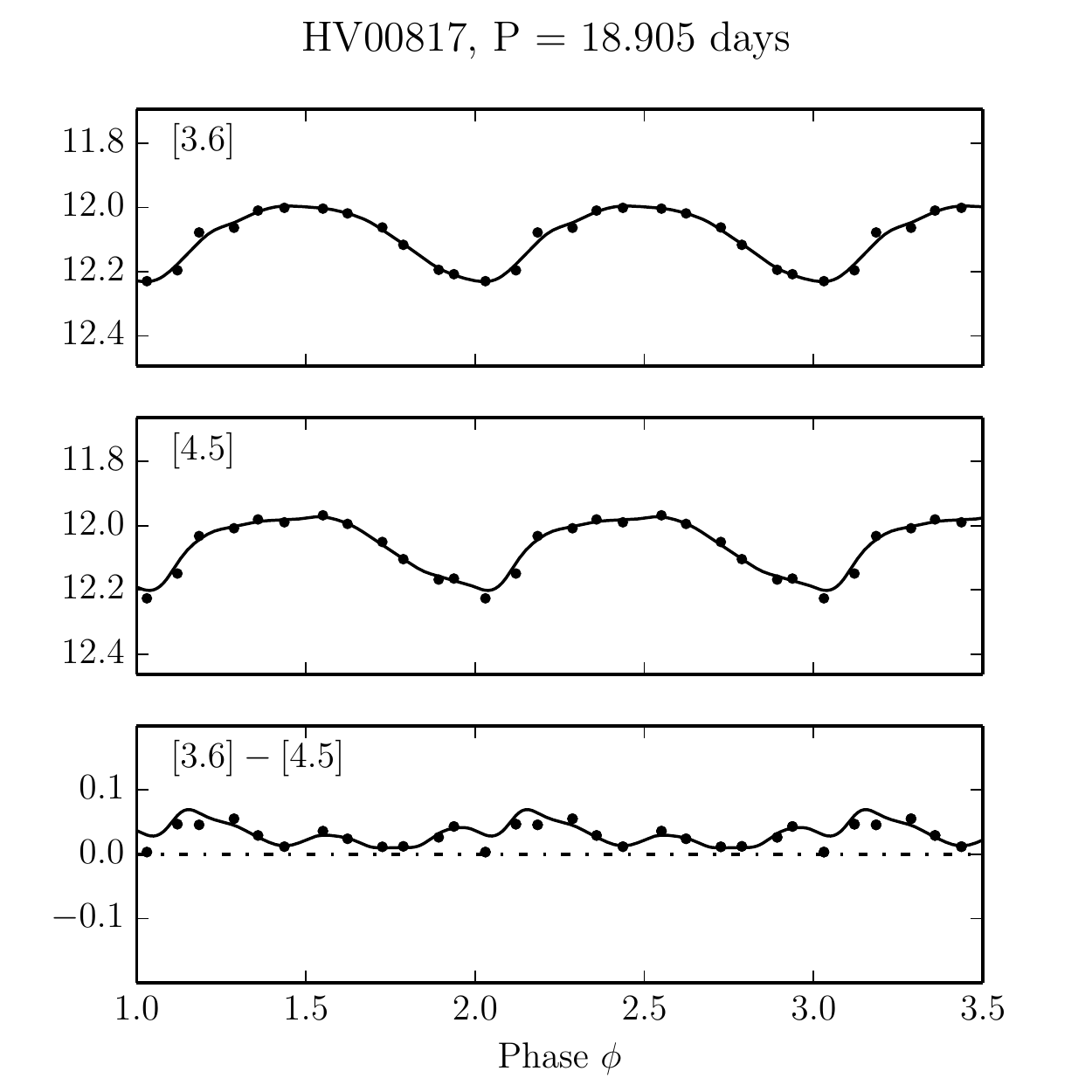} &
\includegraphics[width=50mm]{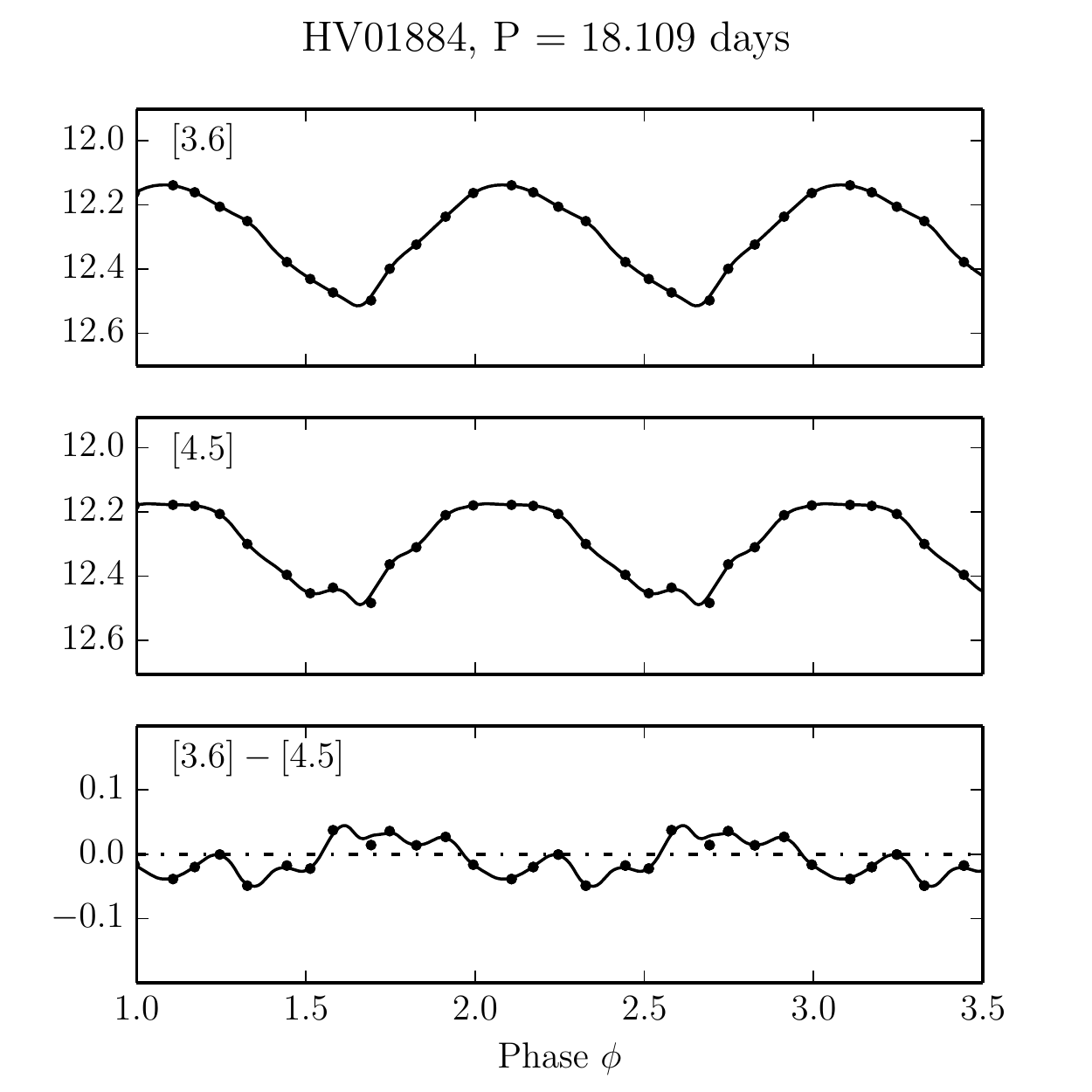} &
\includegraphics[width=50mm]{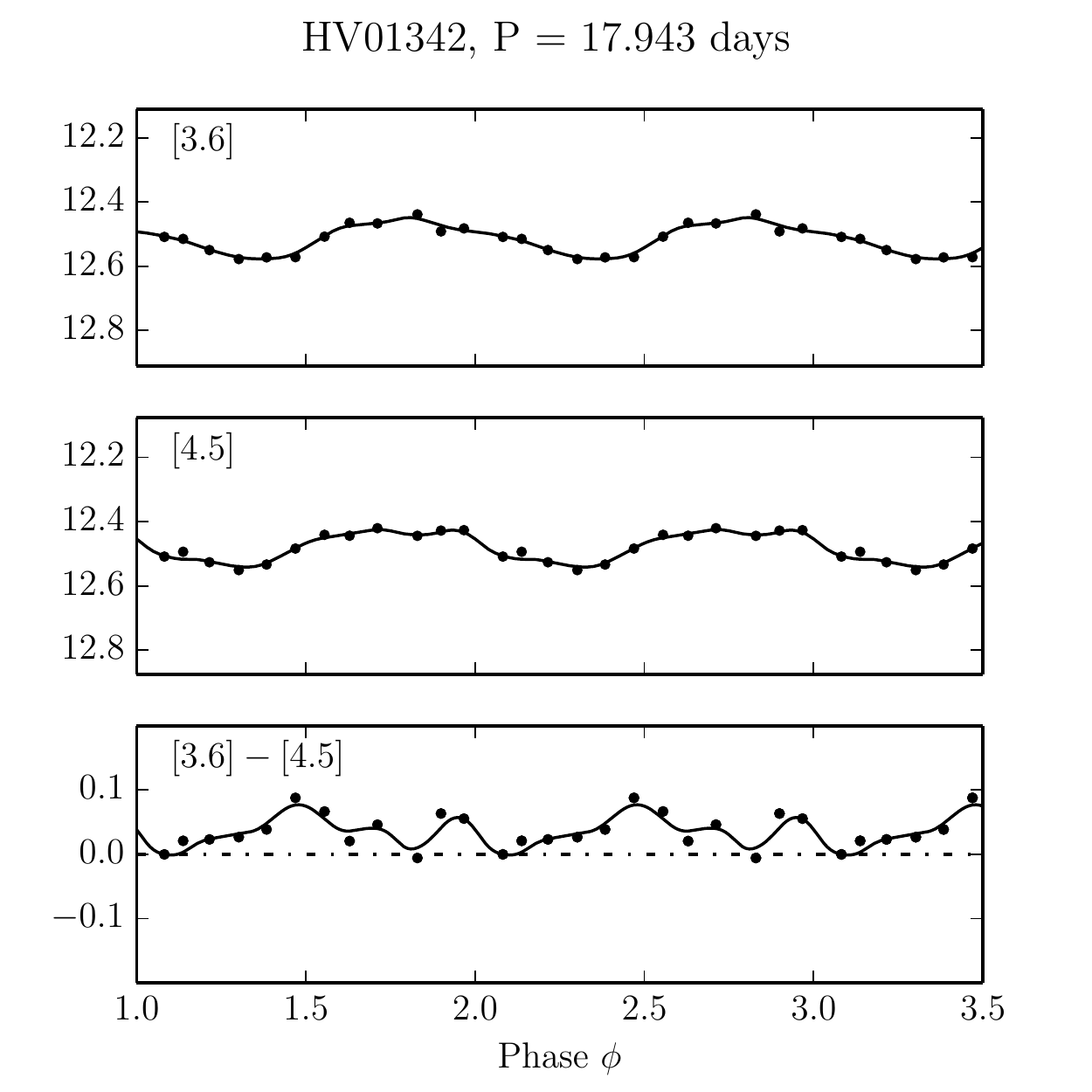} \\ 
\includegraphics[width=50mm]{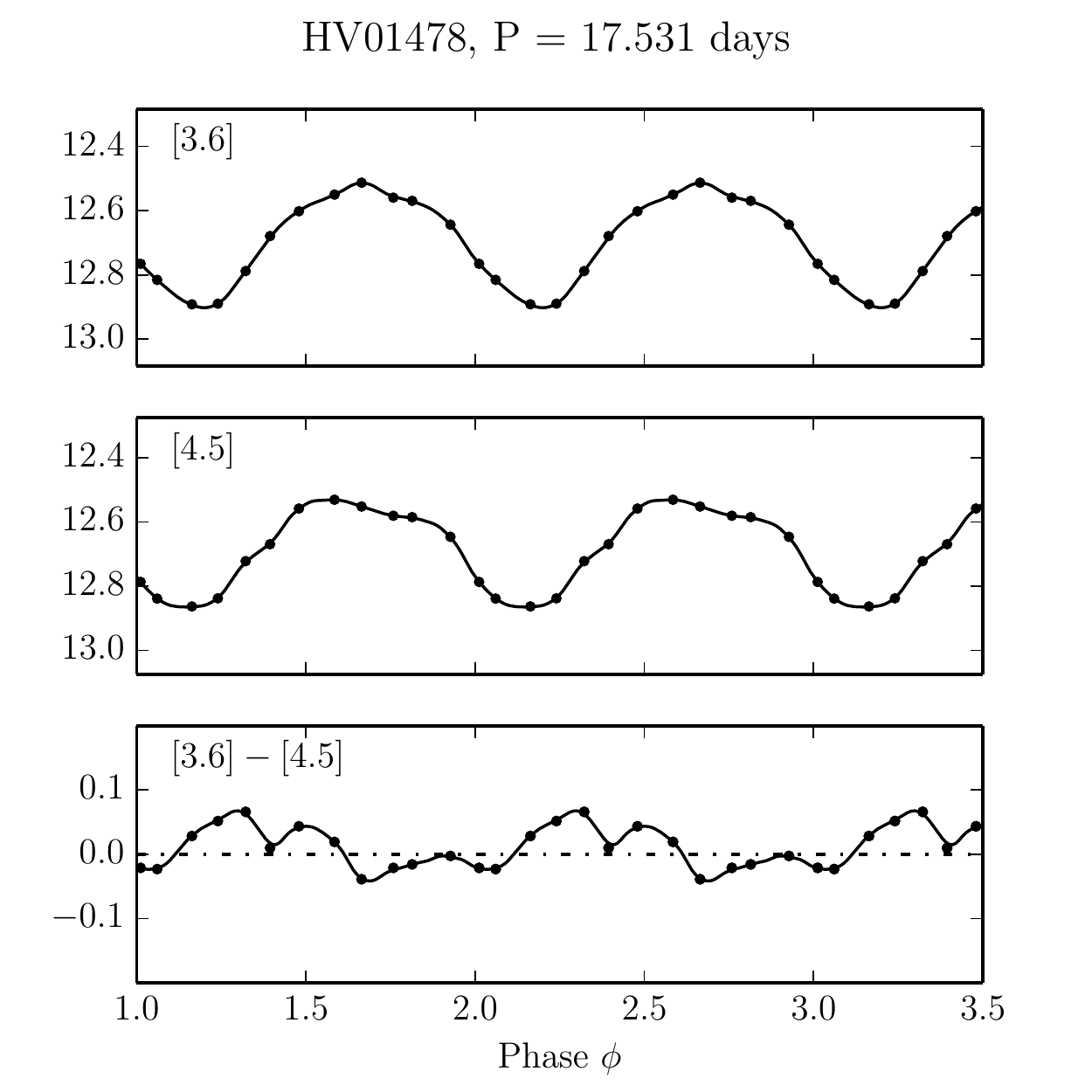} & 
\includegraphics[width=50mm]{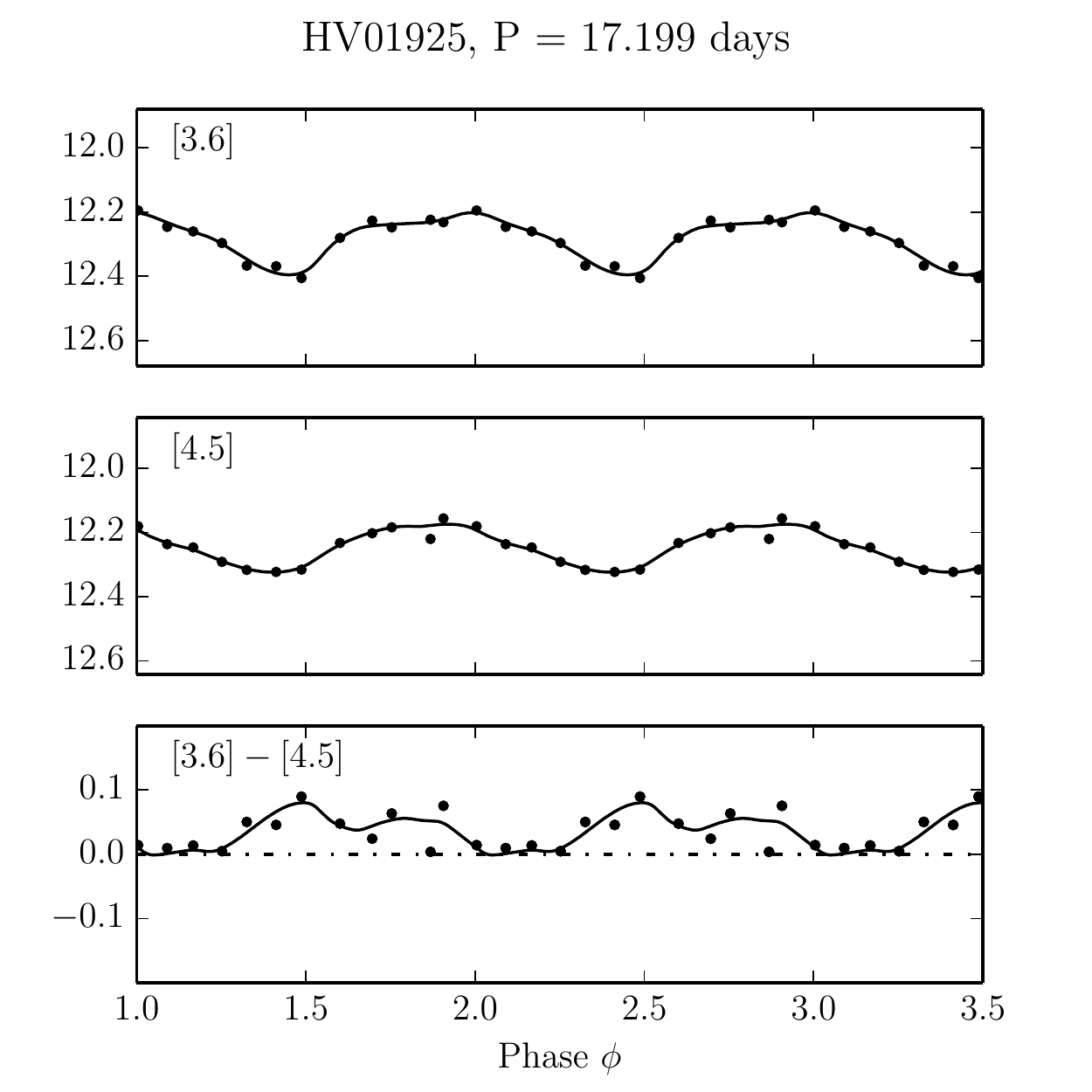} &
\includegraphics[width=50mm]{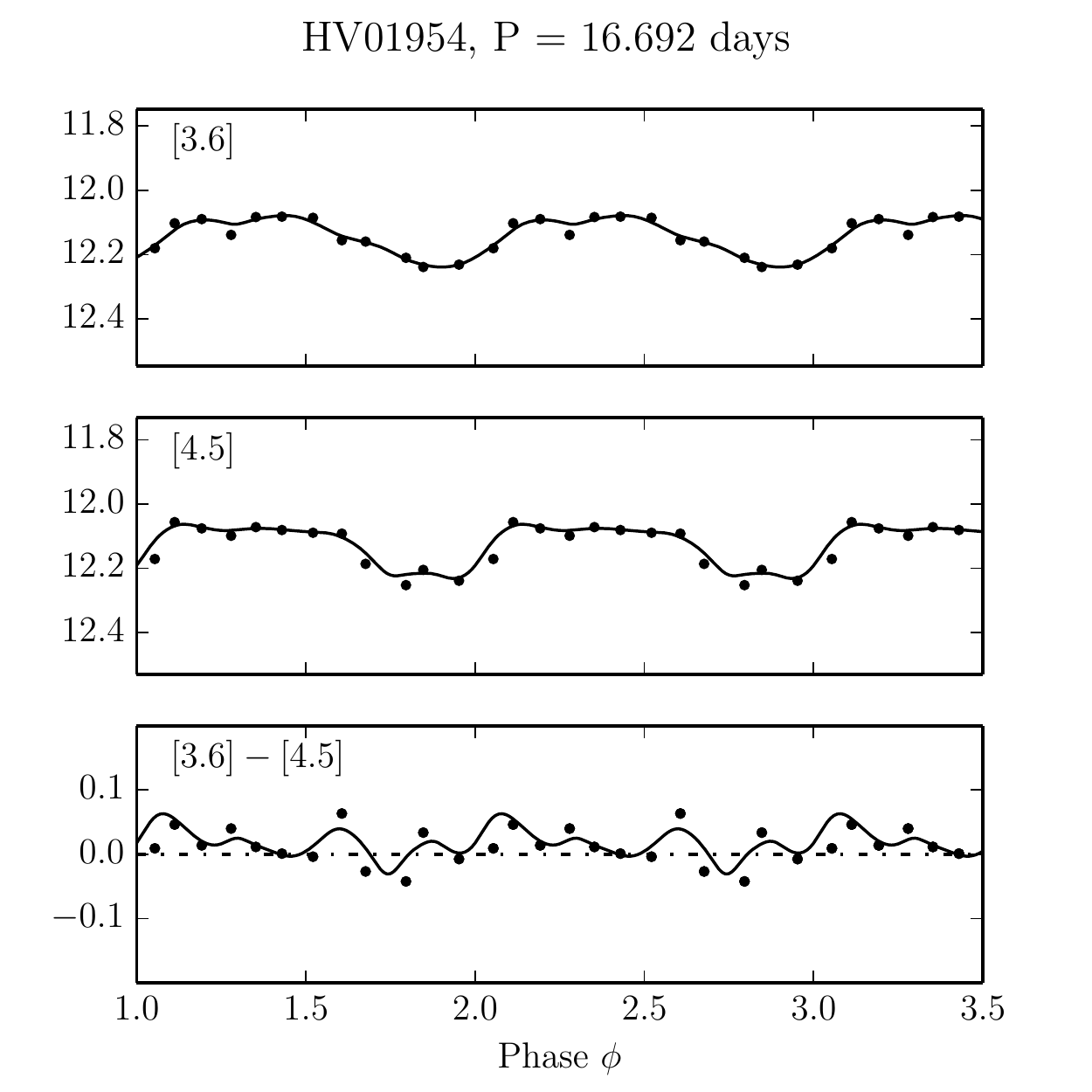} \\ 
\includegraphics[width=50mm]{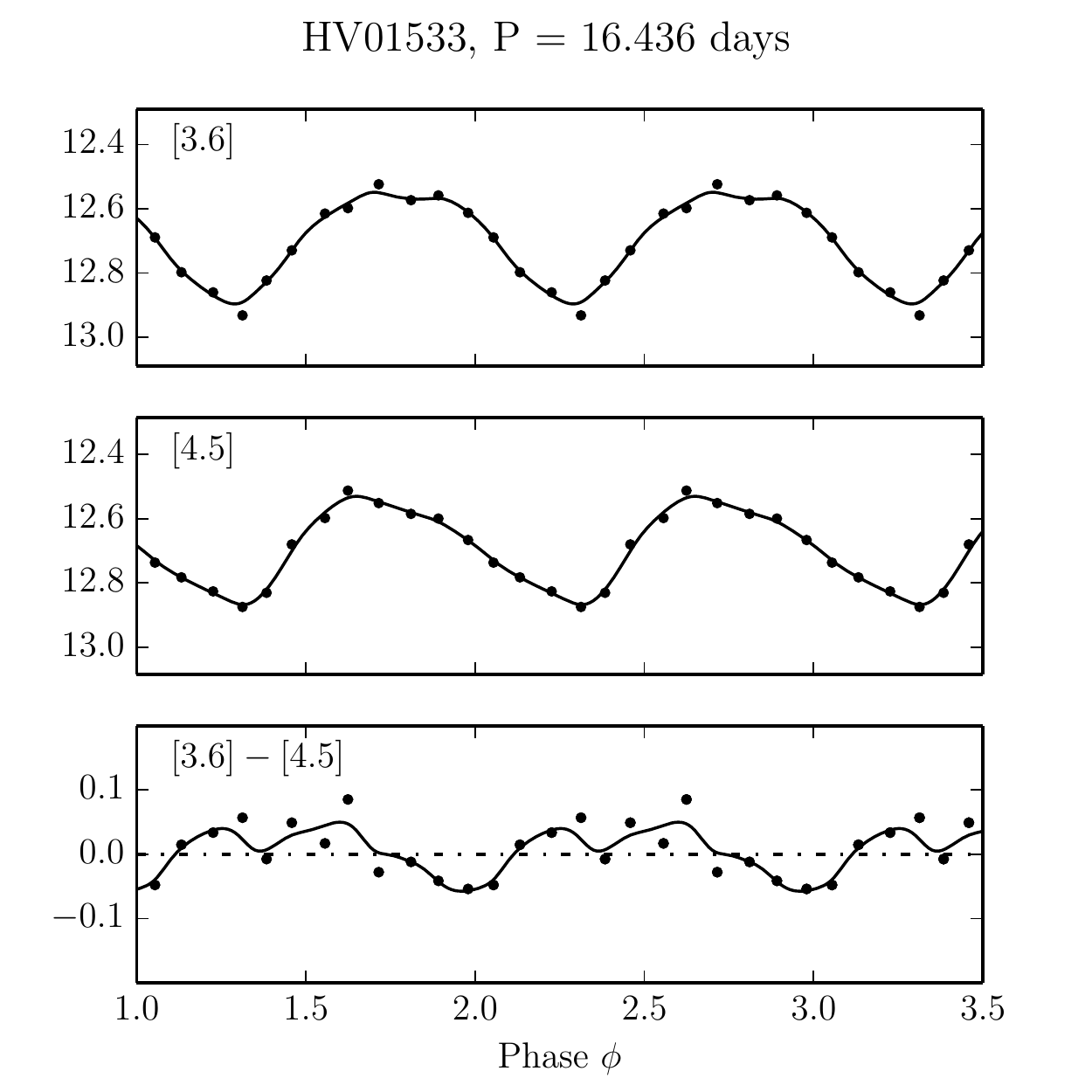} &
\includegraphics[width=50mm]{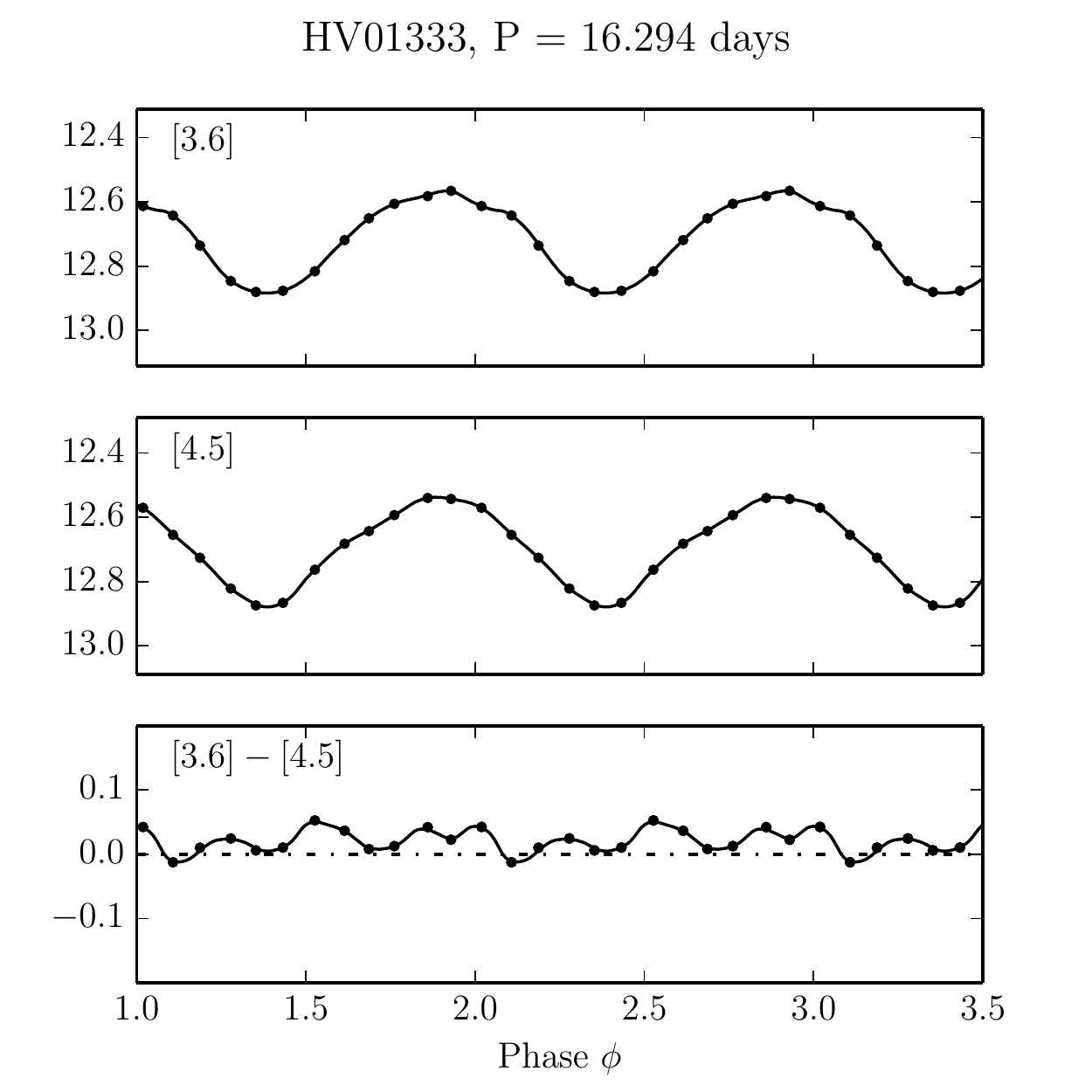} &
\includegraphics[width=50mm]{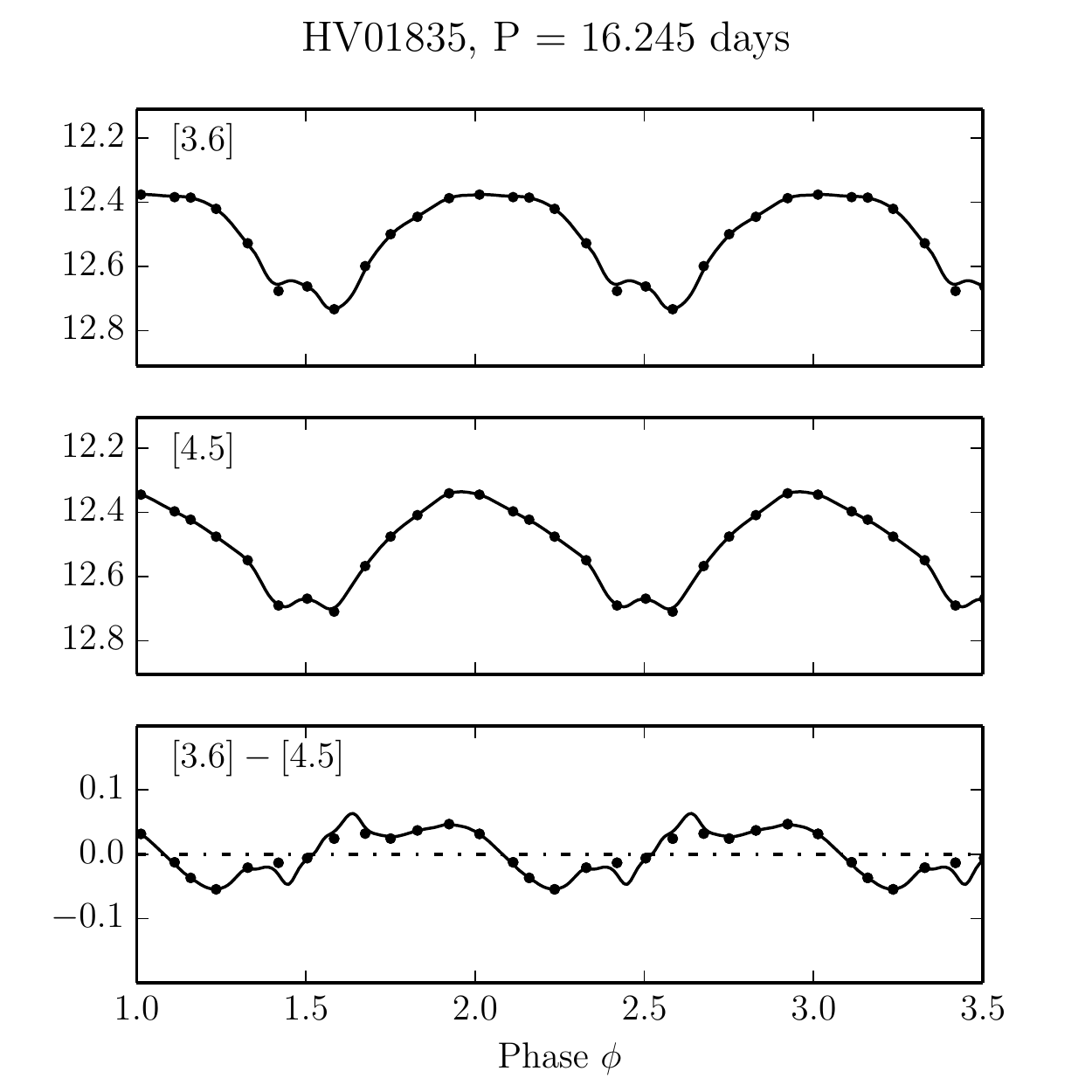} \\

\end{array}$ 
\end{center} 
\end{figure}
\begin{figure} 
 \begin{center}$ 
 \begin{array}{ccc} 
\includegraphics[width=50mm]{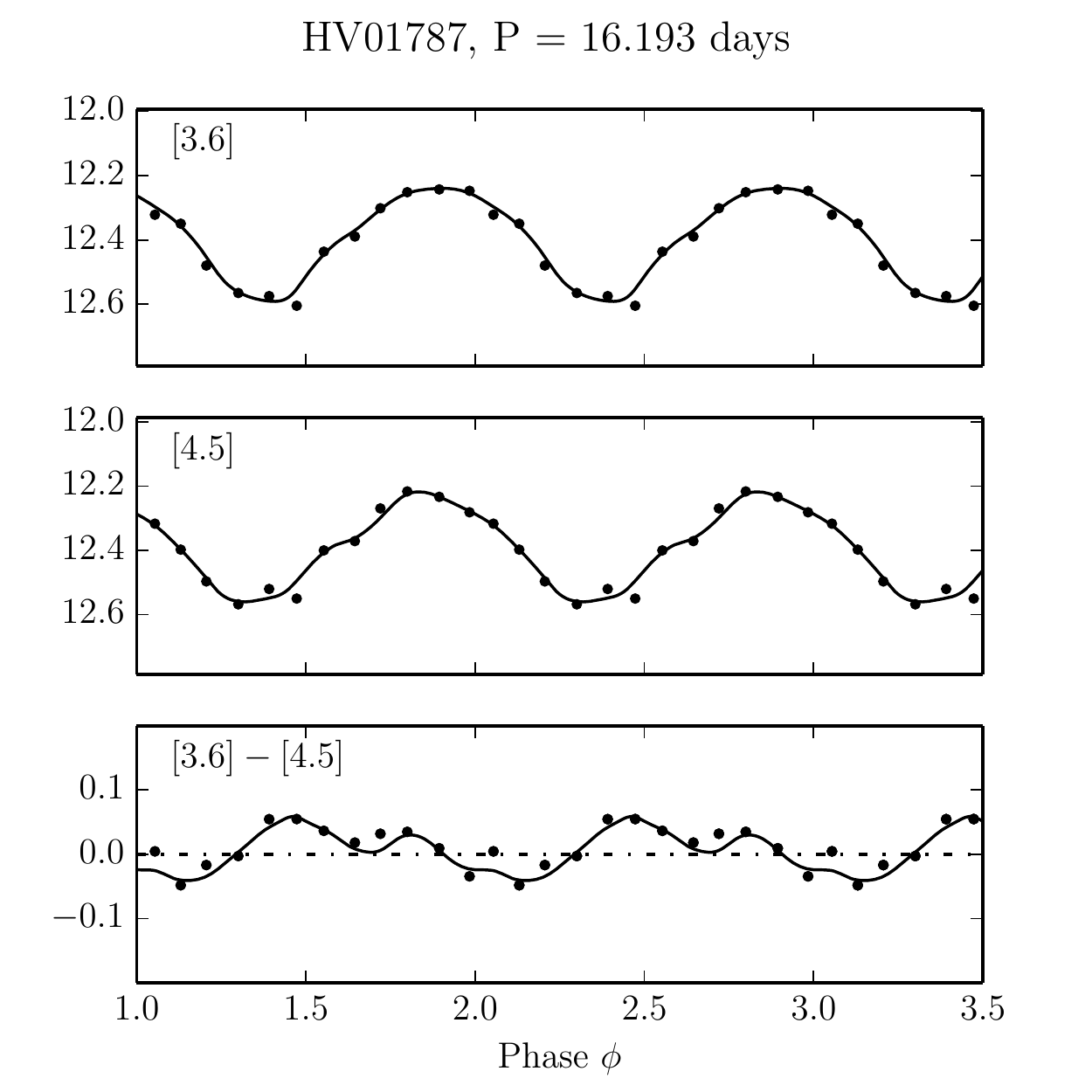} &
\includegraphics[width=50mm]{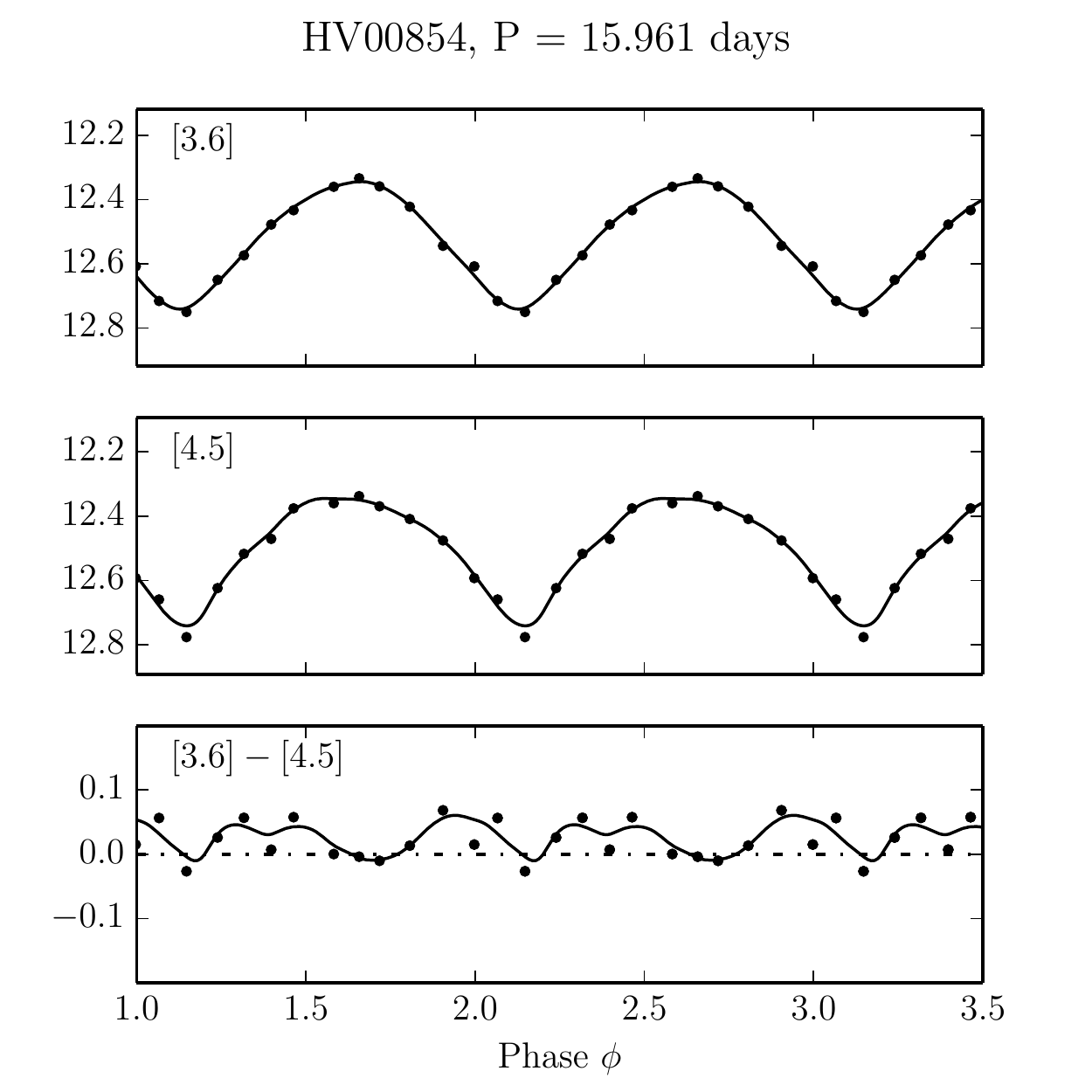} &
\includegraphics[width=50mm]{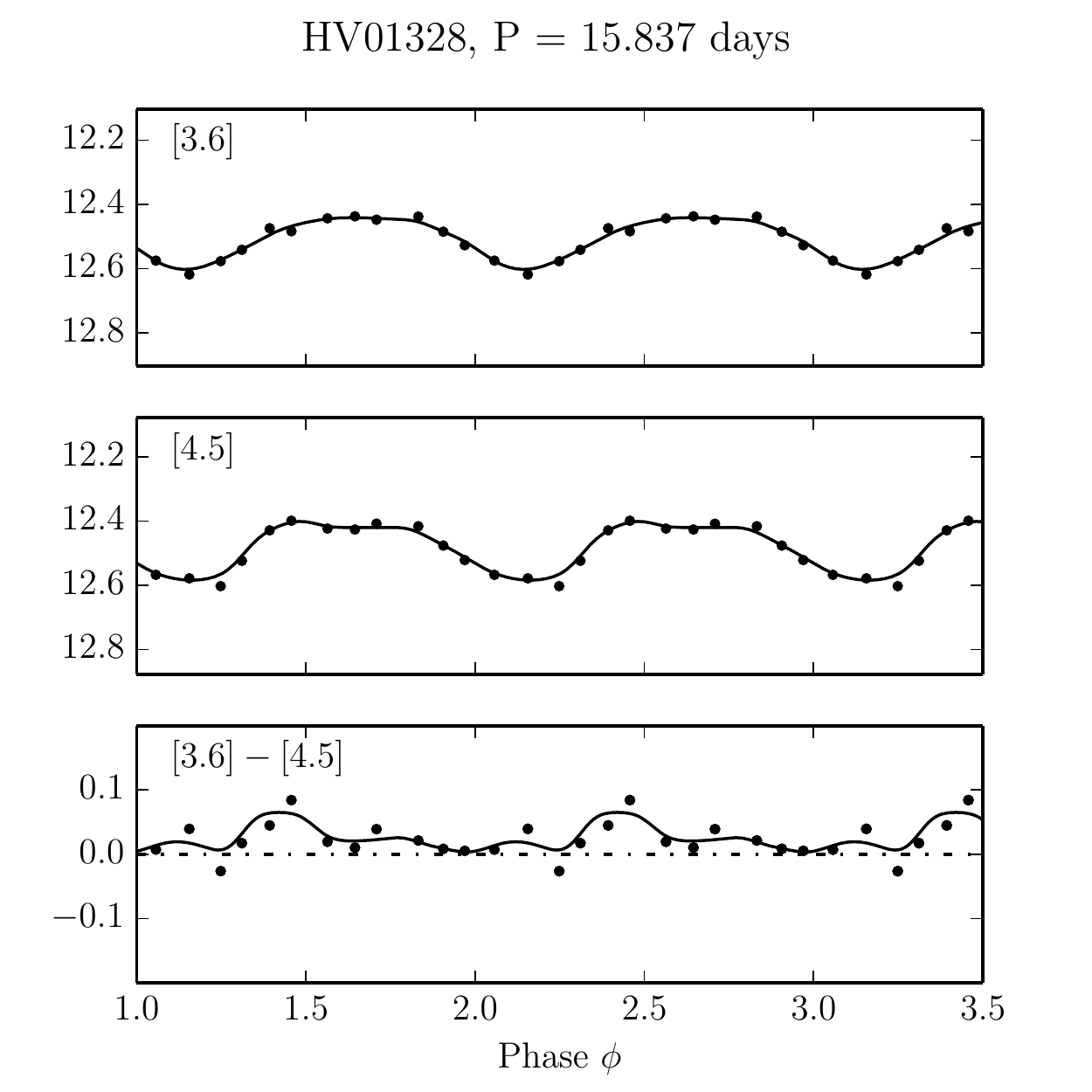} \\
\includegraphics[width=50mm]{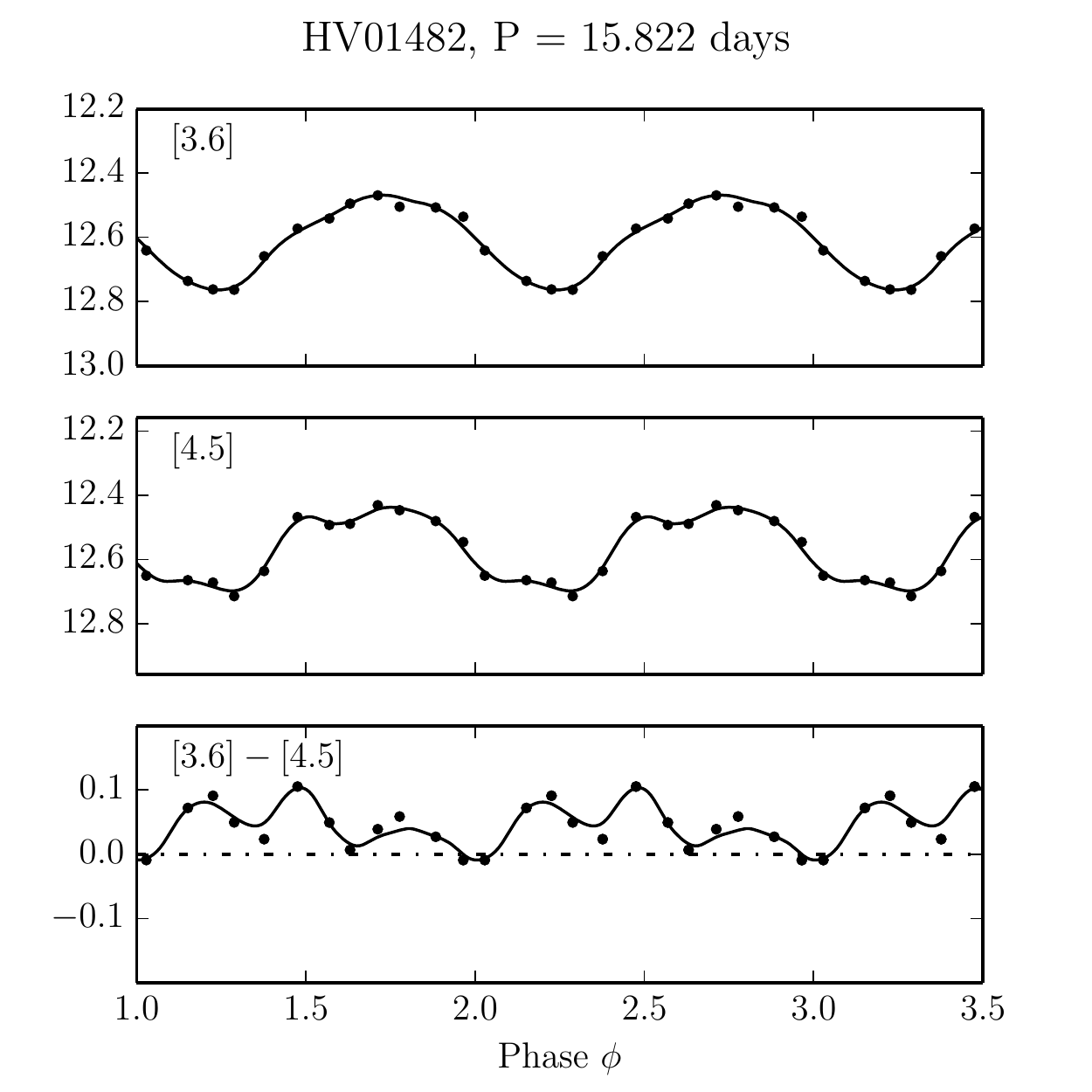}& 
\includegraphics[width=50mm]{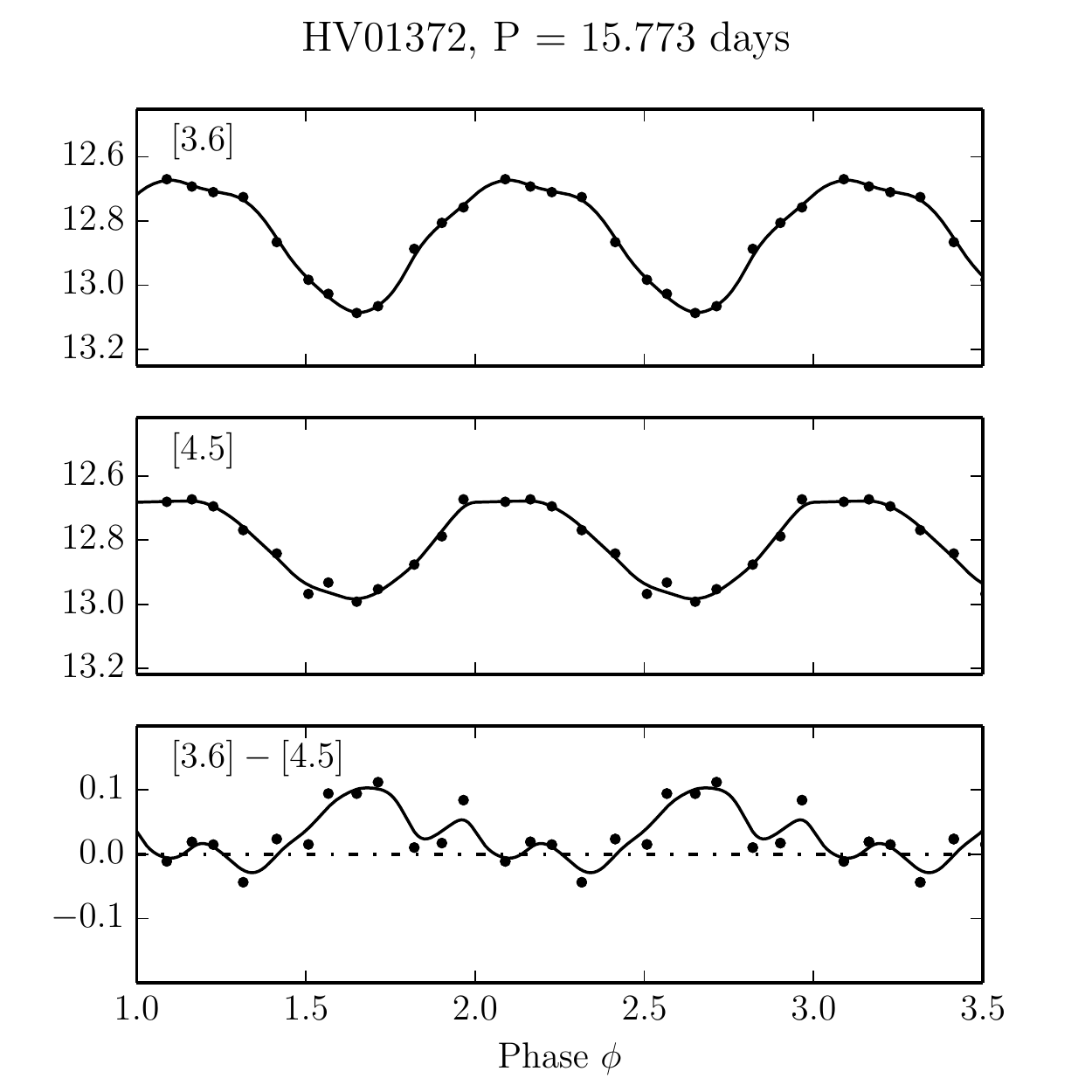}  &
\includegraphics[width=50mm]{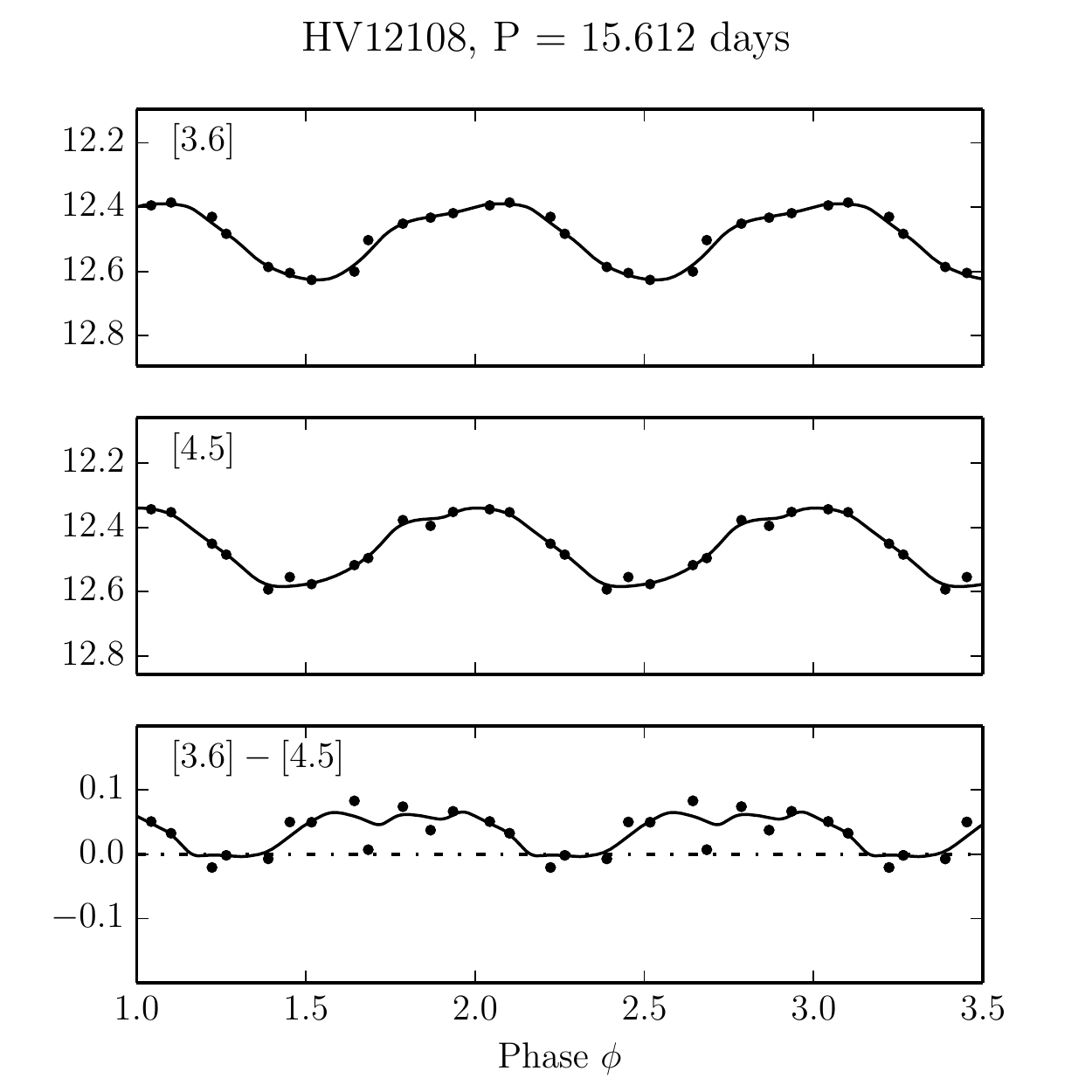} \\
\includegraphics[width=50mm]{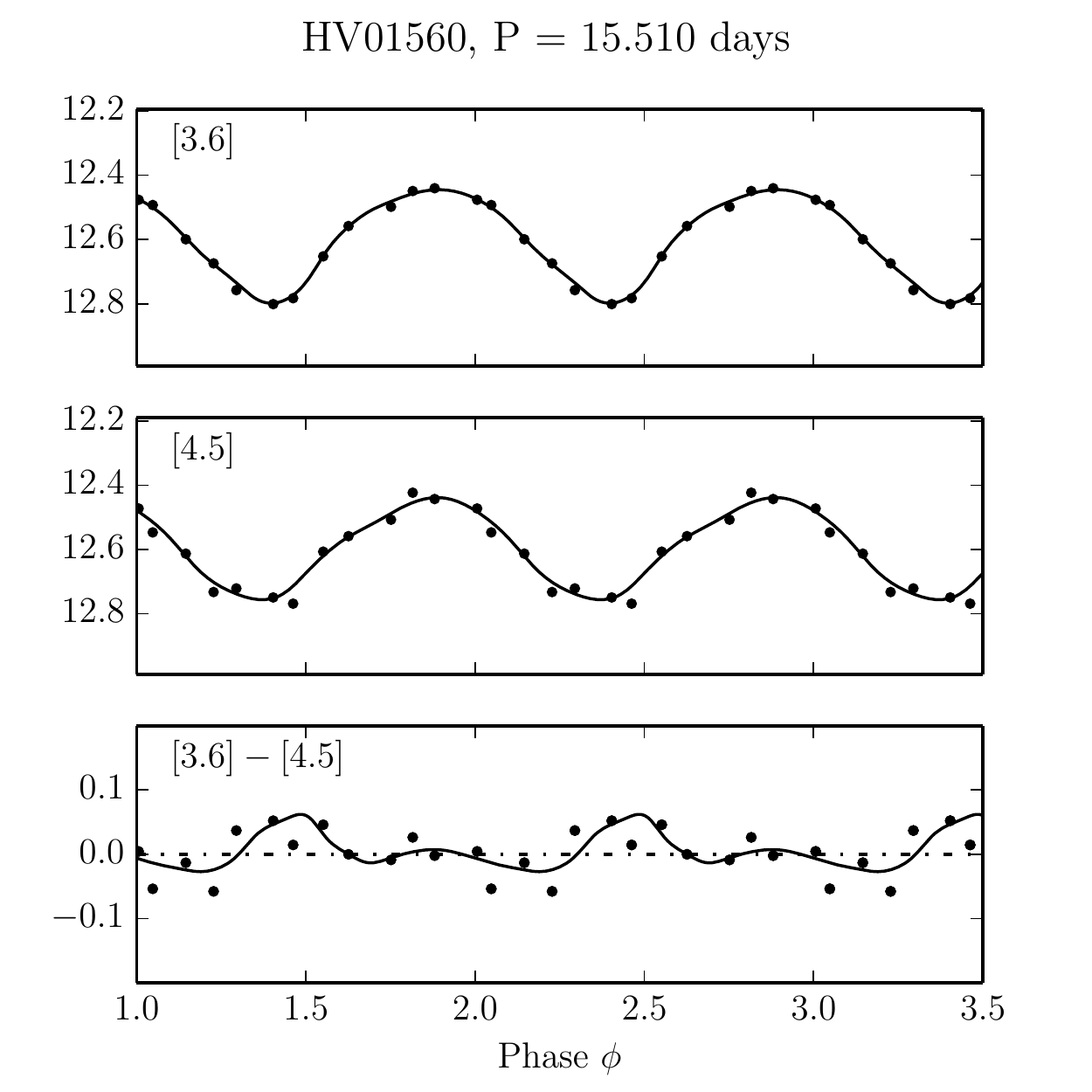}& 
\includegraphics[width=50mm]{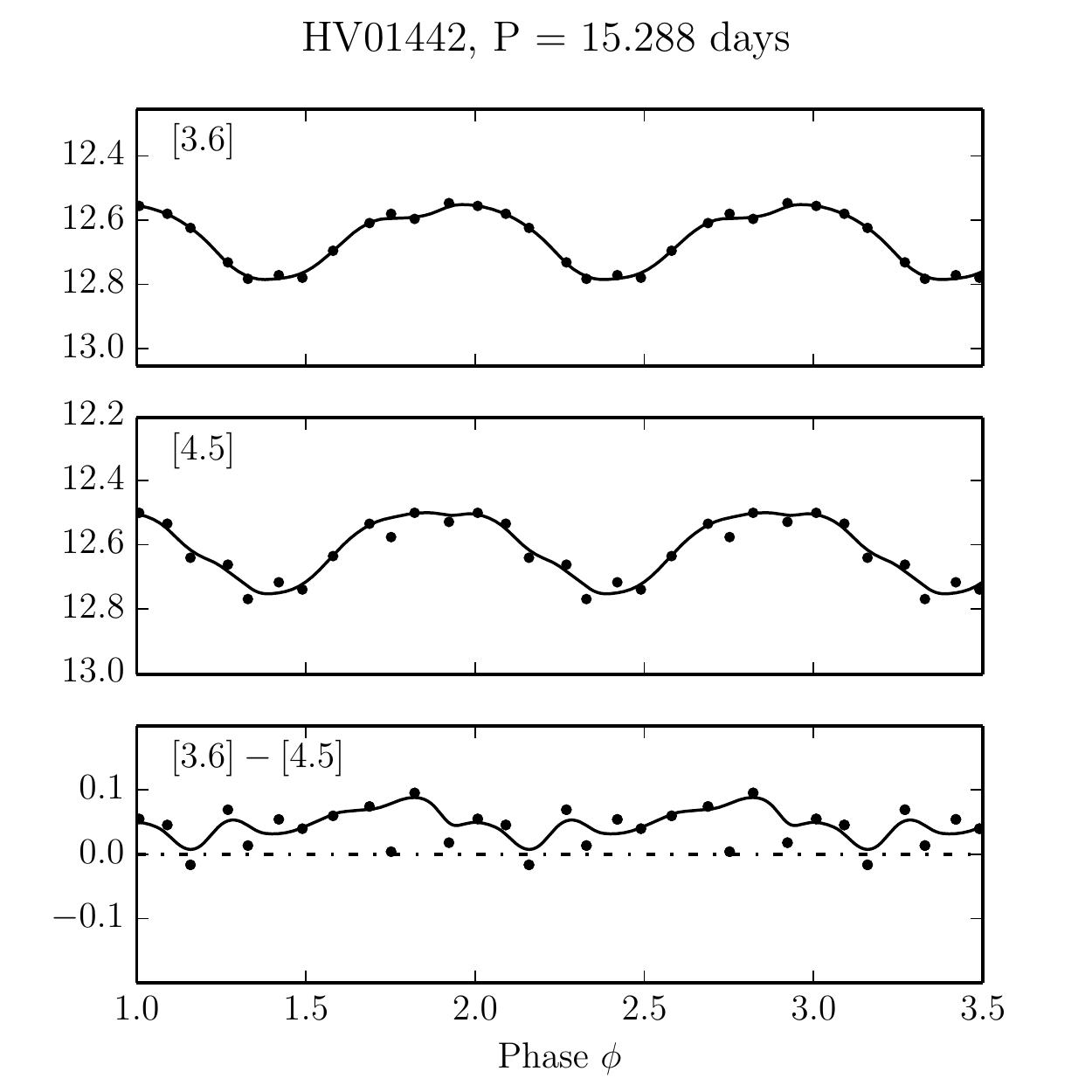} & 
\includegraphics[width=50mm]{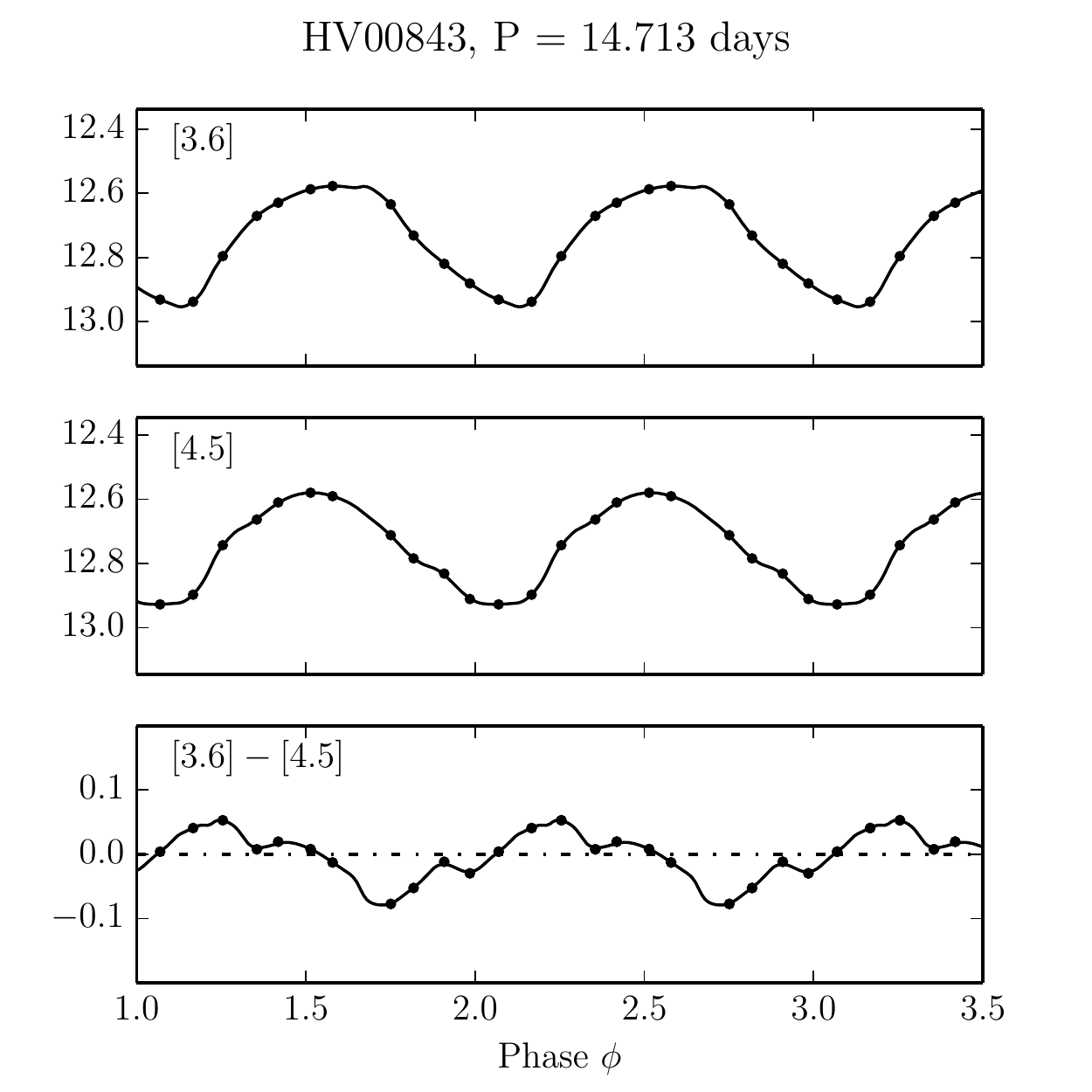} \\
\includegraphics[width=50mm]{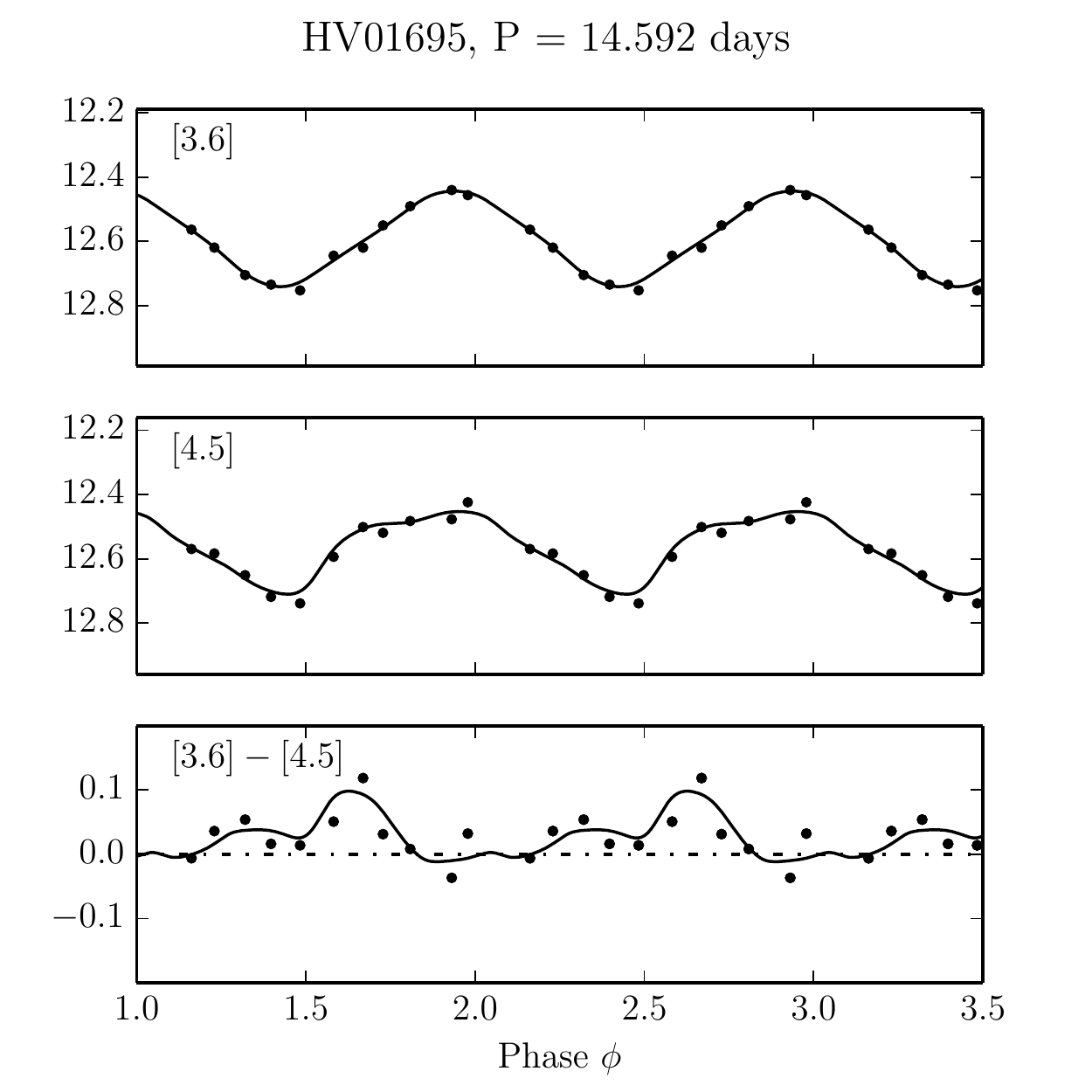} &
\includegraphics[width=50mm]{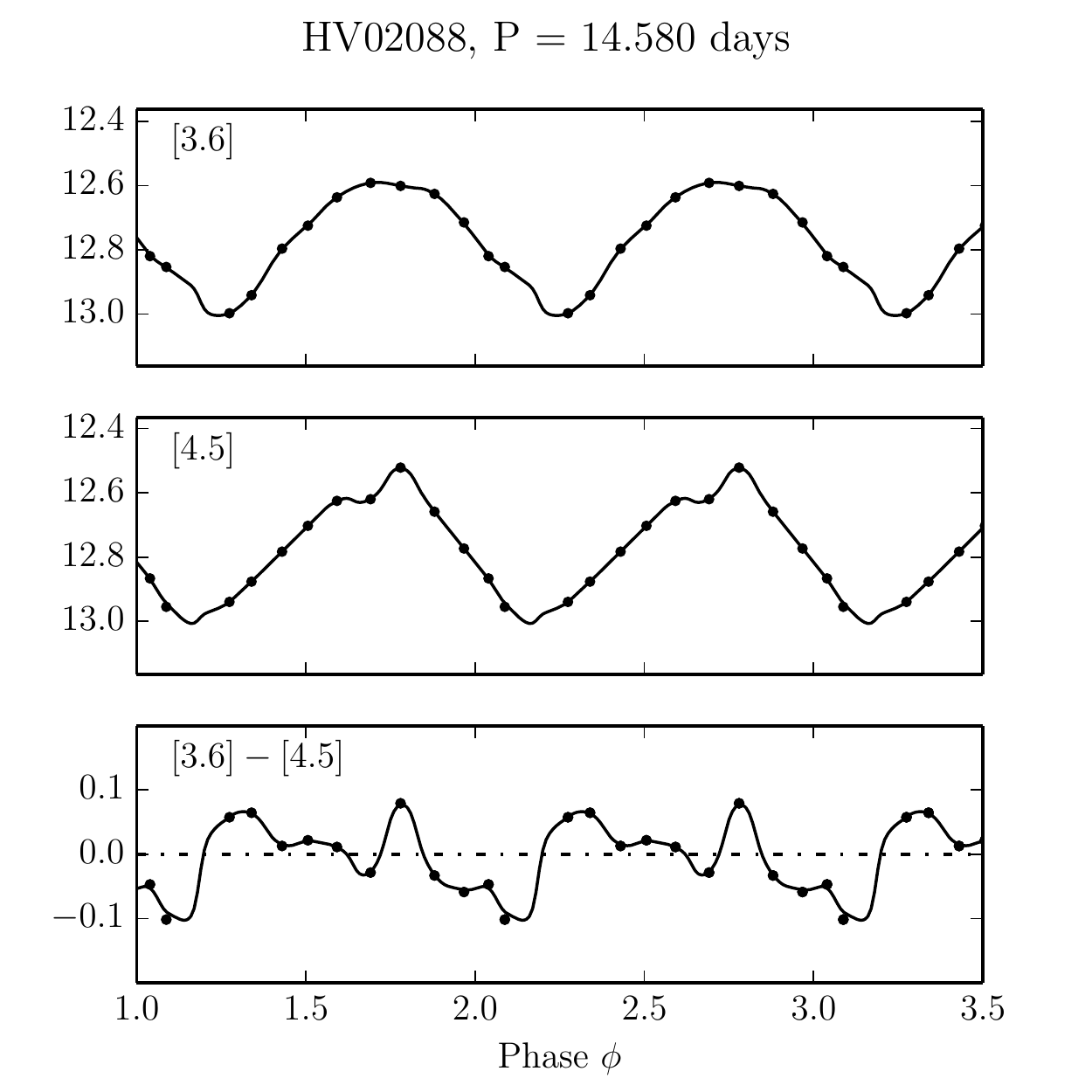} &
\includegraphics[width=50mm]{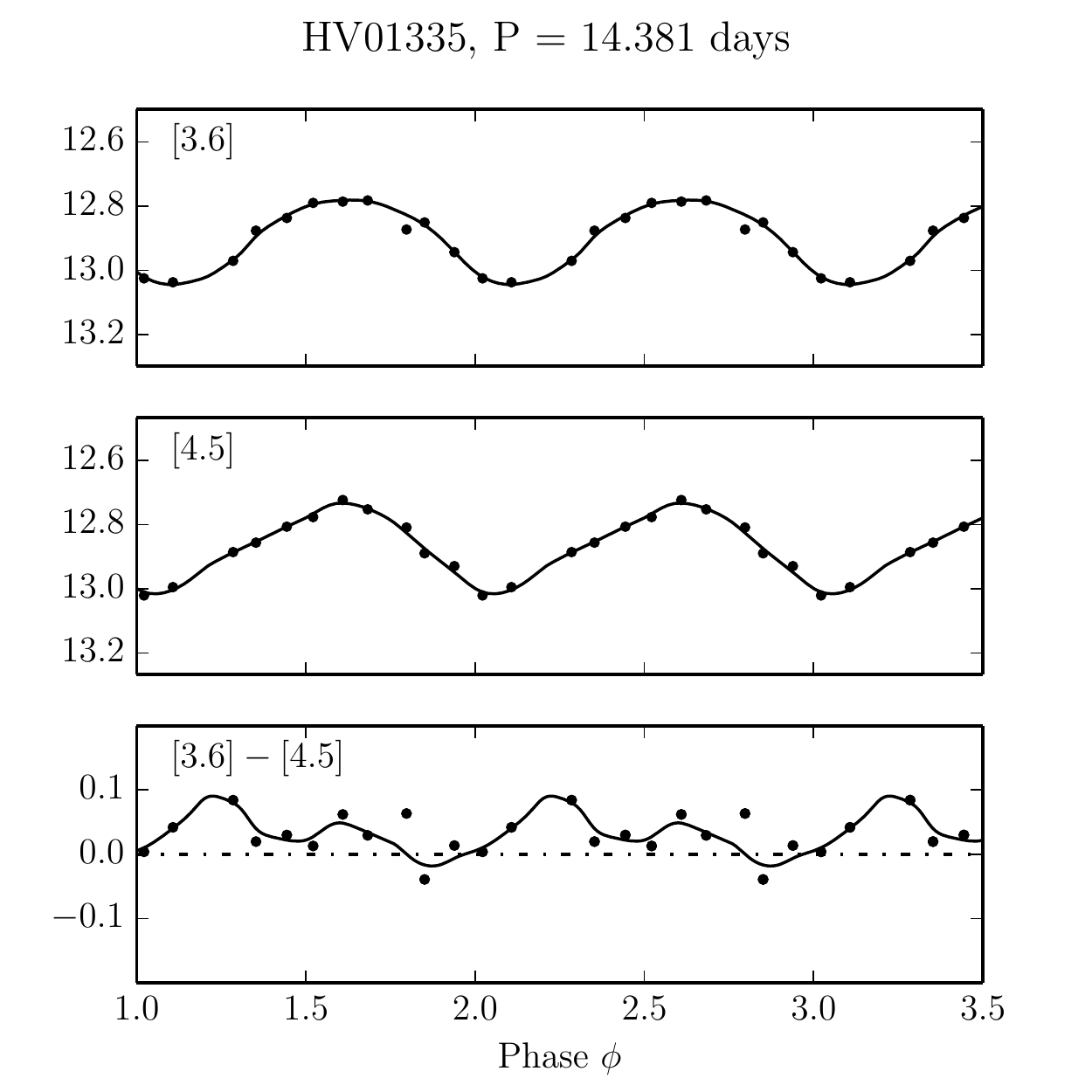} \\ 

\end{array}$ 
\end{center} 
\end{figure}
\begin{figure} 
 \begin{center}$ 
 \begin{array}{ccc} 
\includegraphics[width=50mm]{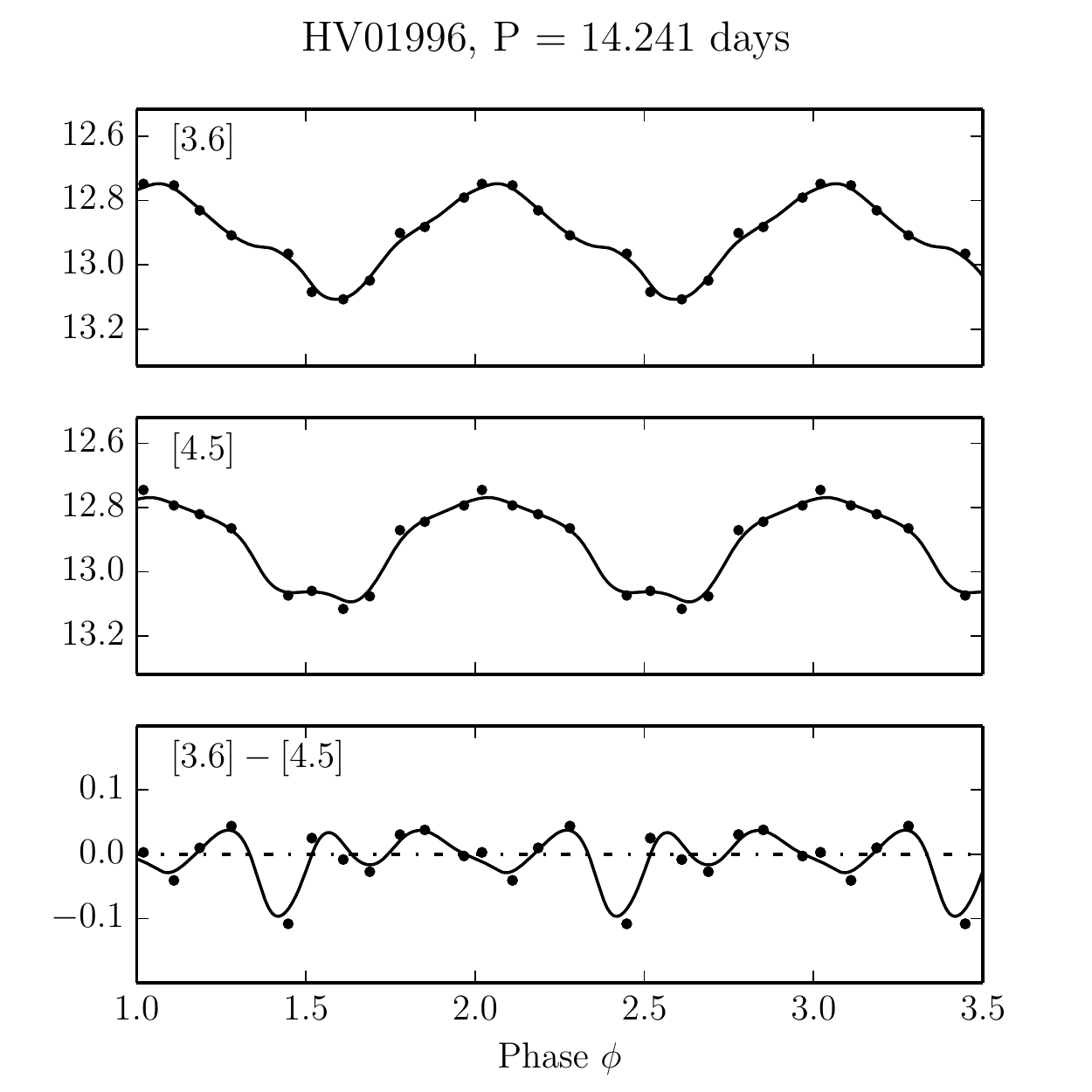}&
\includegraphics[width=50mm]{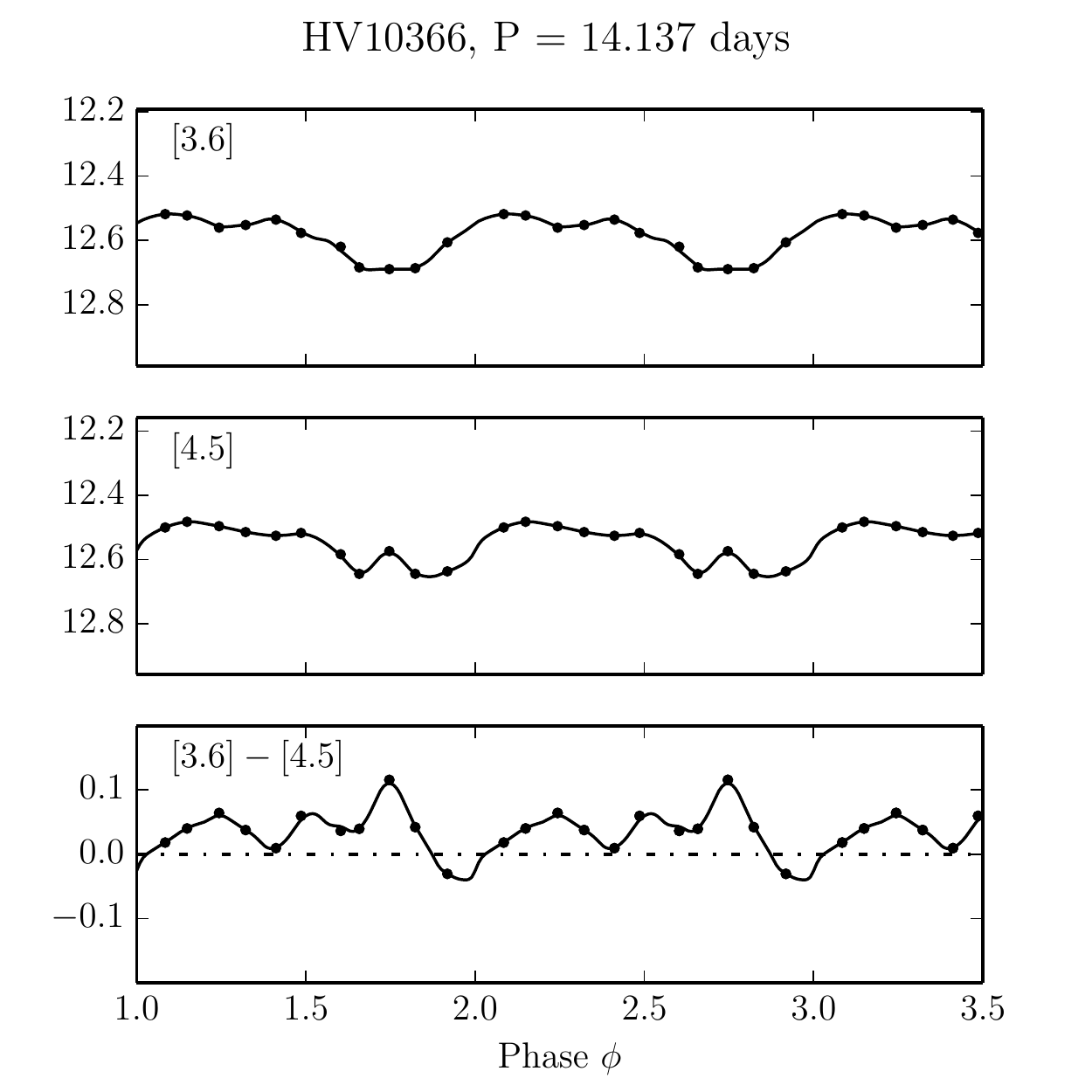} &
\includegraphics[width=50mm]{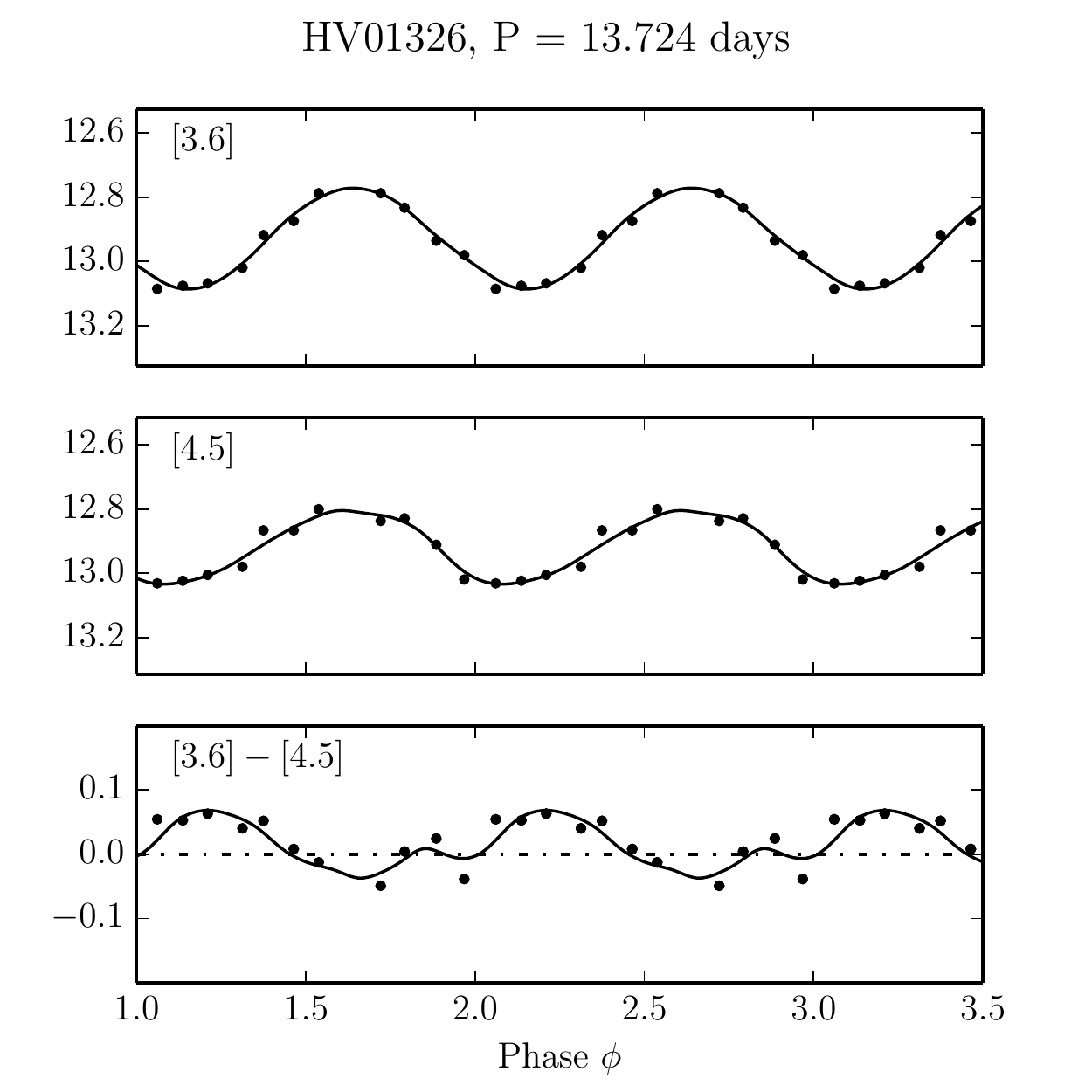} \\ 
\includegraphics[width=50mm]{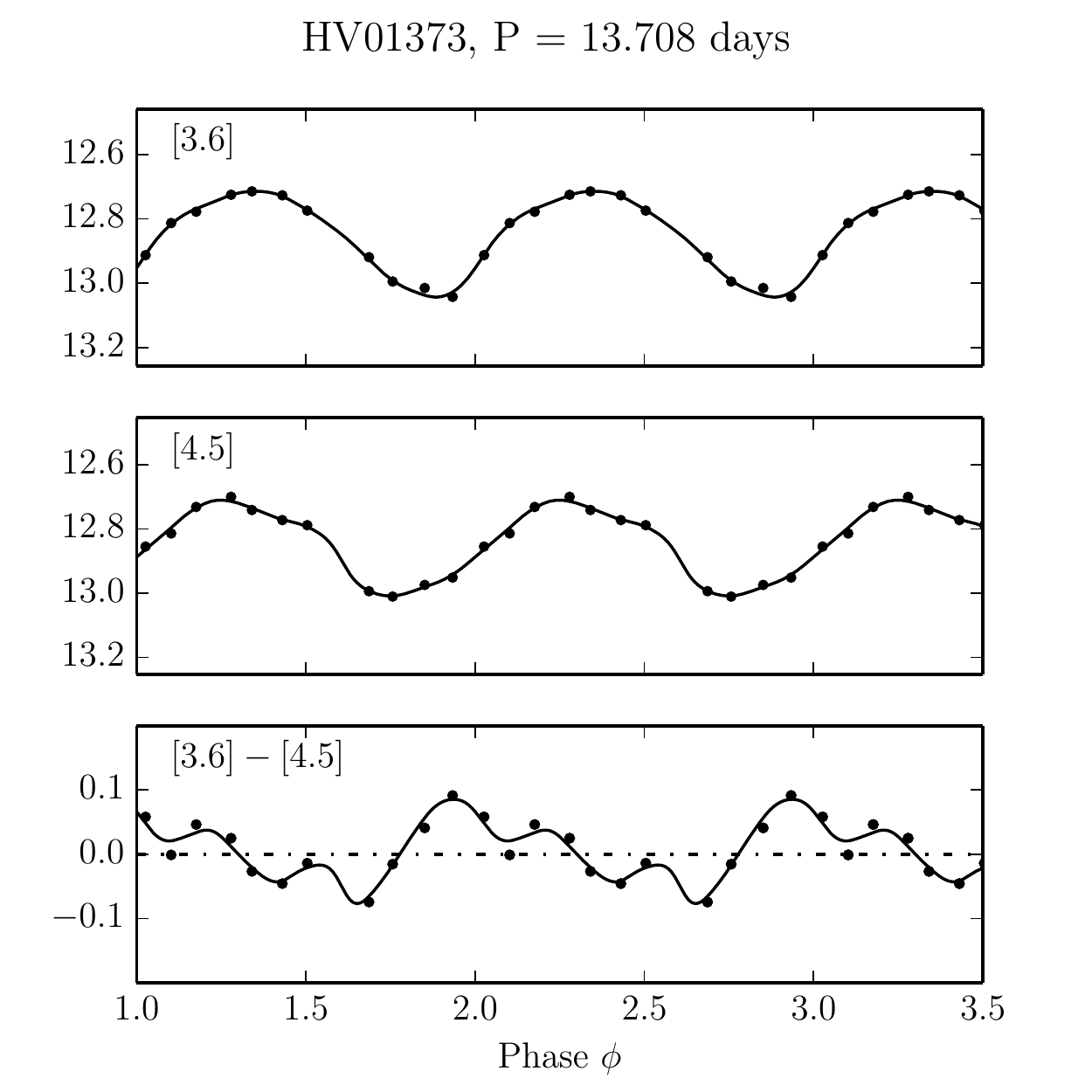} & 
\includegraphics[width=50mm]{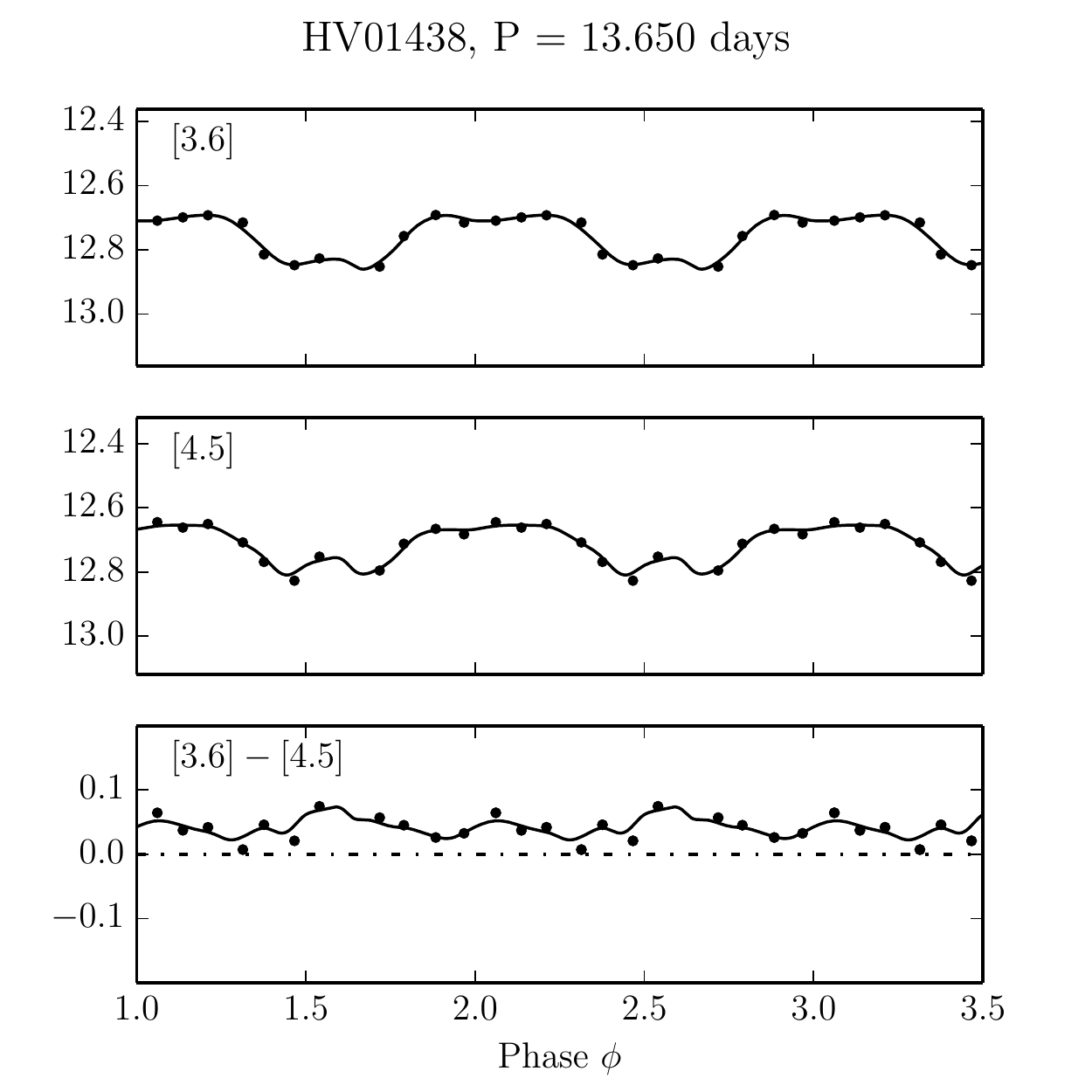} &
\includegraphics[width=50mm]{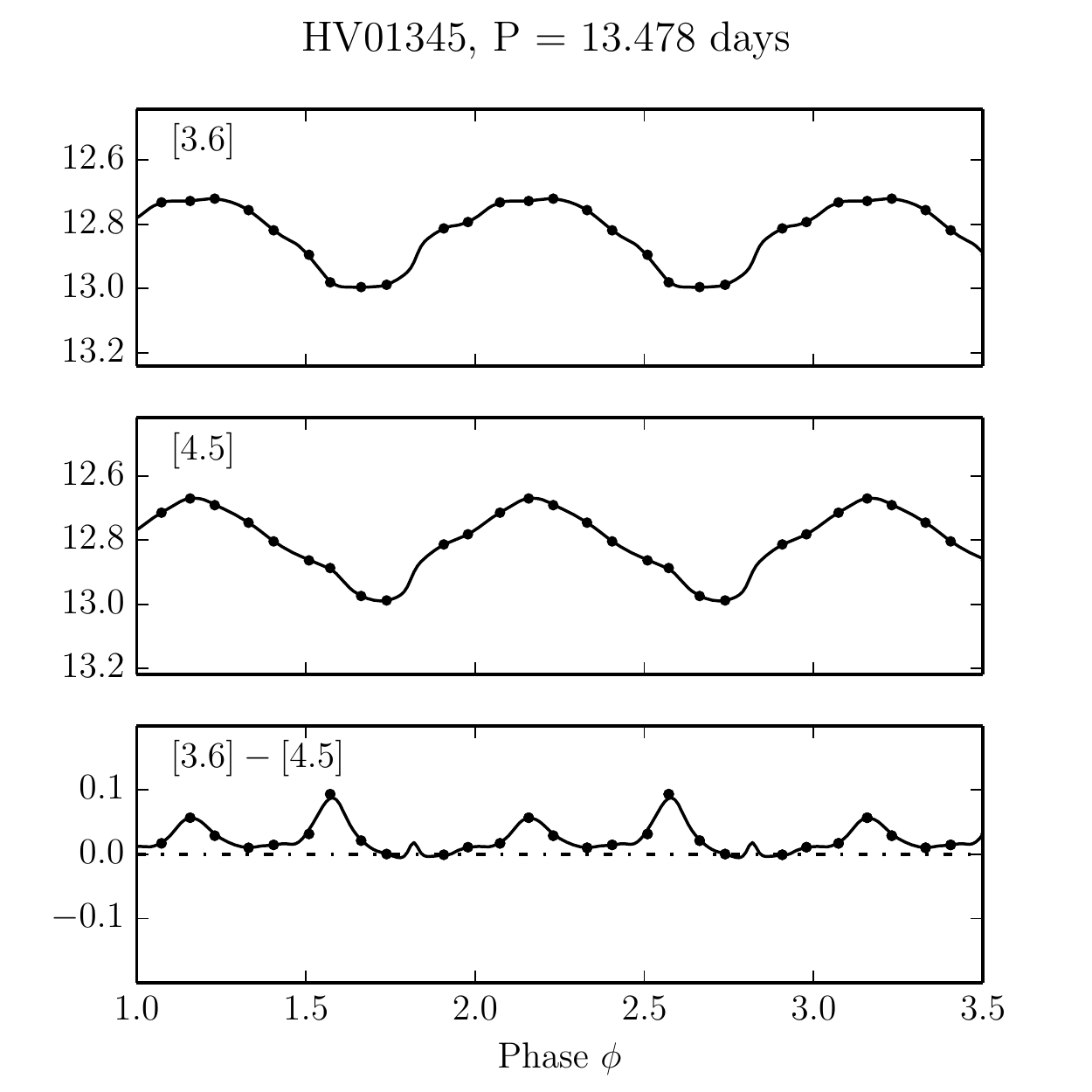} \\
\includegraphics[width=50mm]{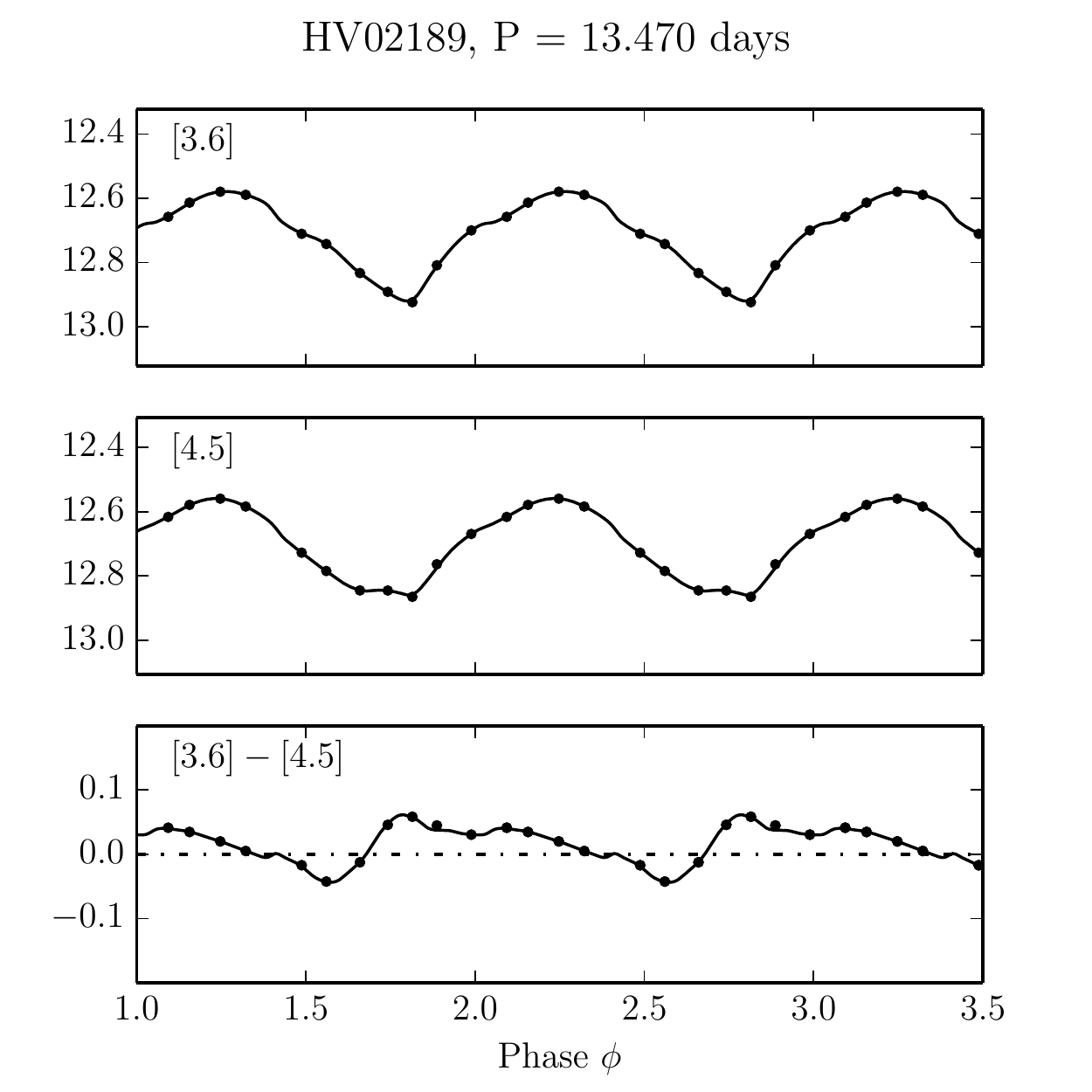} & 
\includegraphics[width=50mm]{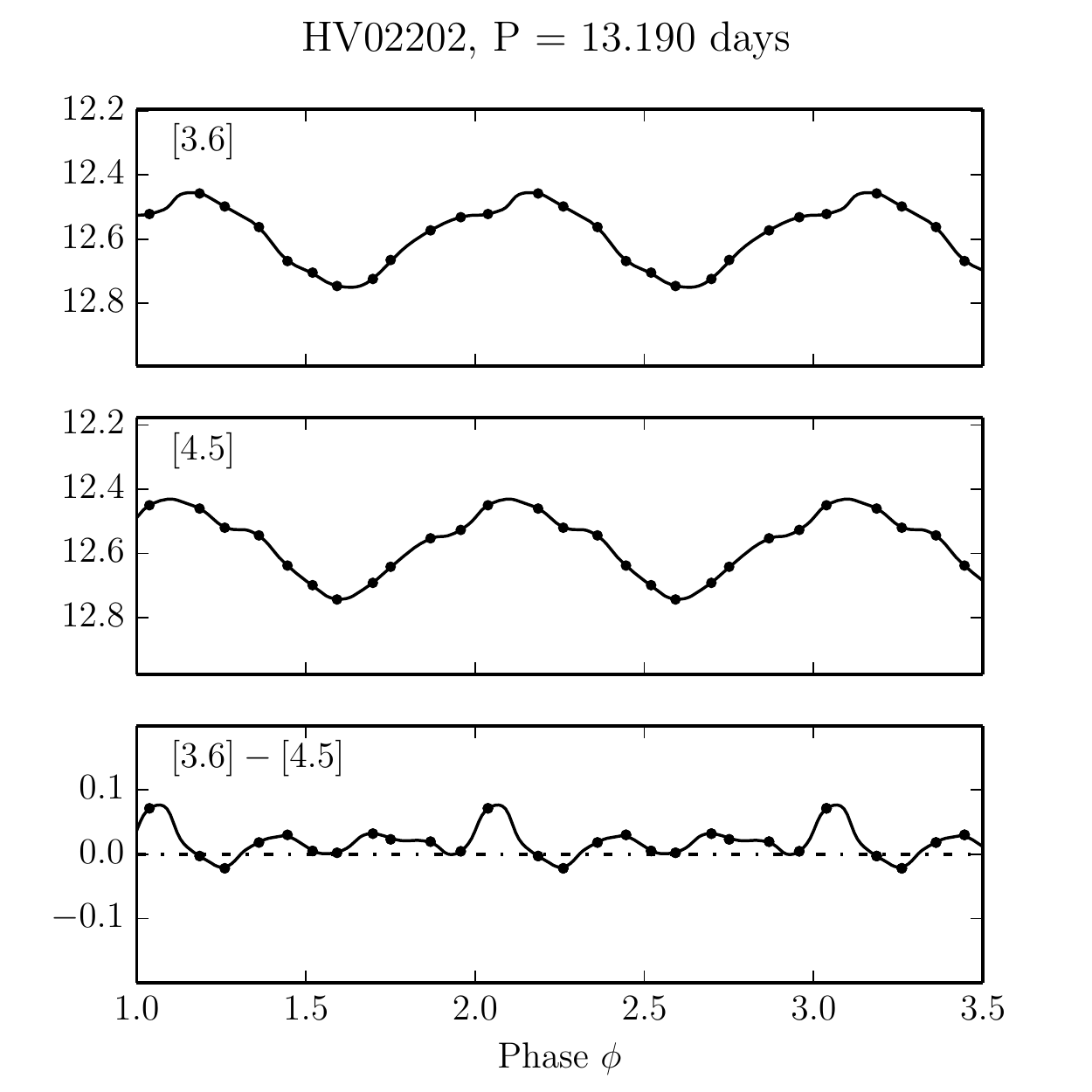} &
\includegraphics[width=50mm]{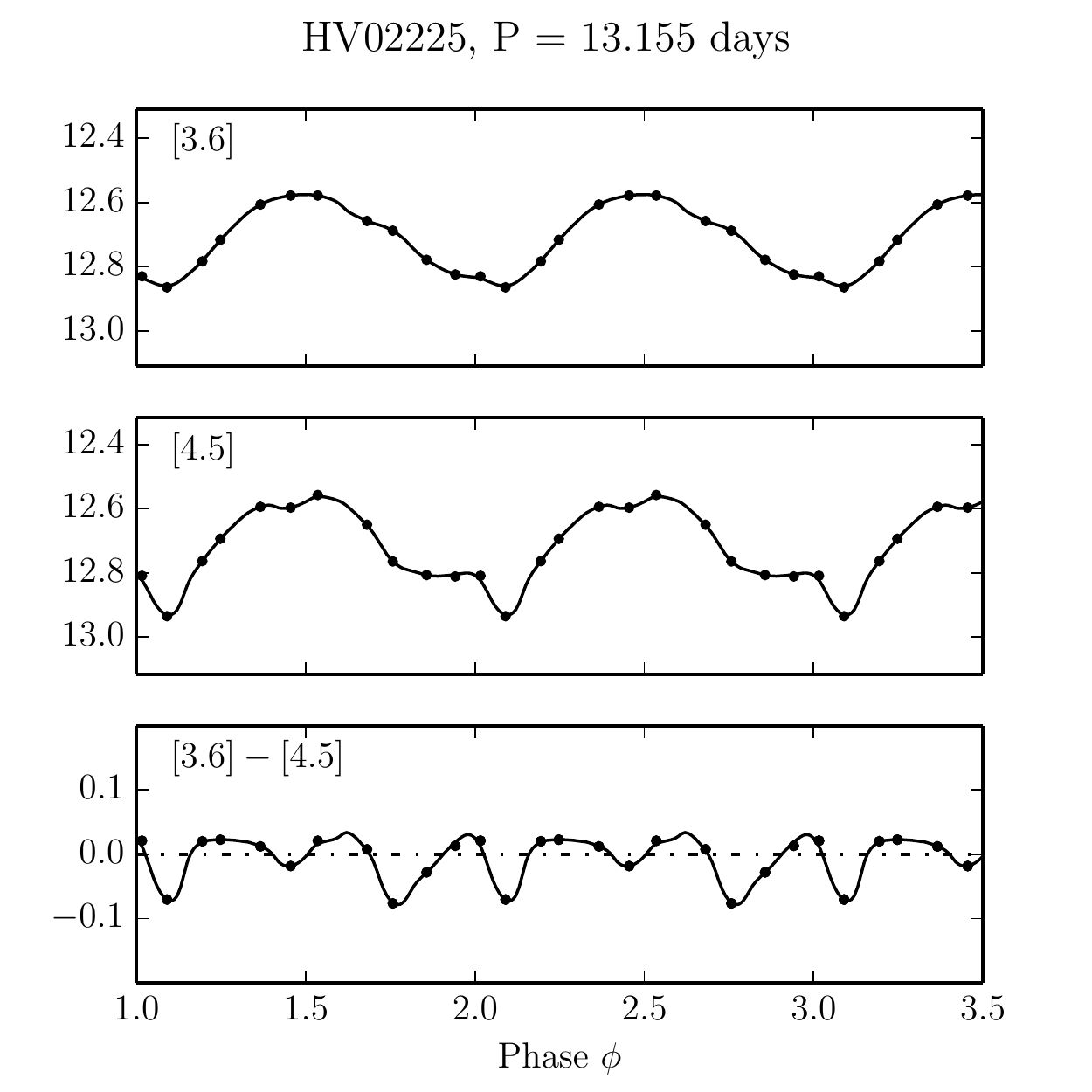} \\
\includegraphics[width=50mm]{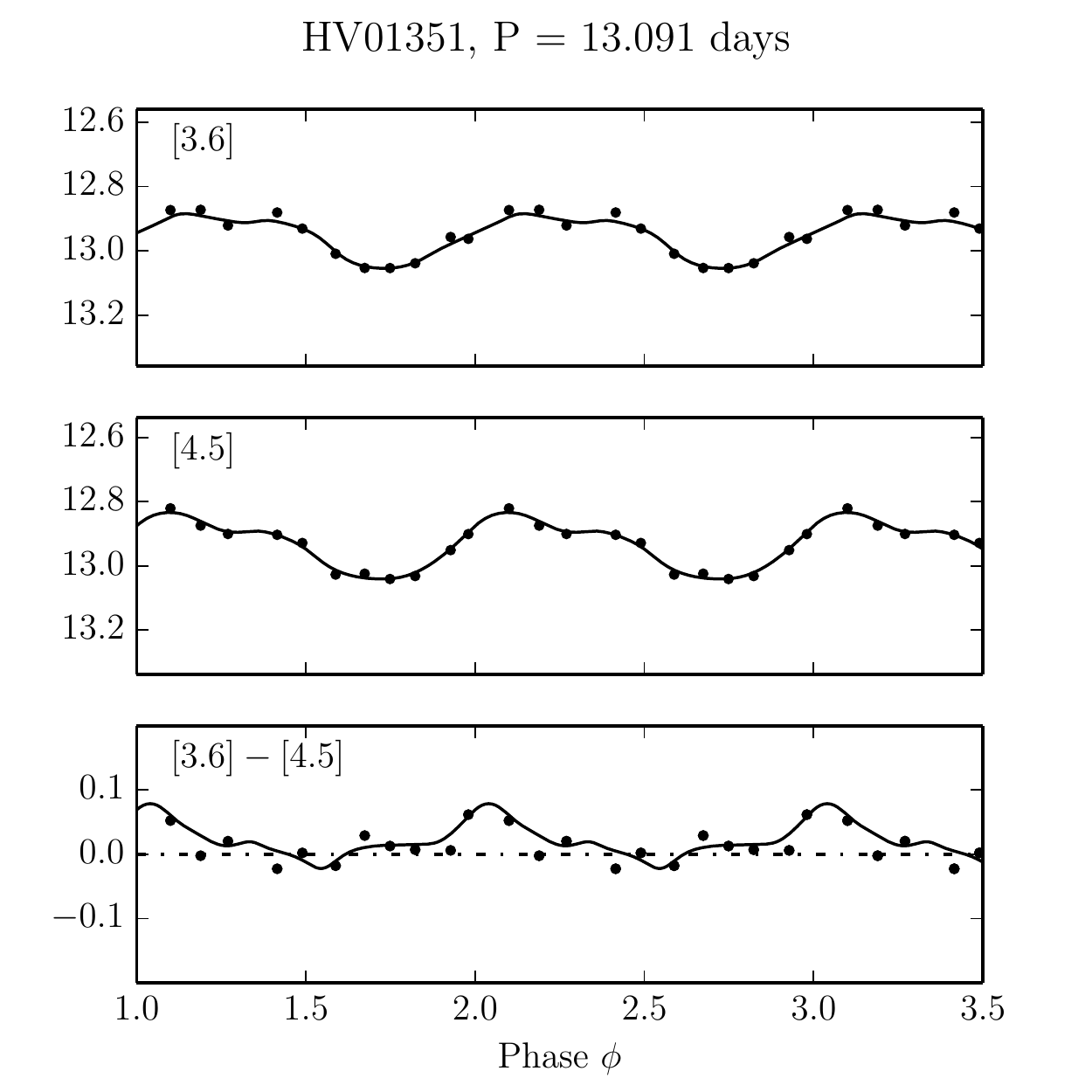} &
\includegraphics[width=50mm]{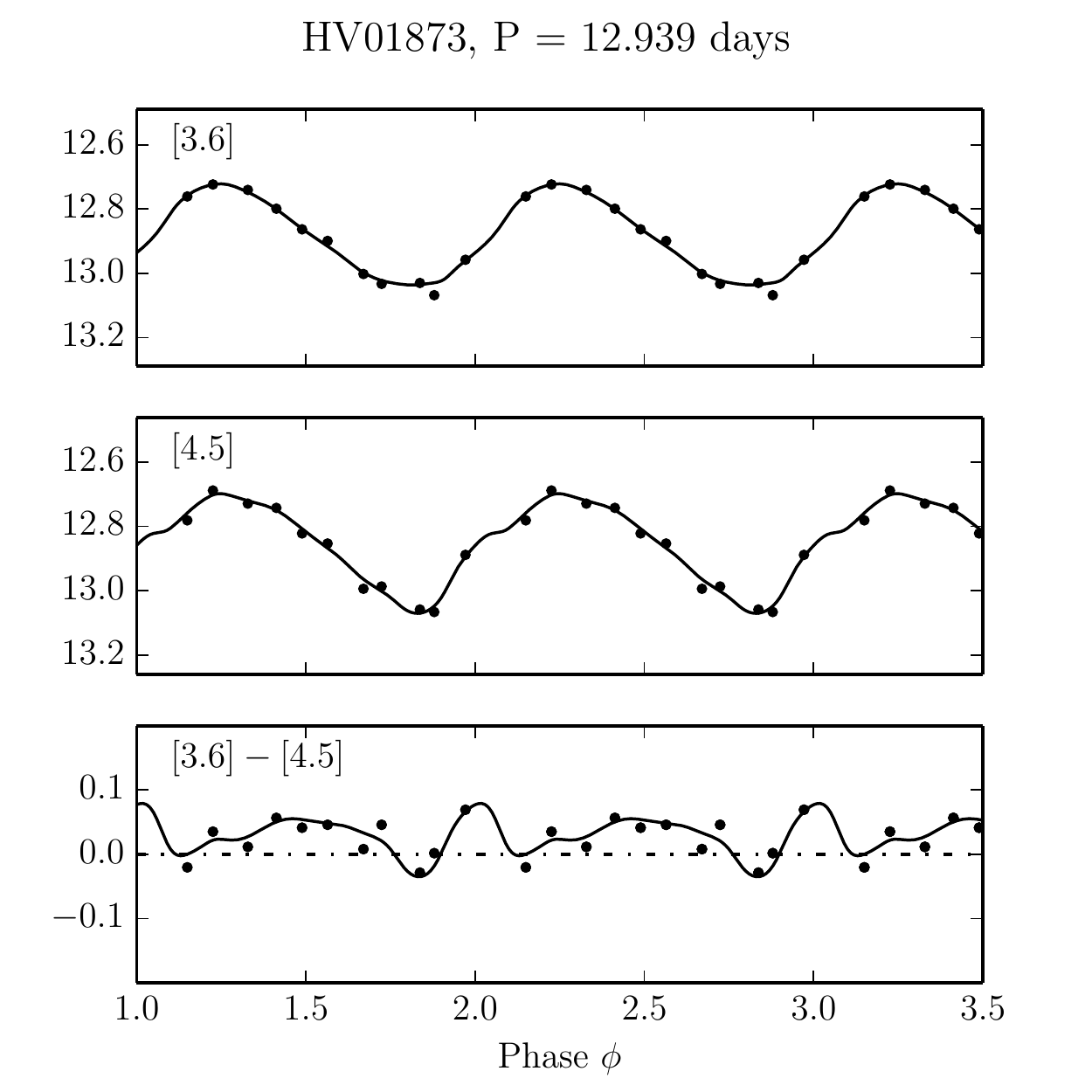} &
\includegraphics[width=50mm]{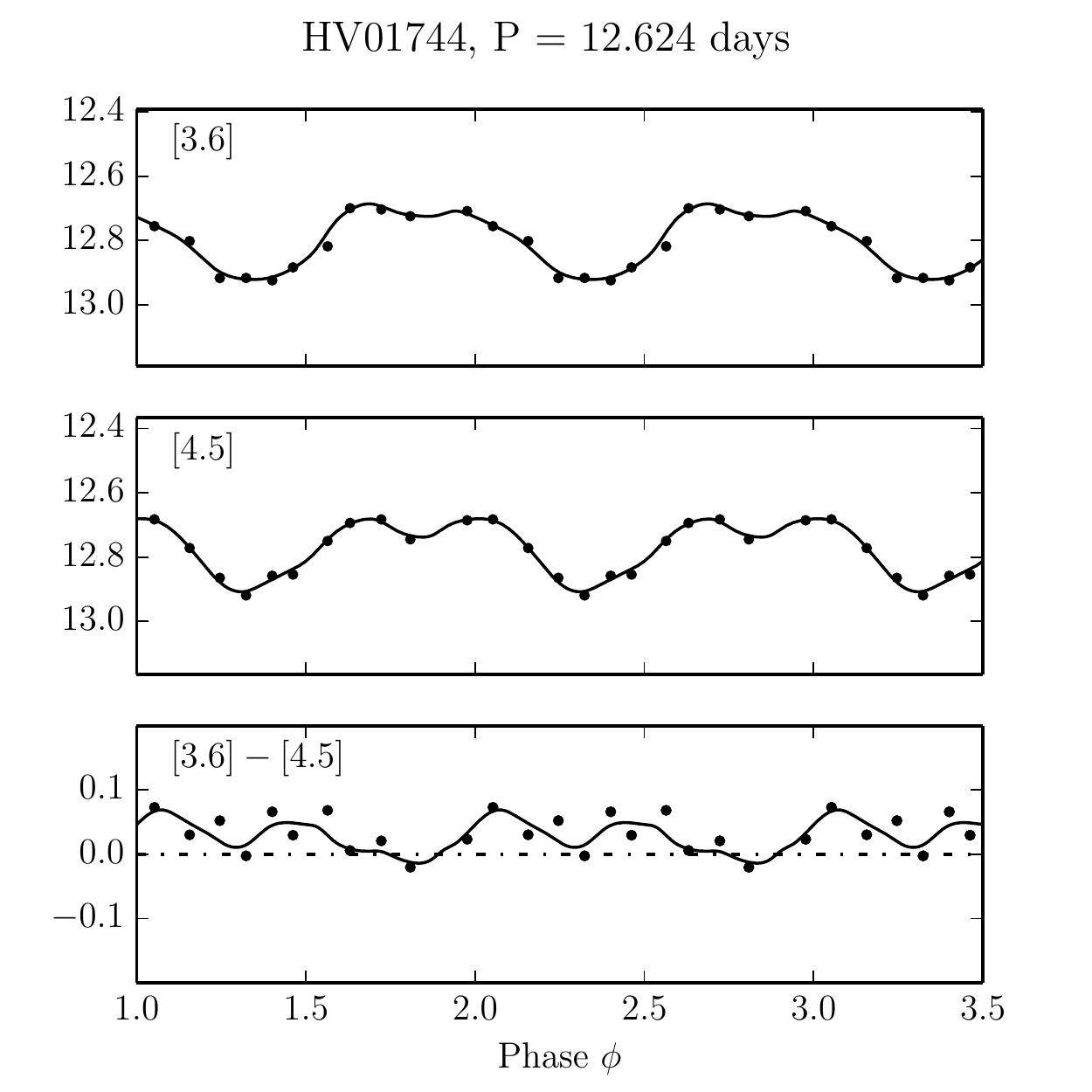}  \\

\end{array}$ 
\end{center} 
\end{figure}
\begin{figure} 
 \begin{center}$ 
 \begin{array}{ccc} 
\includegraphics[width=50mm]{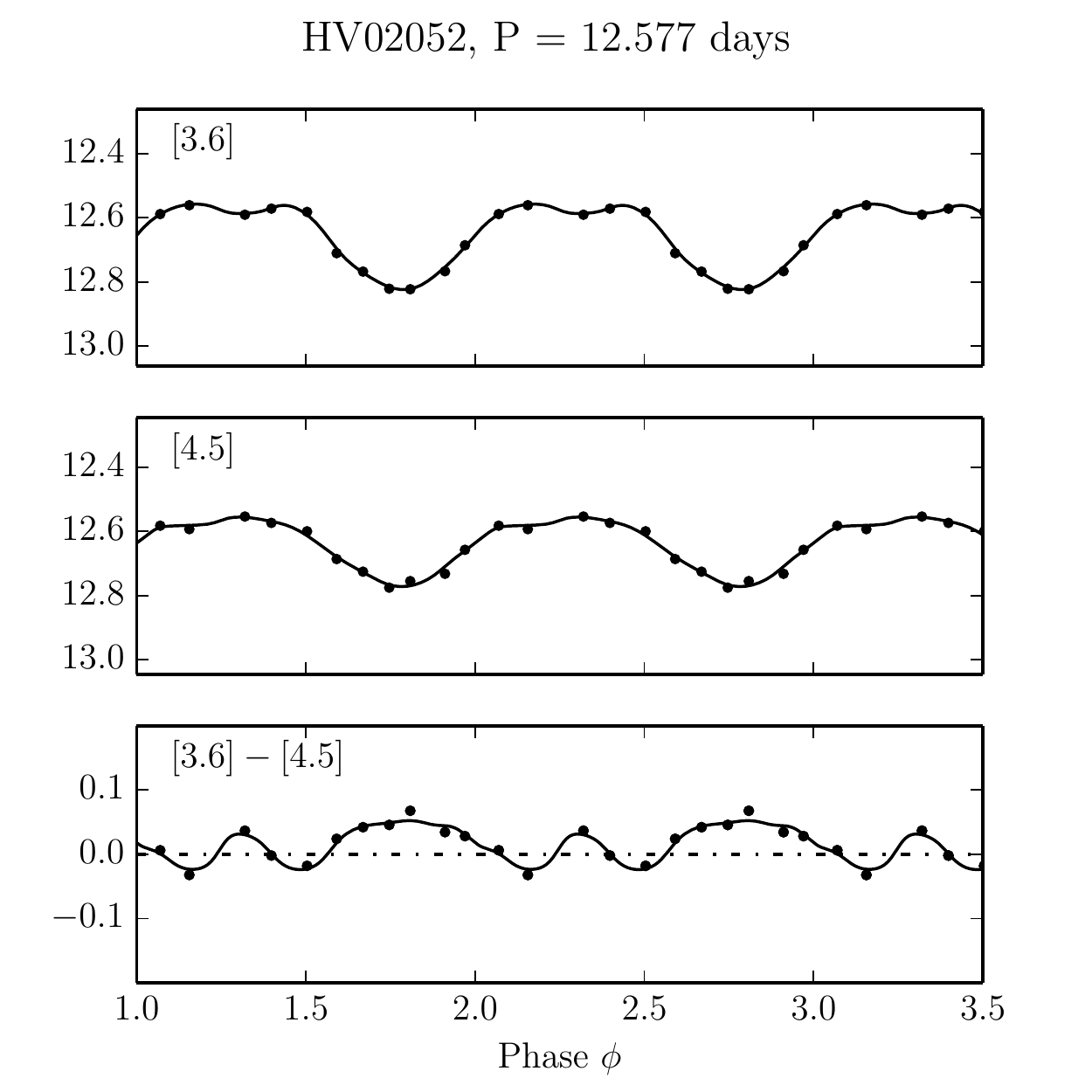}&
\includegraphics[width=50mm]{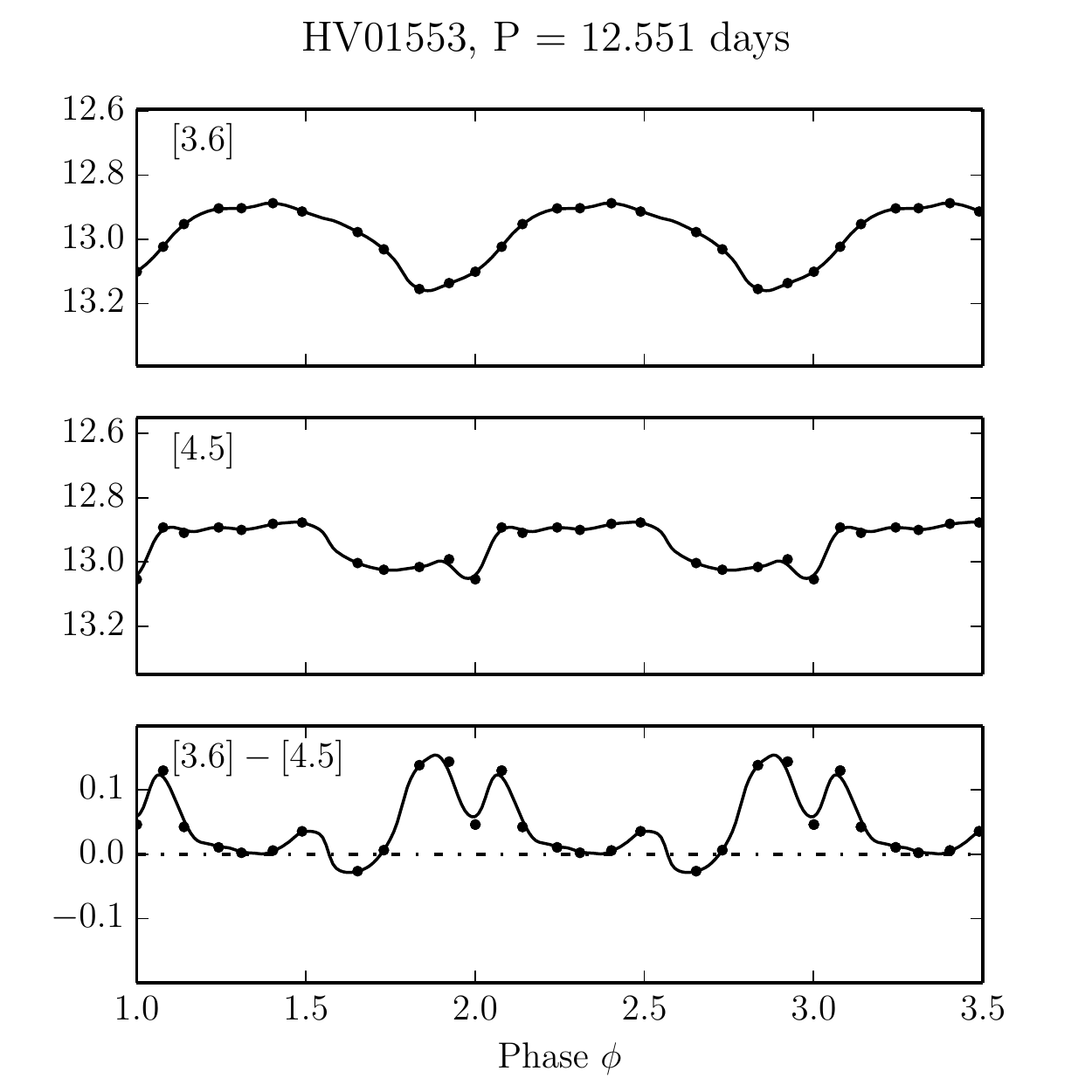} &
\includegraphics[width=50mm]{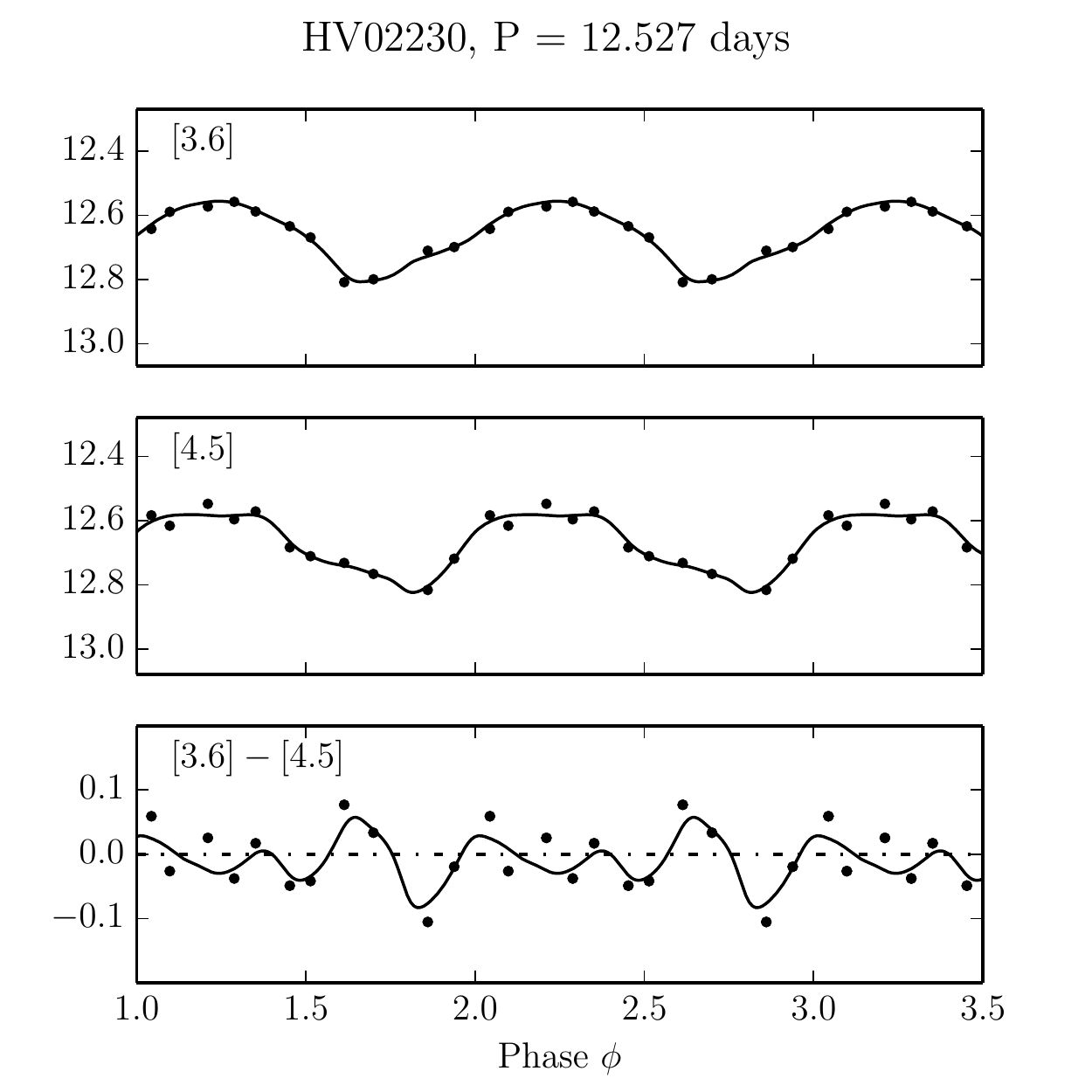} \\ 
\includegraphics[width=50mm]{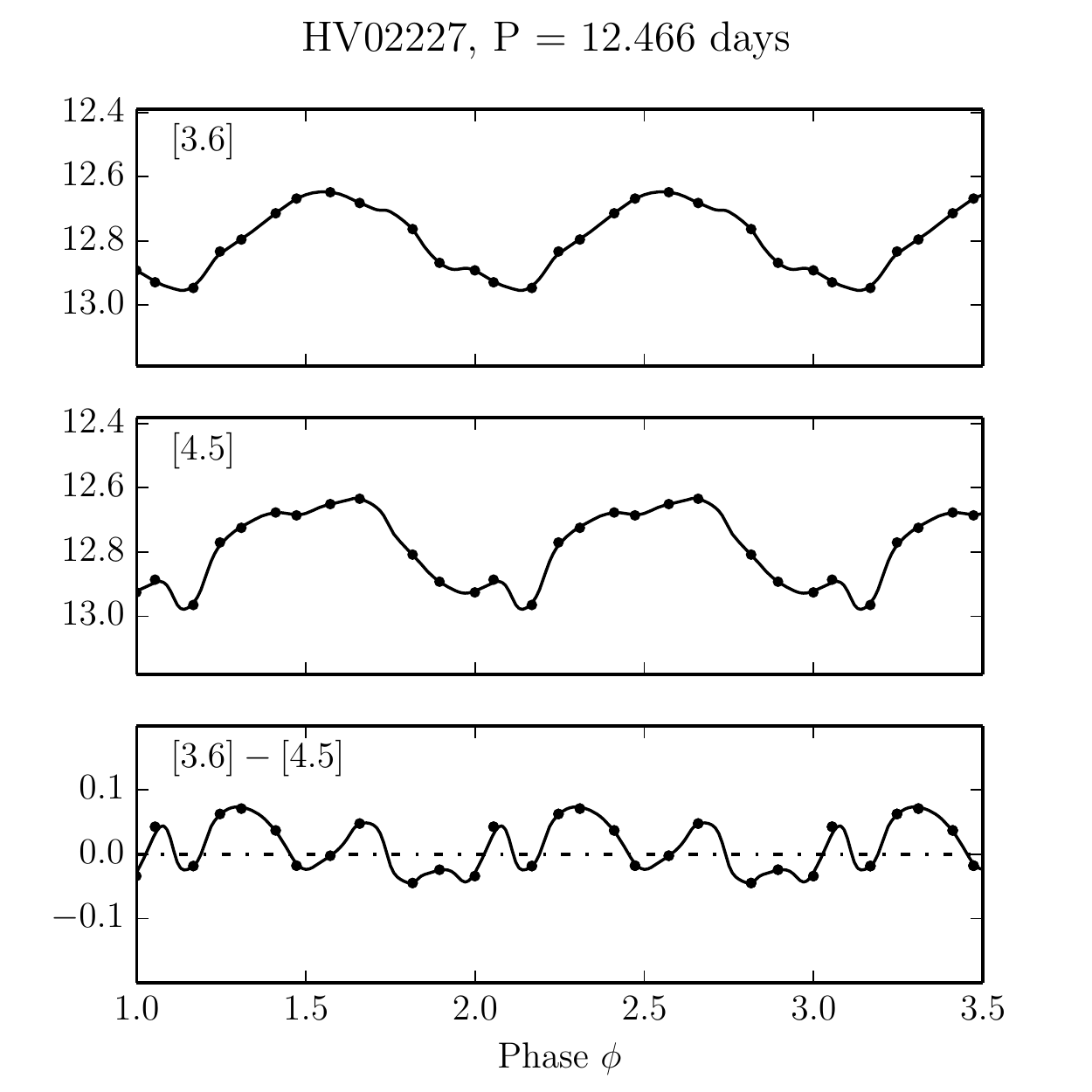} & 
\includegraphics[width=50mm]{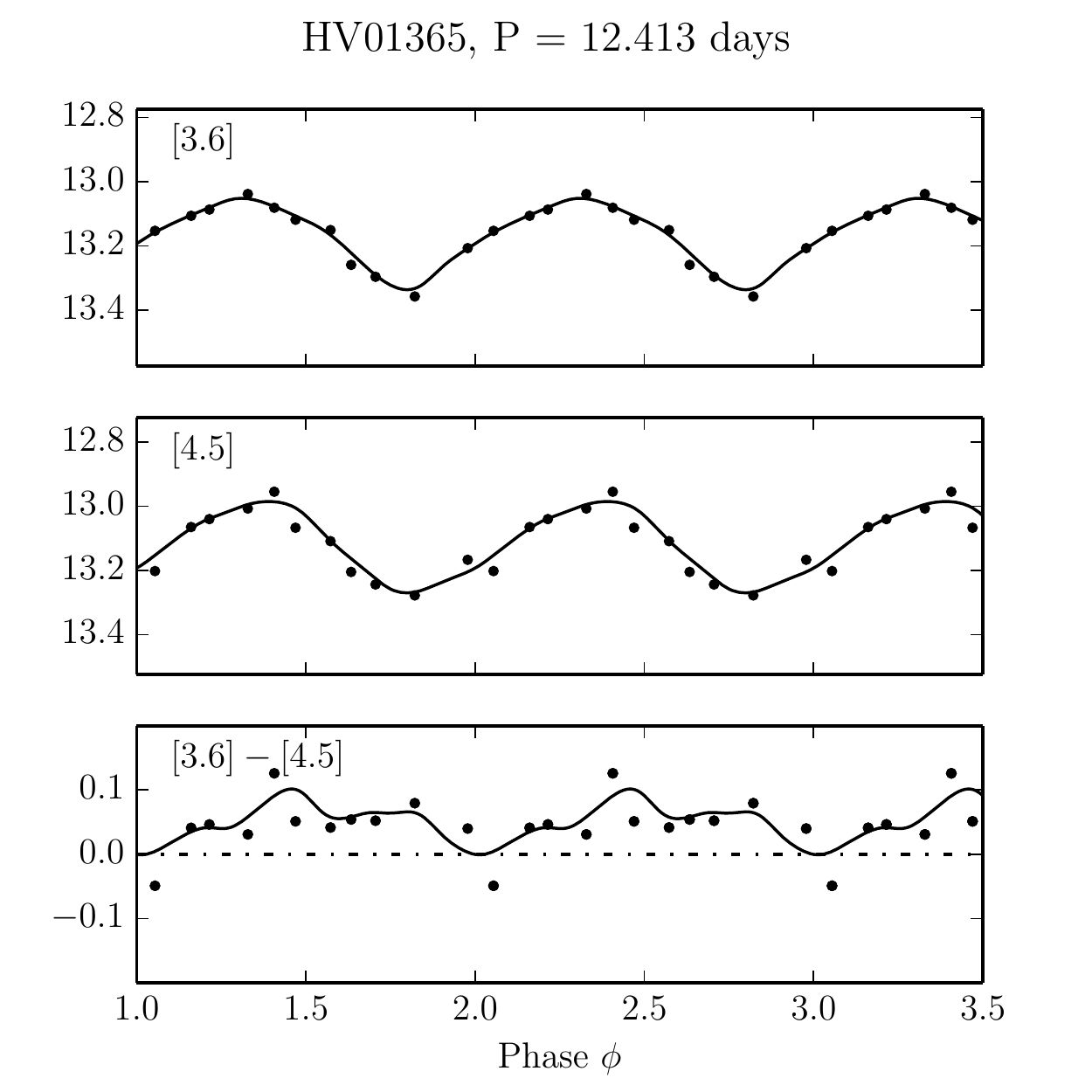} &
\includegraphics[width=50mm]{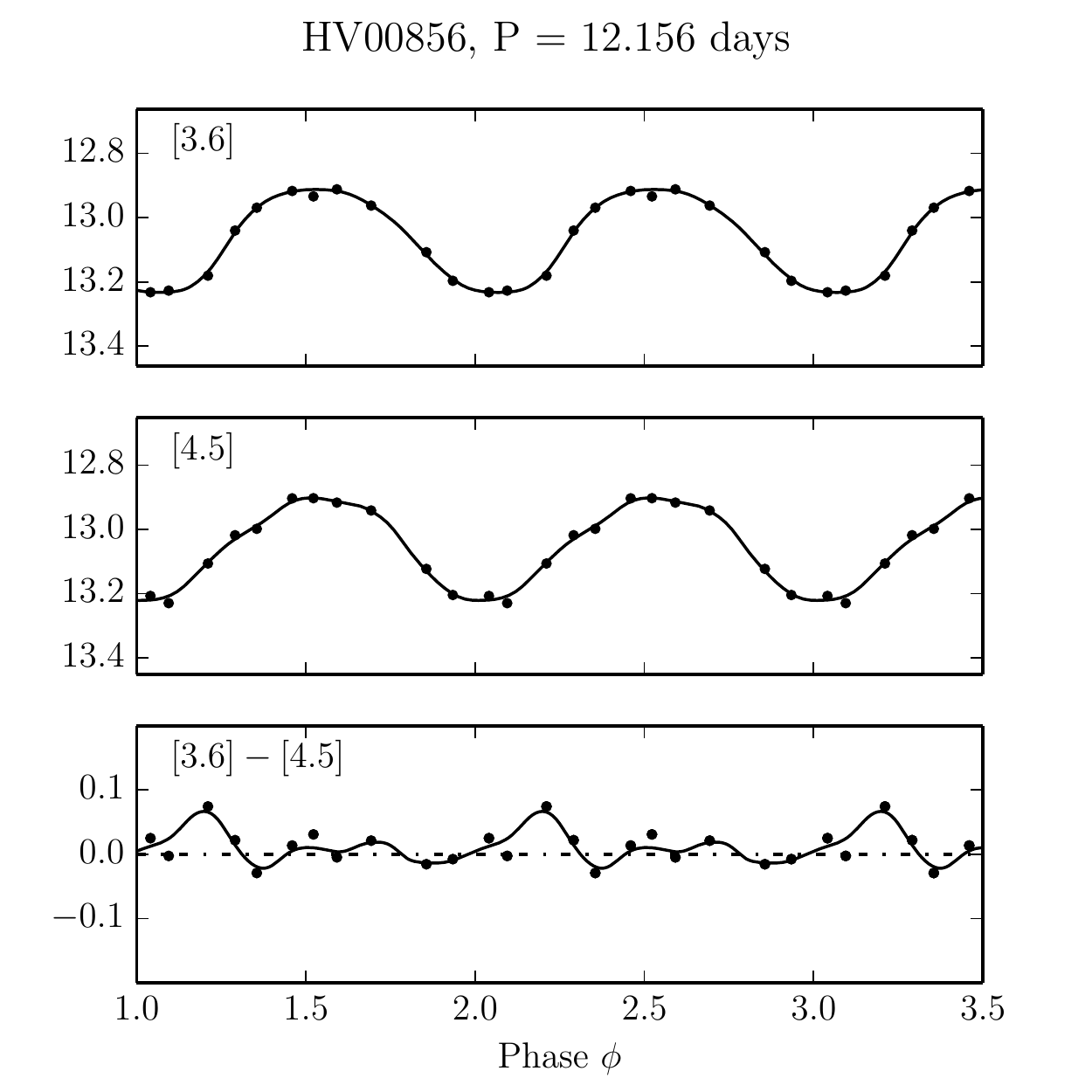} \\
\includegraphics[width=50mm]{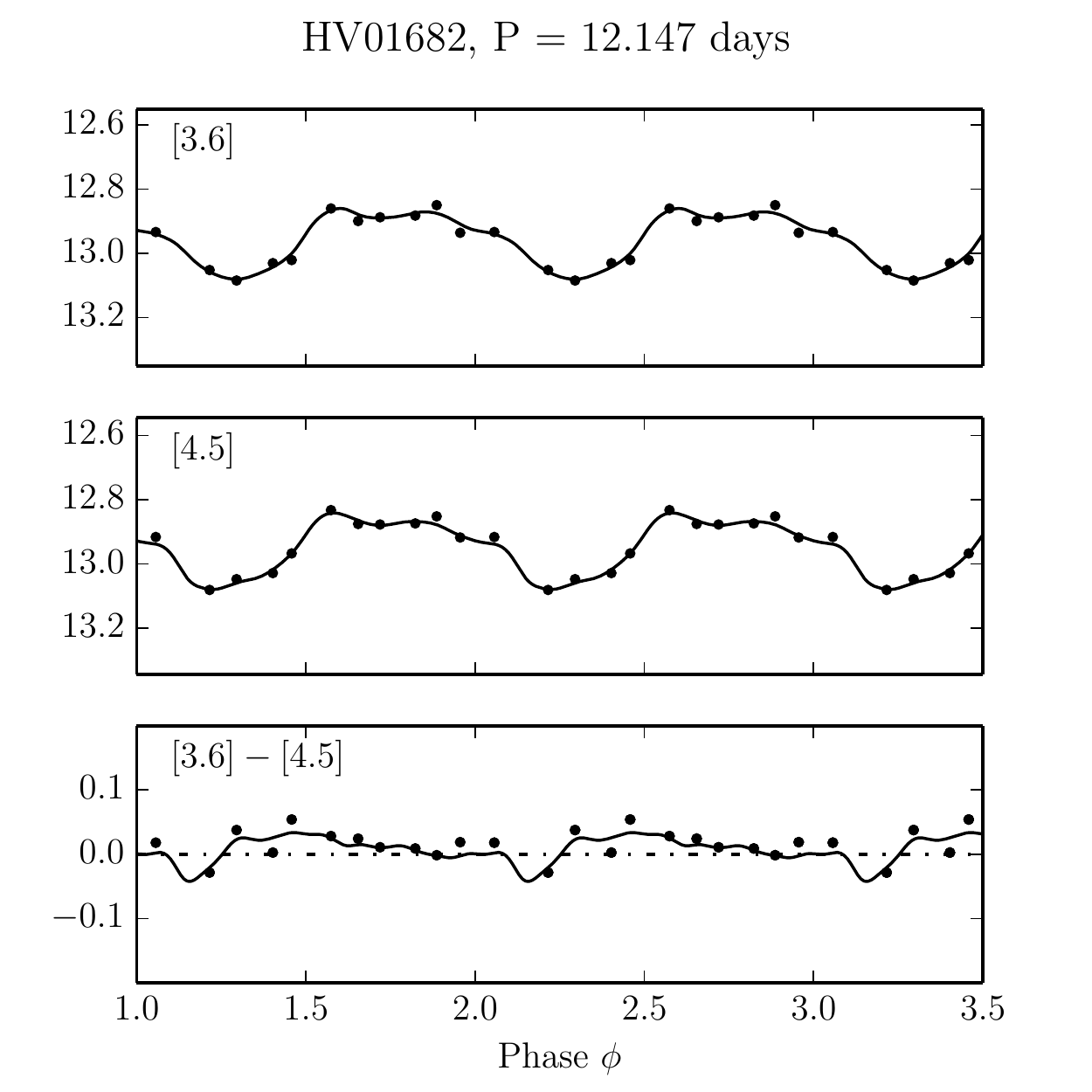} & 
\includegraphics[width=50mm]{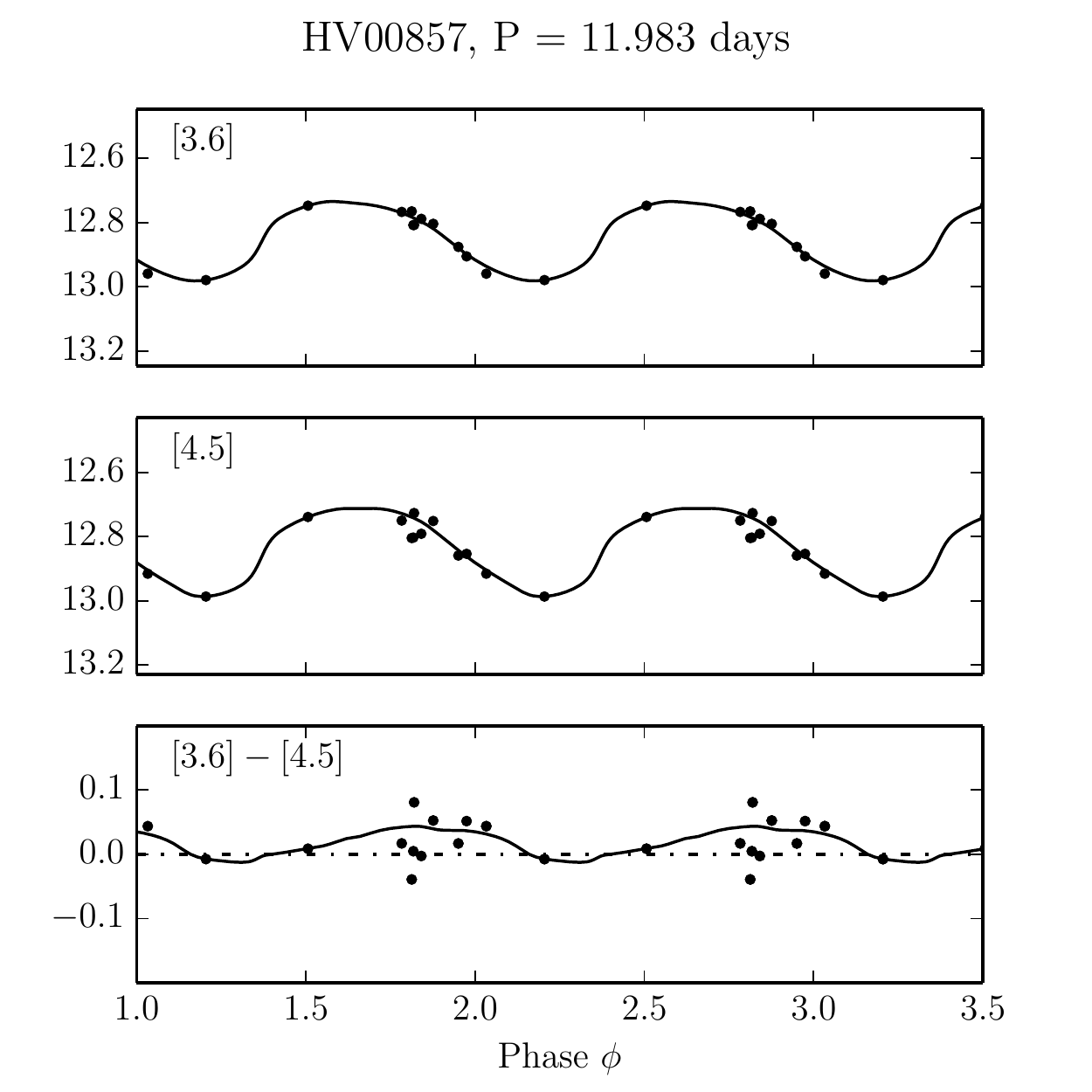} &
\includegraphics[width=50mm]{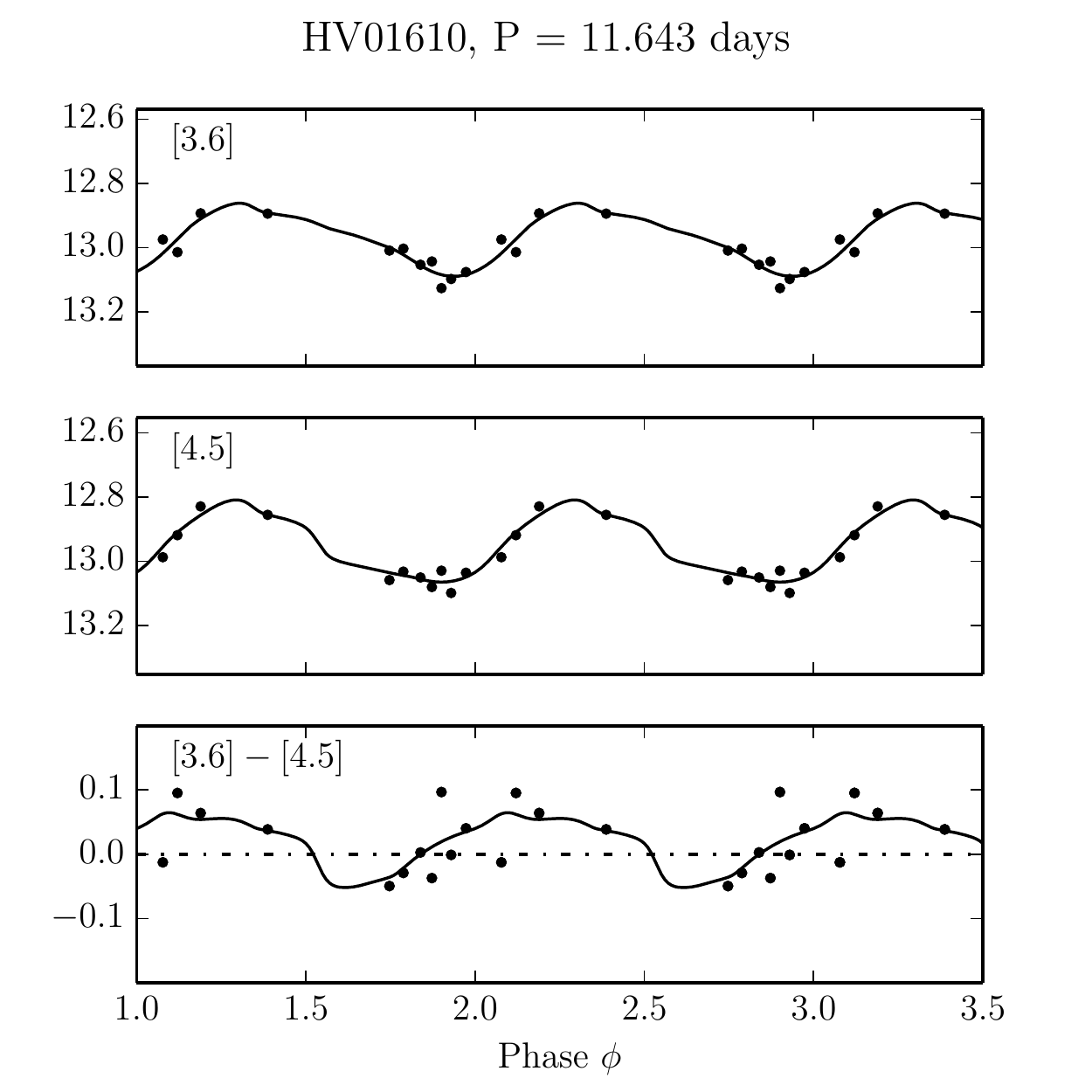} \\
\includegraphics[width=50mm]{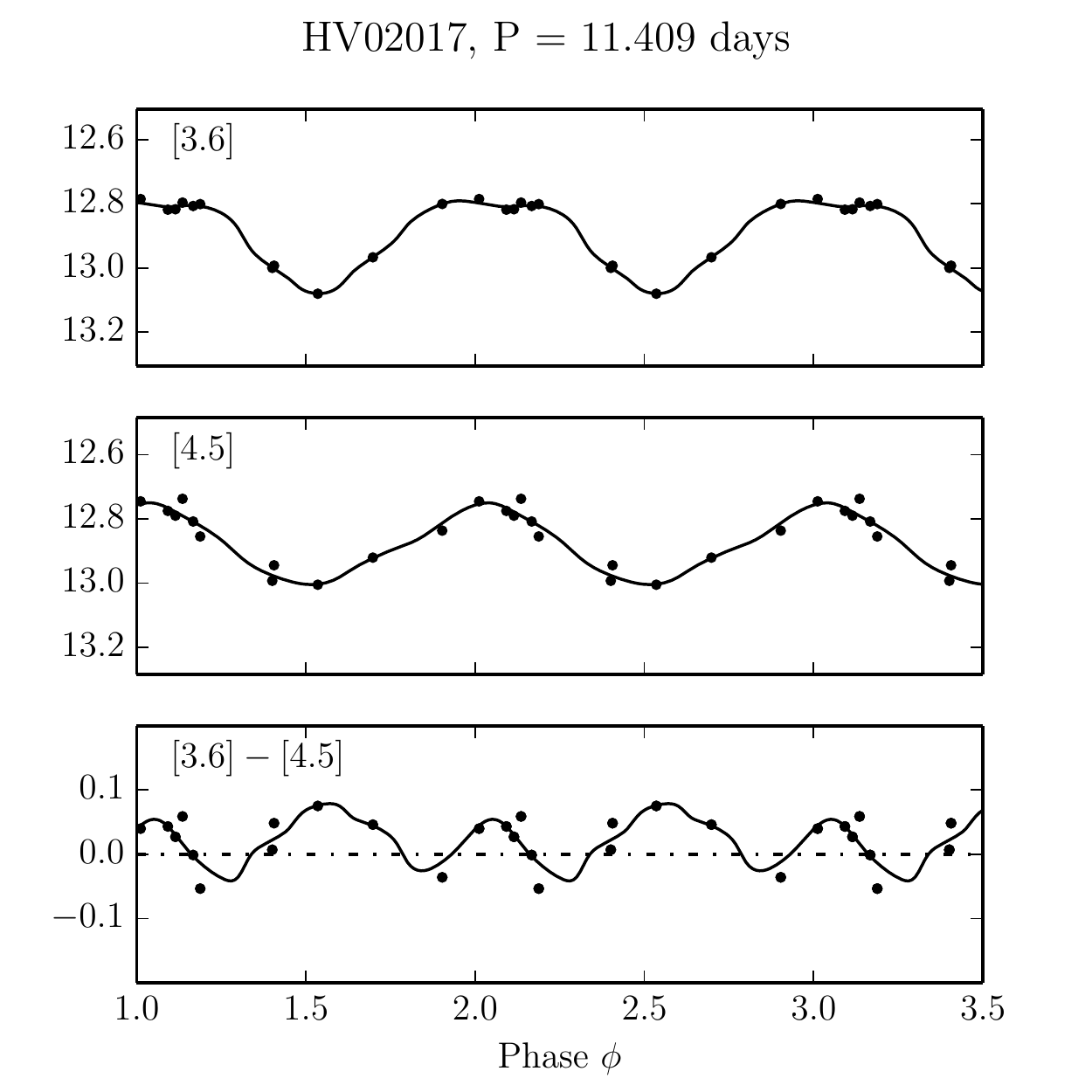} &
\includegraphics[width=50mm]{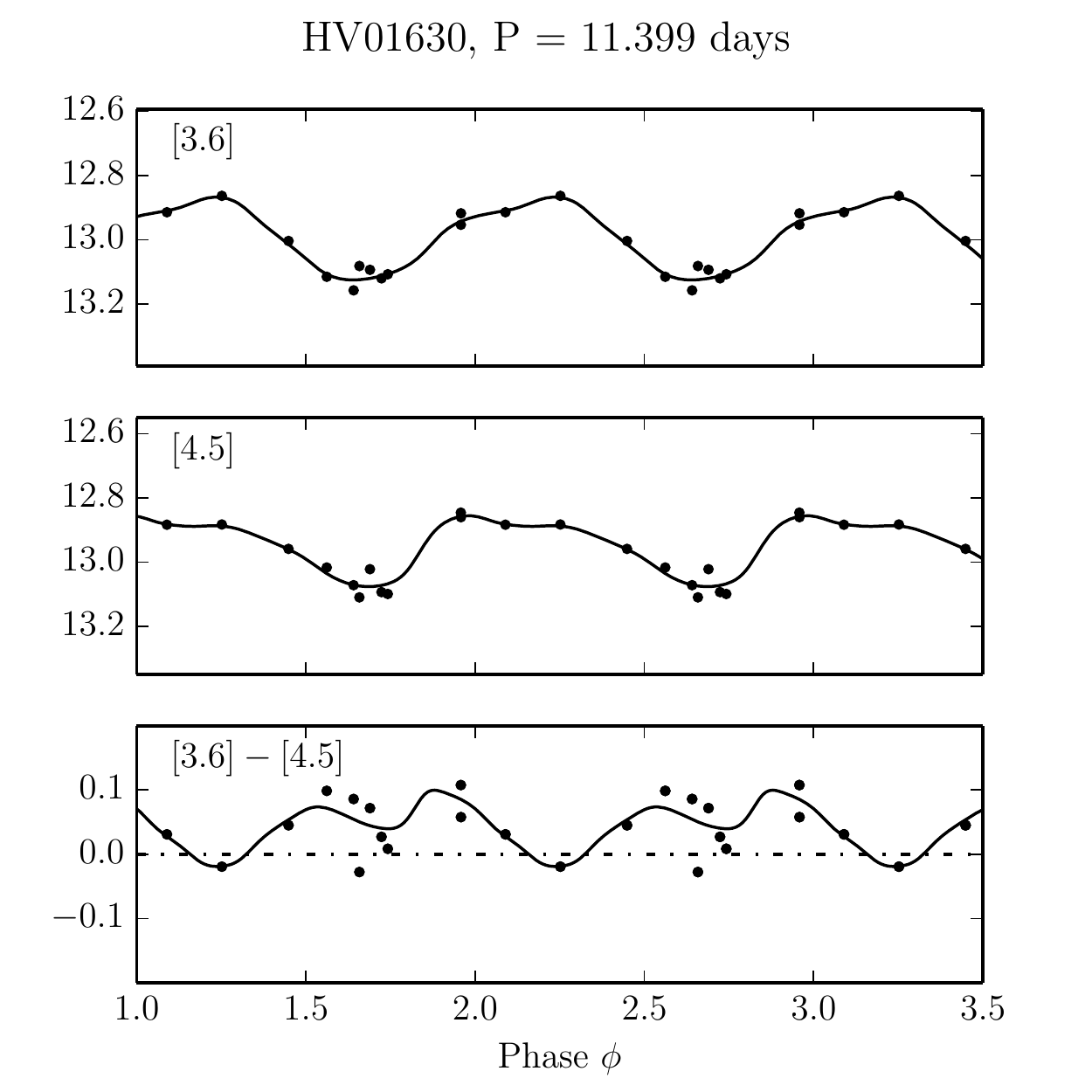} &
\includegraphics[width=50mm]{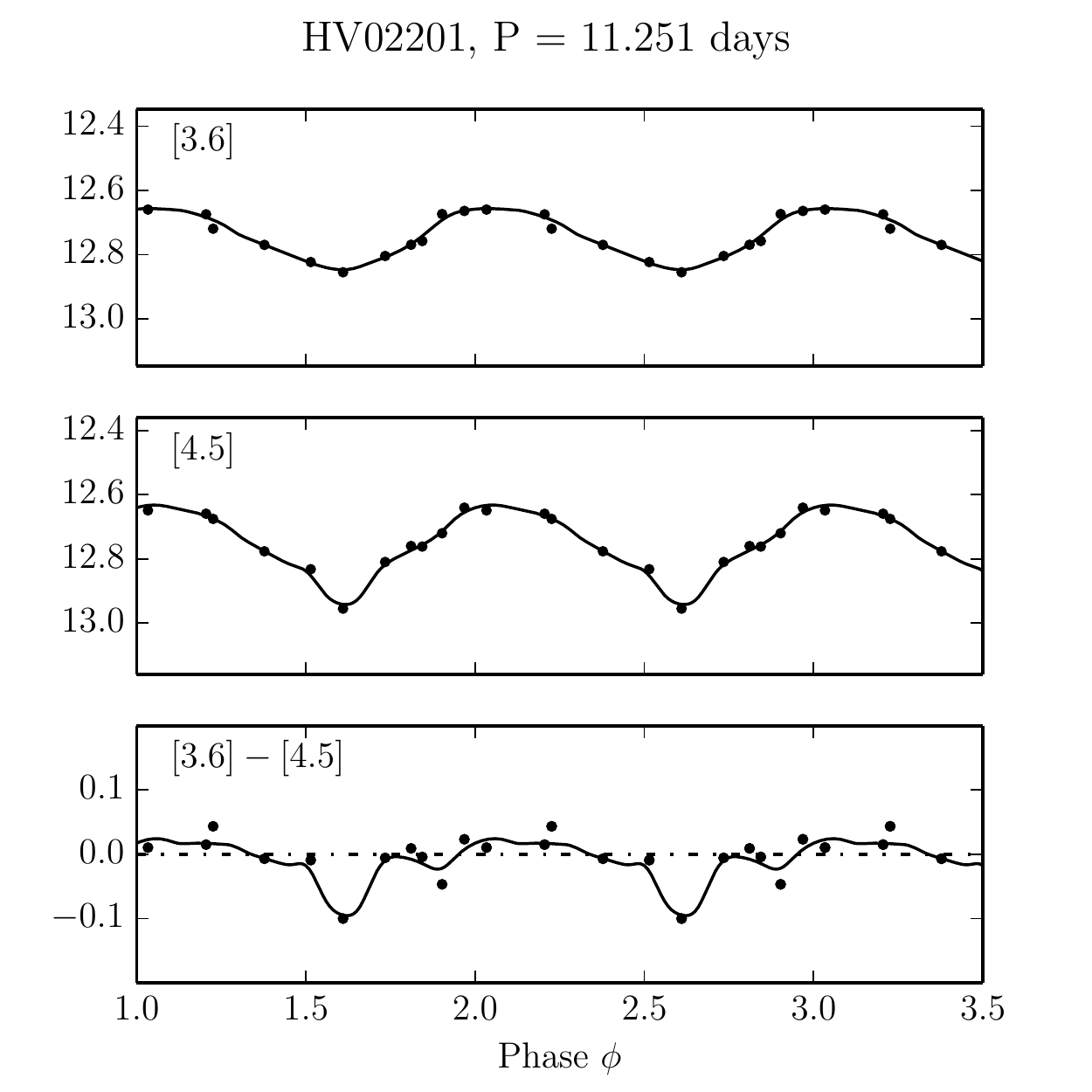}  \\

\end{array}$ 
\end{center} 
\end{figure}
\begin{figure} 
 \begin{center}$ 
 \begin{array}{ccc} 
\includegraphics[width=50mm]{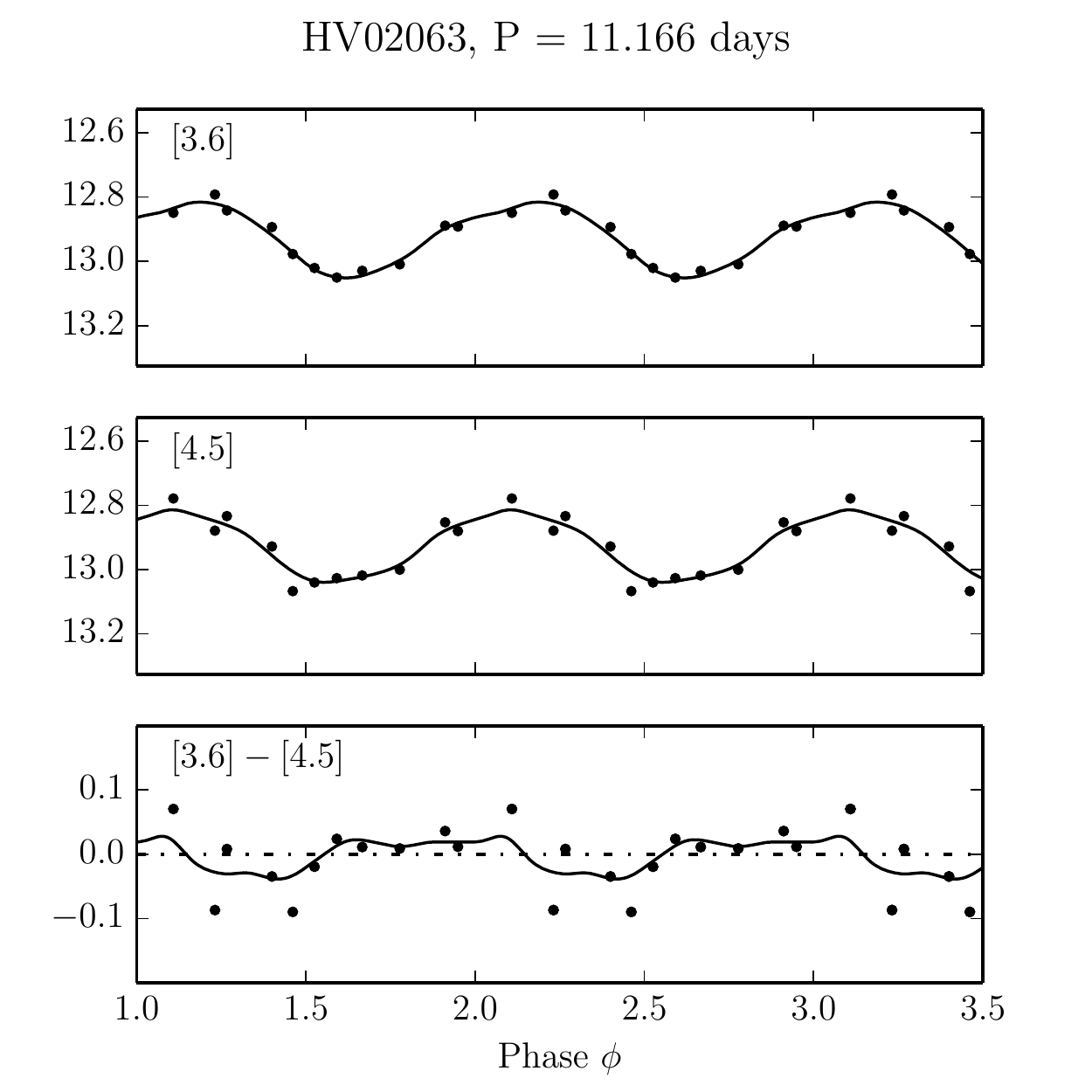}&
\includegraphics[width=50mm]{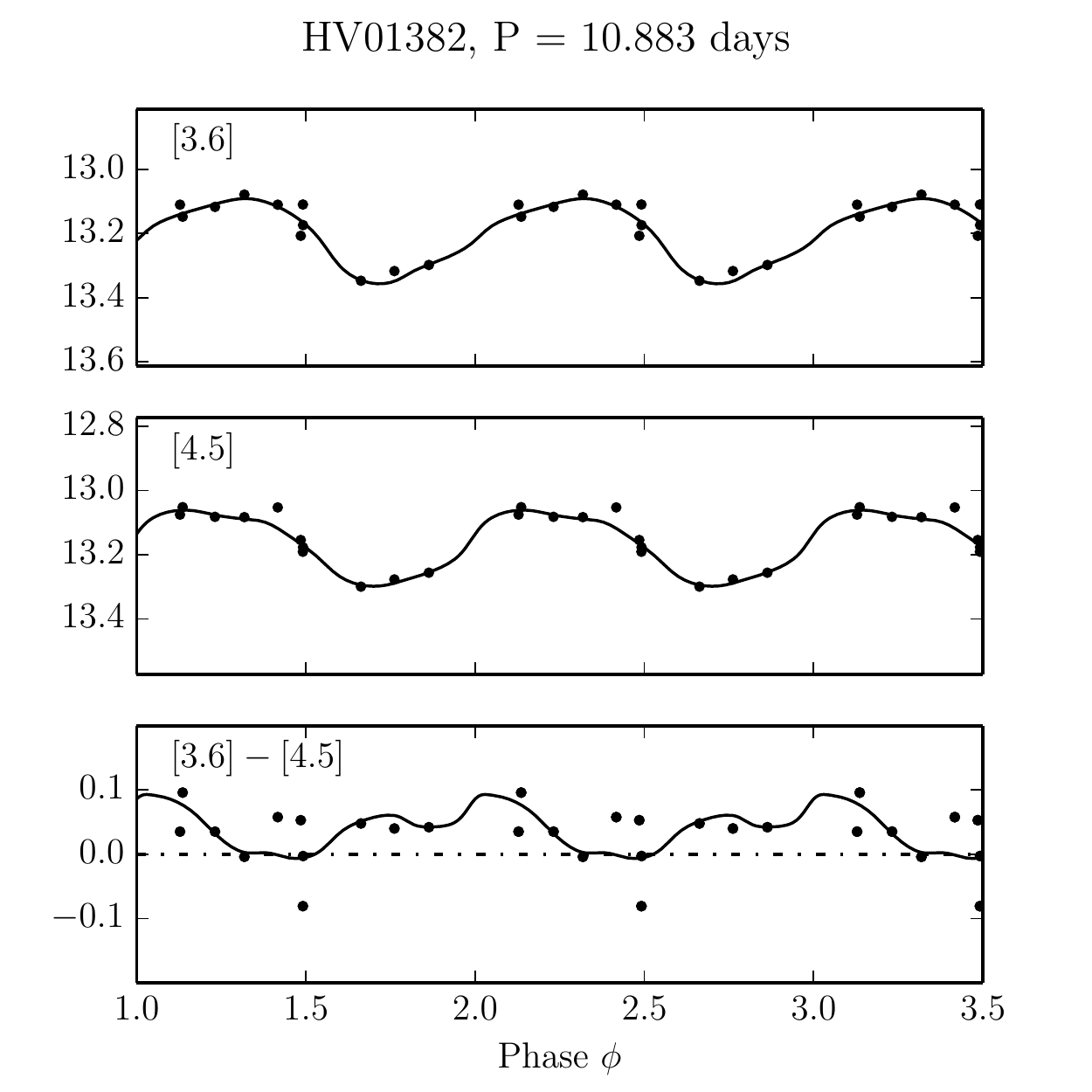} &
\includegraphics[width=50mm]{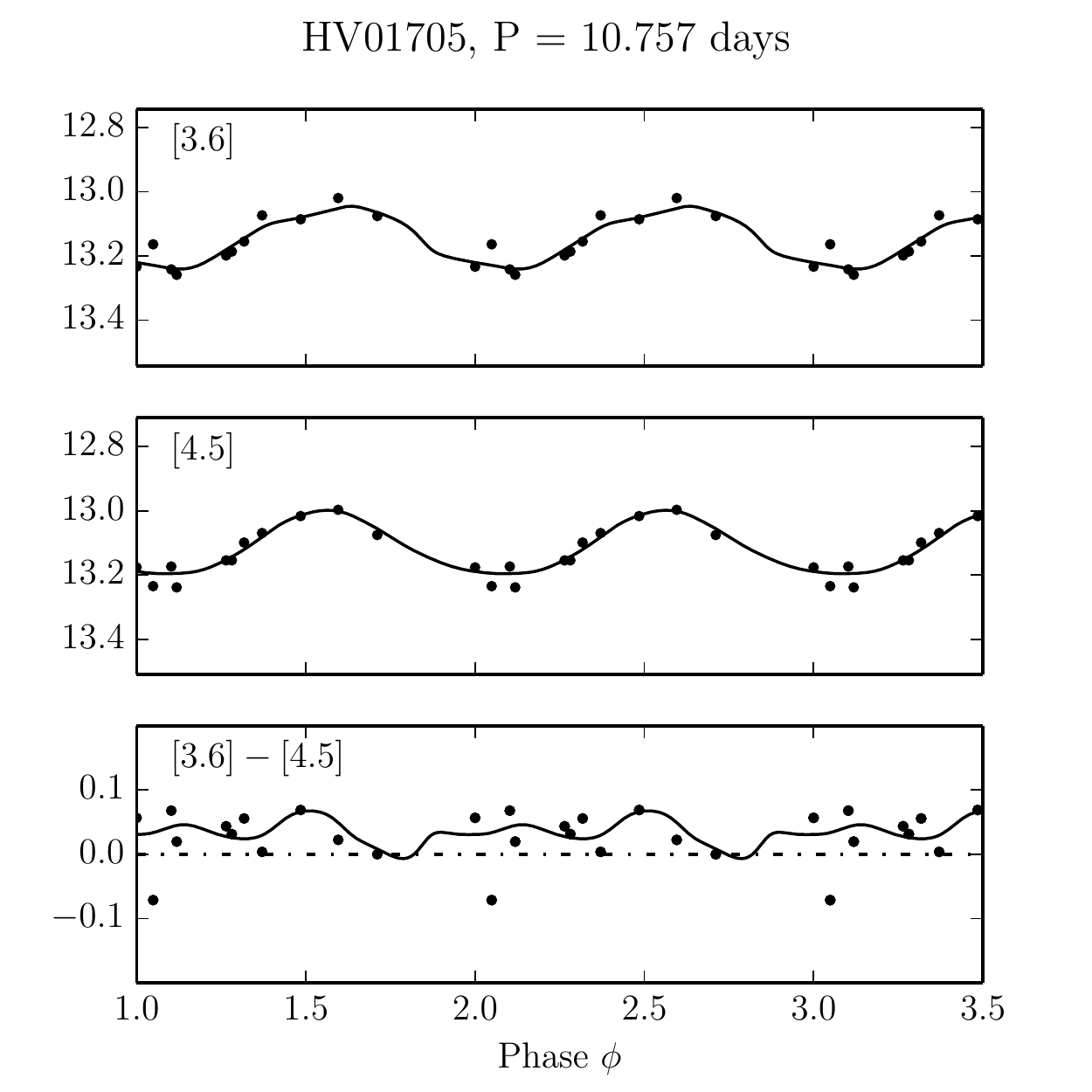} \\ 
\includegraphics[width=50mm]{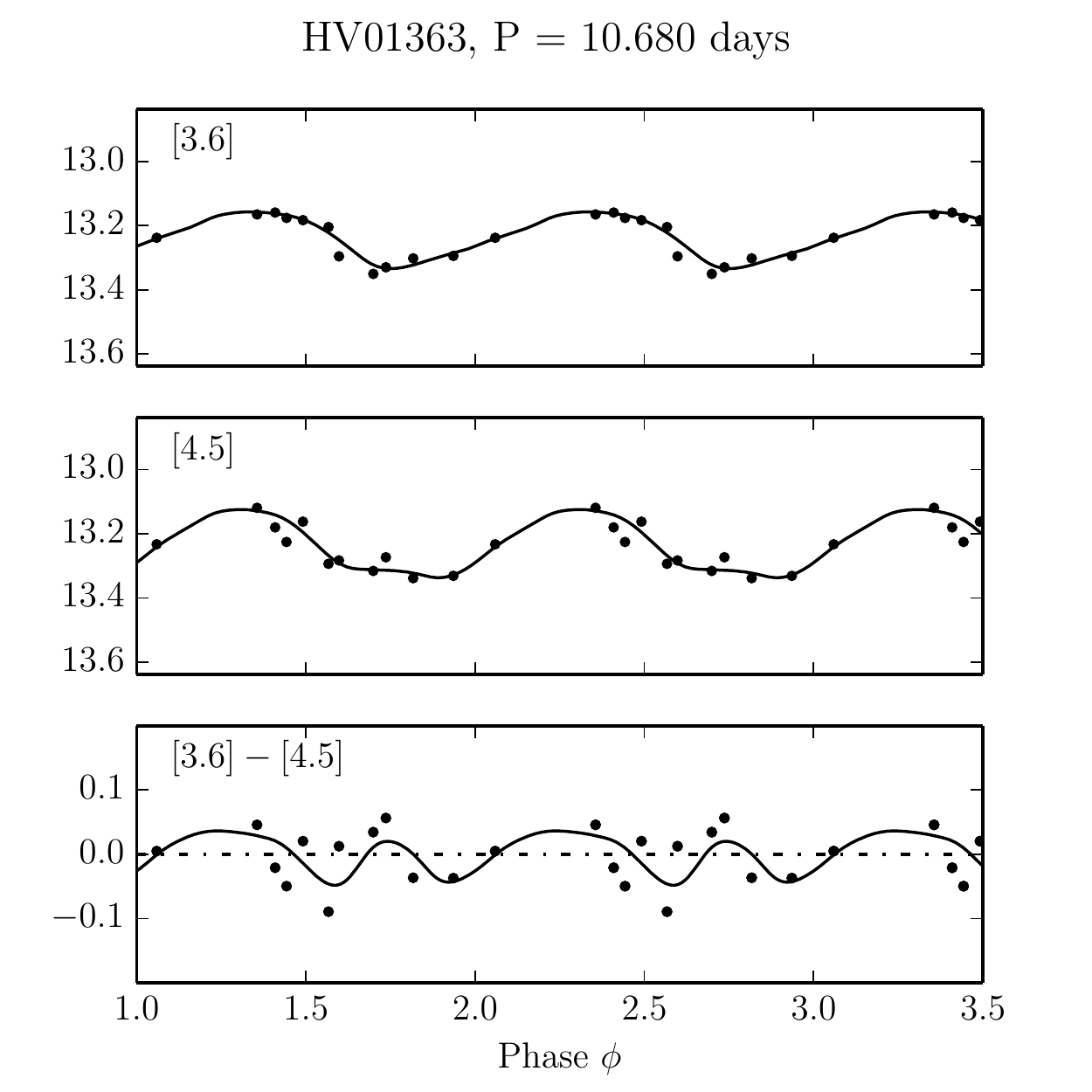} & 
\includegraphics[width=50mm]{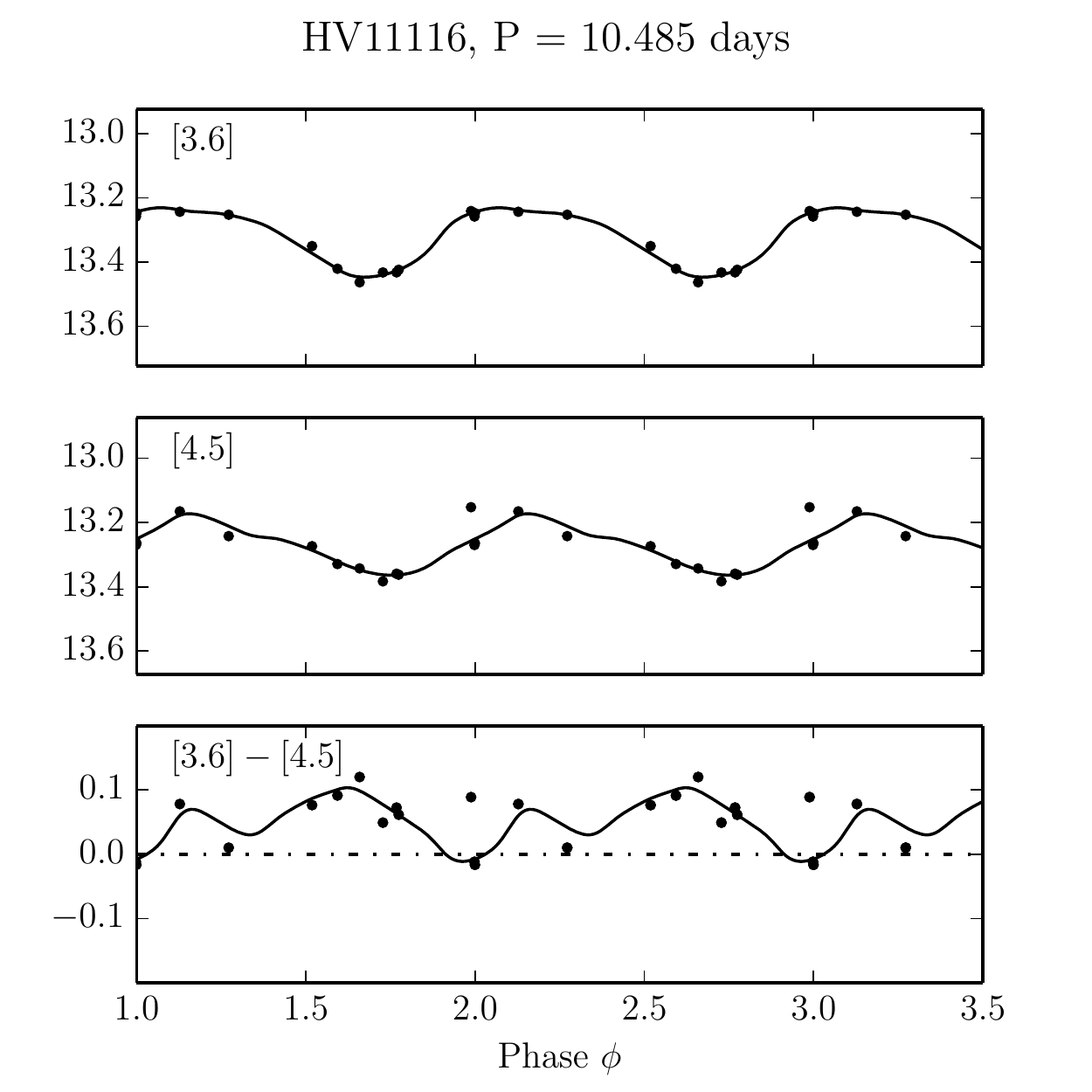} &
\includegraphics[width=50mm]{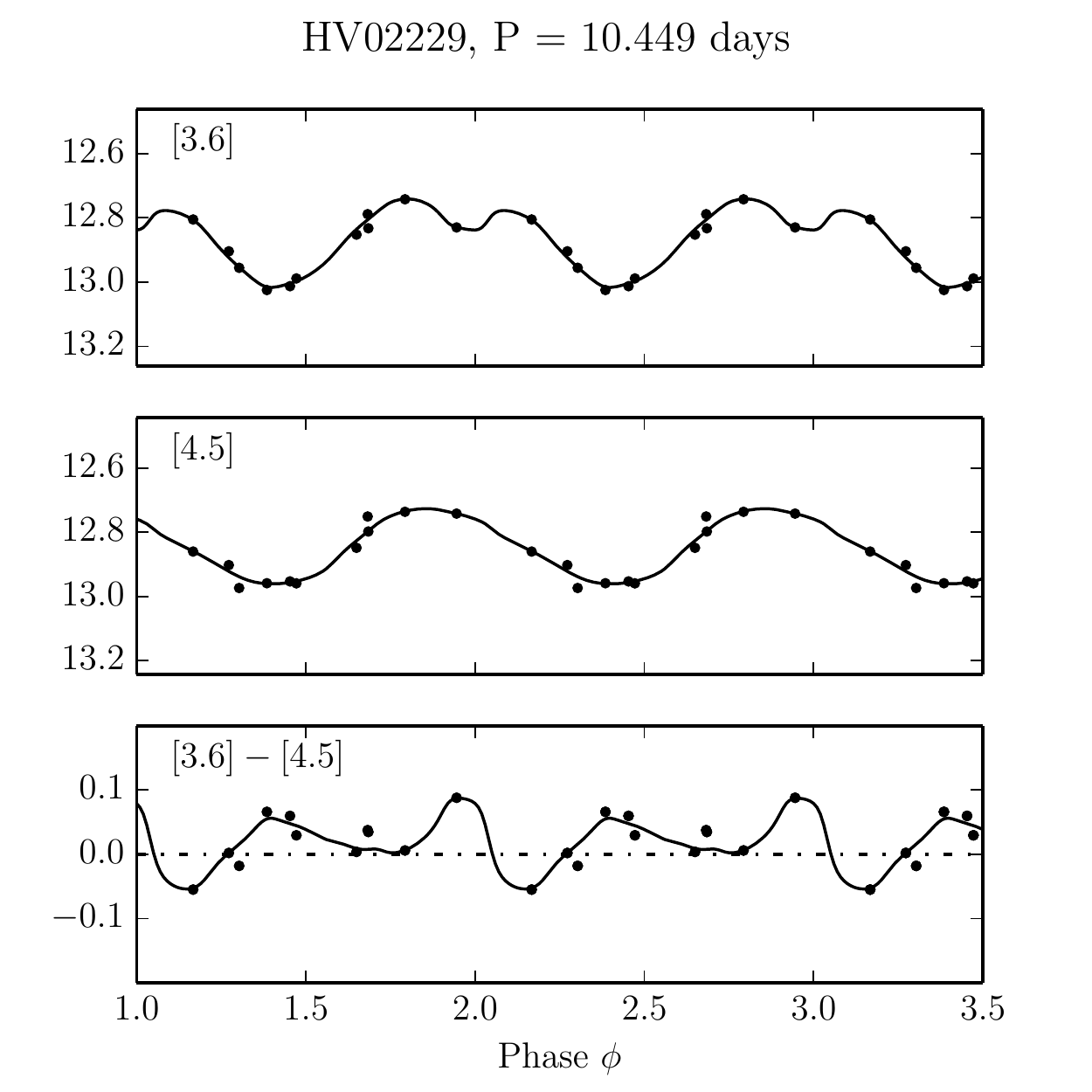} \\
\includegraphics[width=50mm]{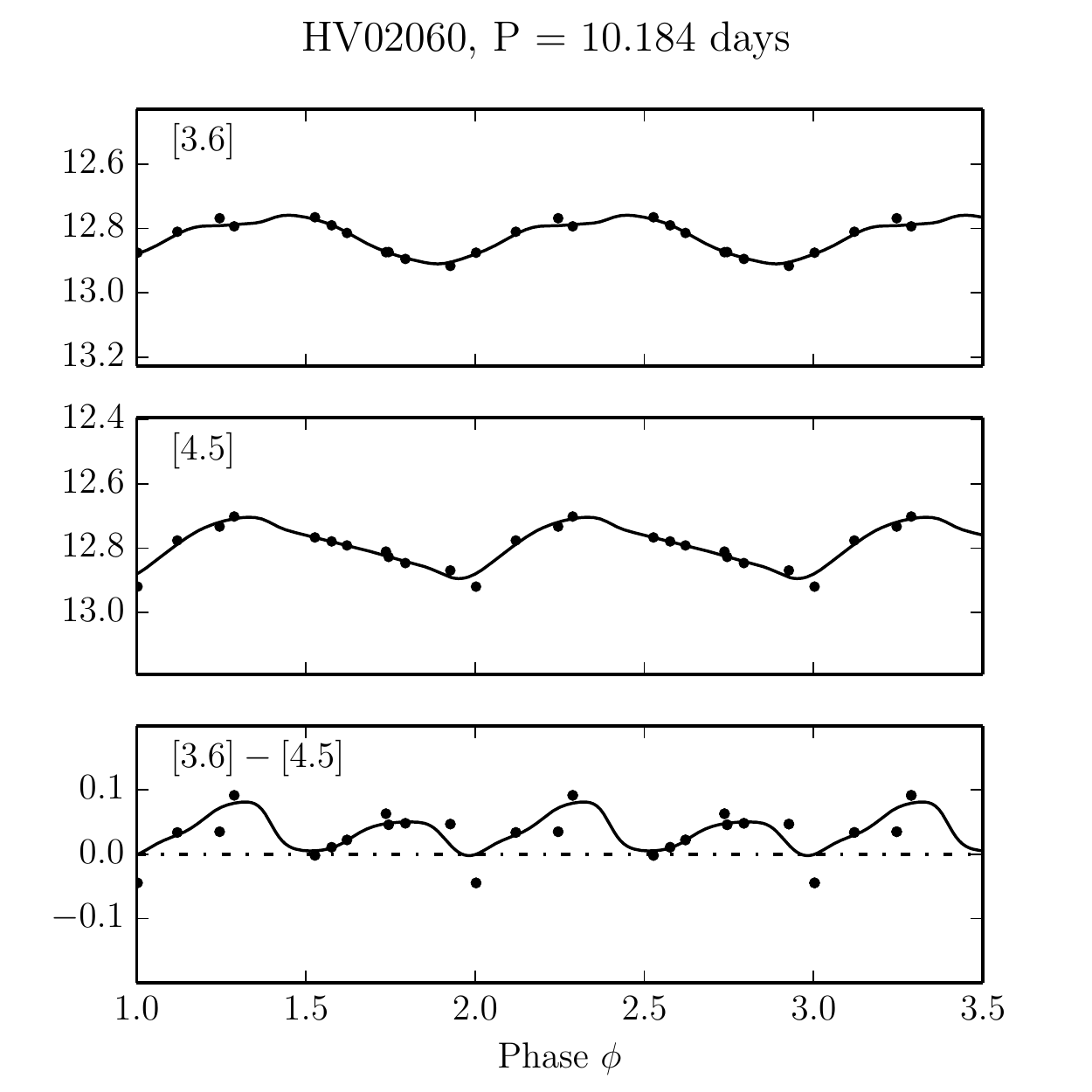} & 
\includegraphics[width=50mm]{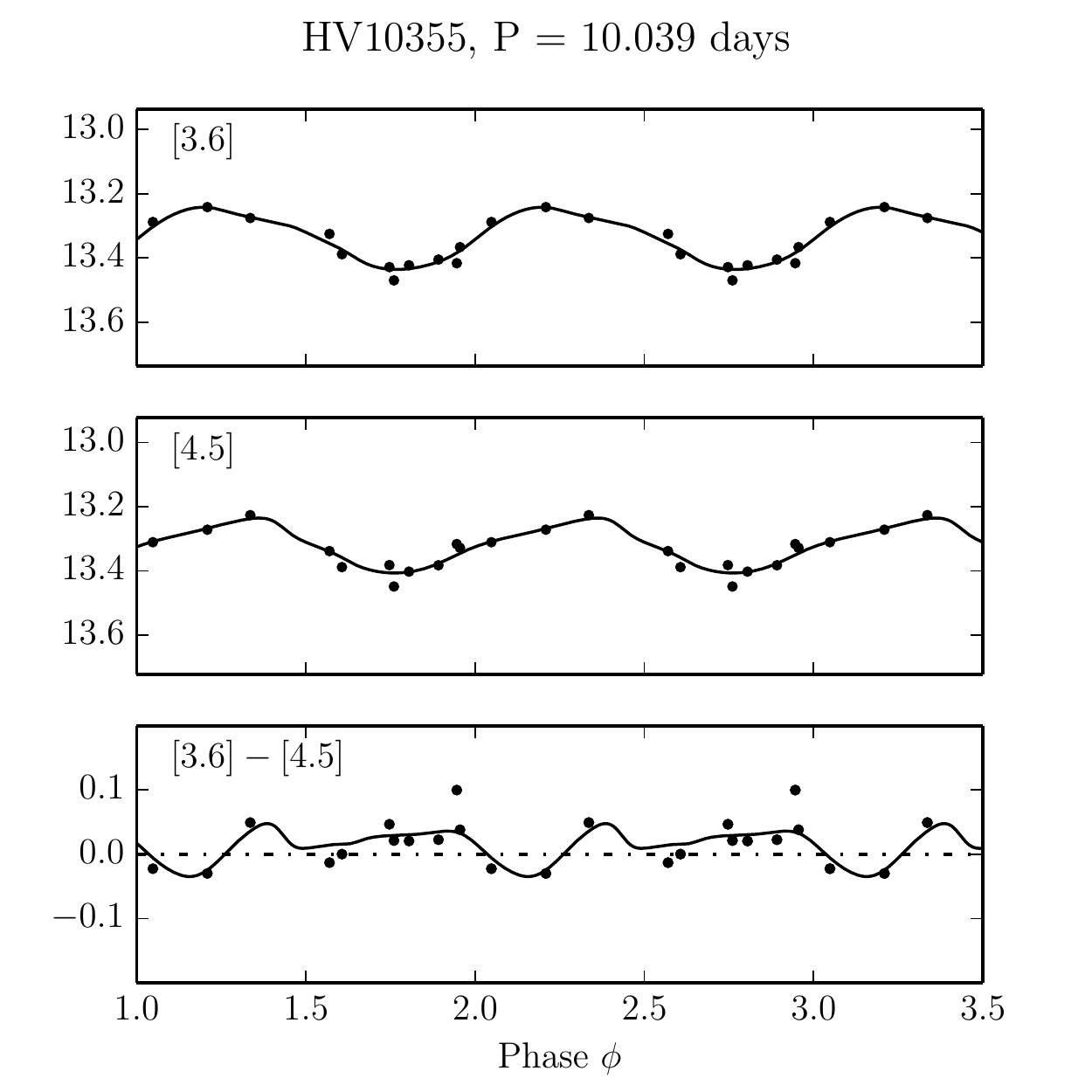} &
\includegraphics[width=50mm]{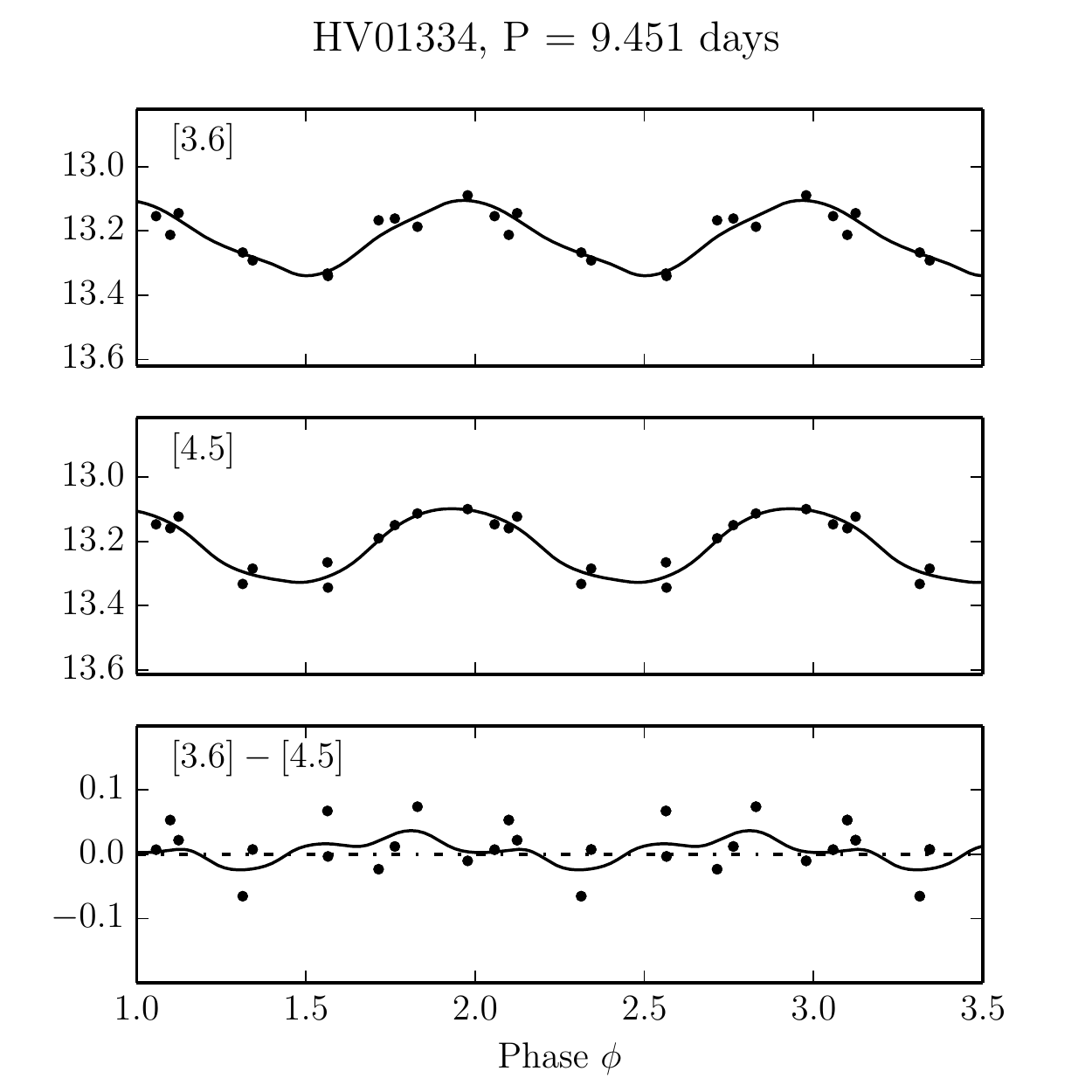} \\
\includegraphics[width=50mm]{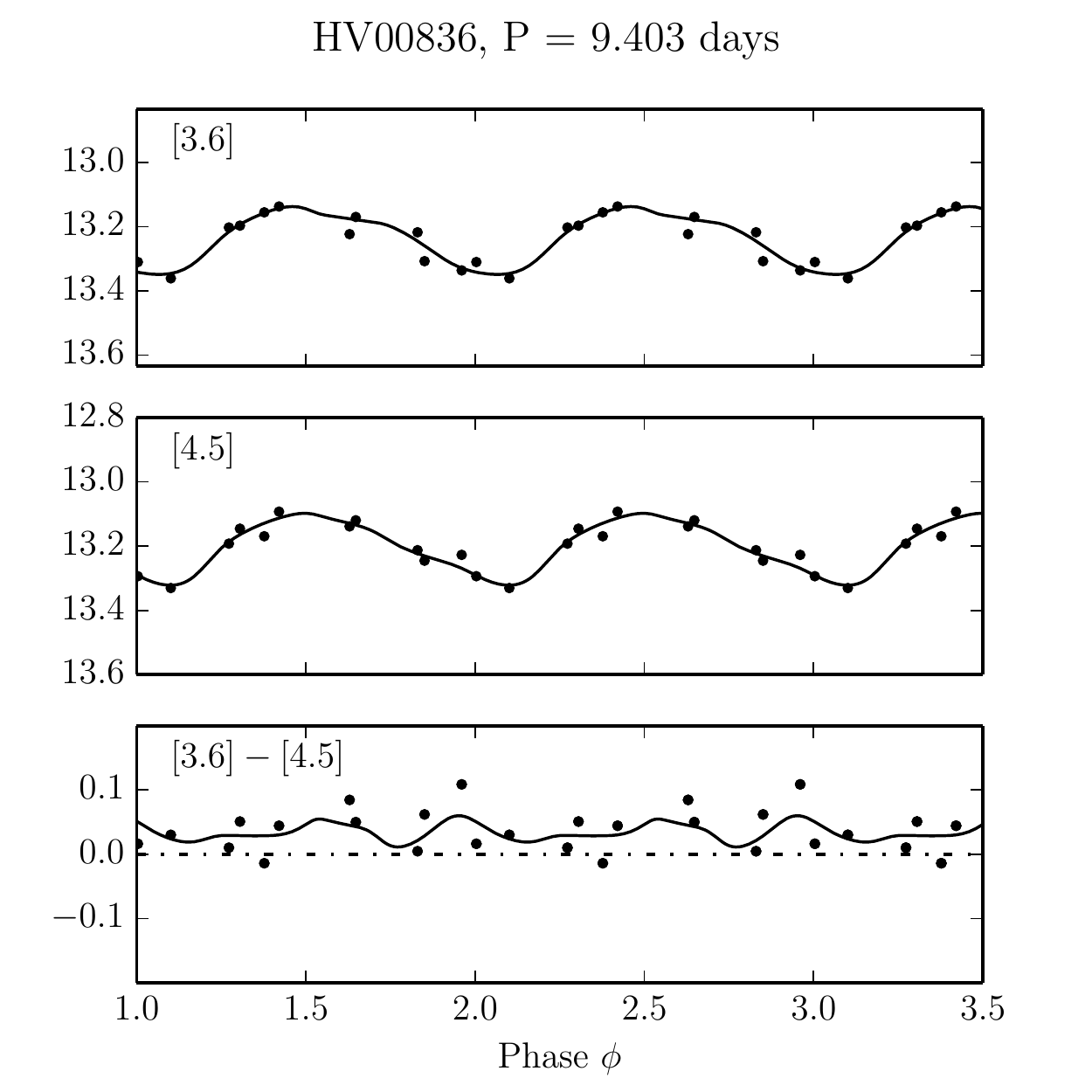} &
\includegraphics[width=50mm]{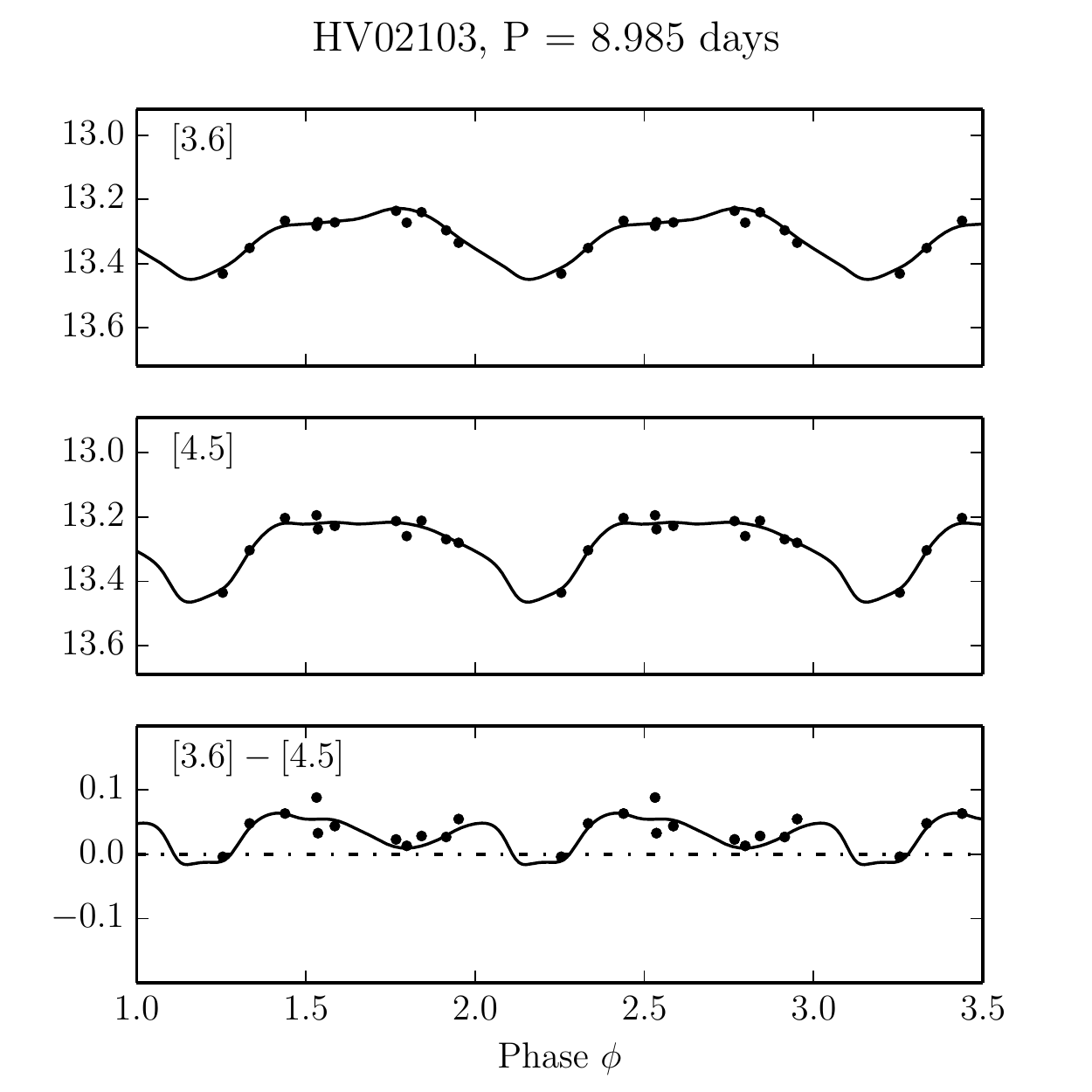} &
\includegraphics[width=50mm]{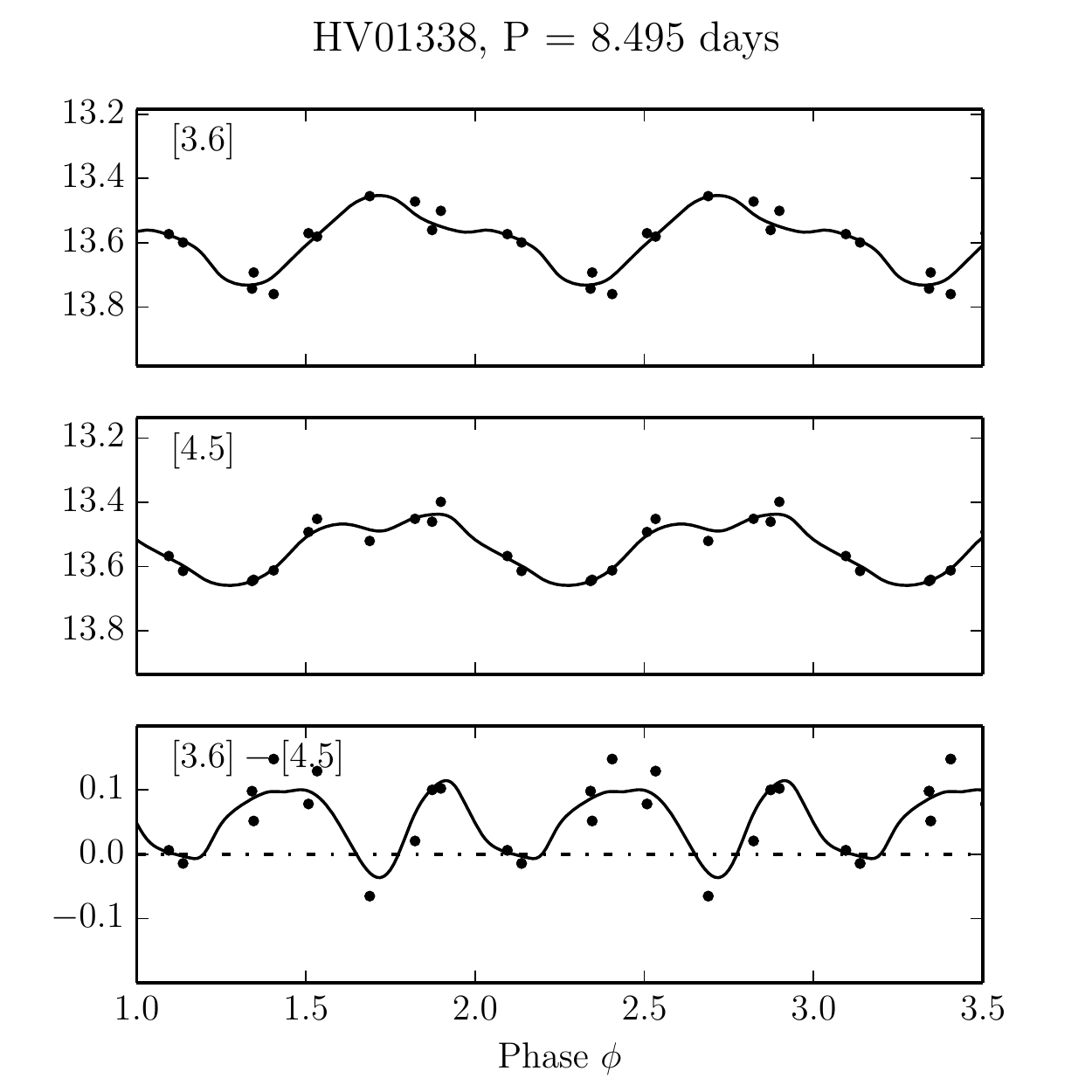}  \\

\end{array}$ 
\end{center} 
\end{figure}
\begin{figure} 
 \begin{center}$ 
 \begin{array}{ccc} 
\includegraphics[width=50mm]{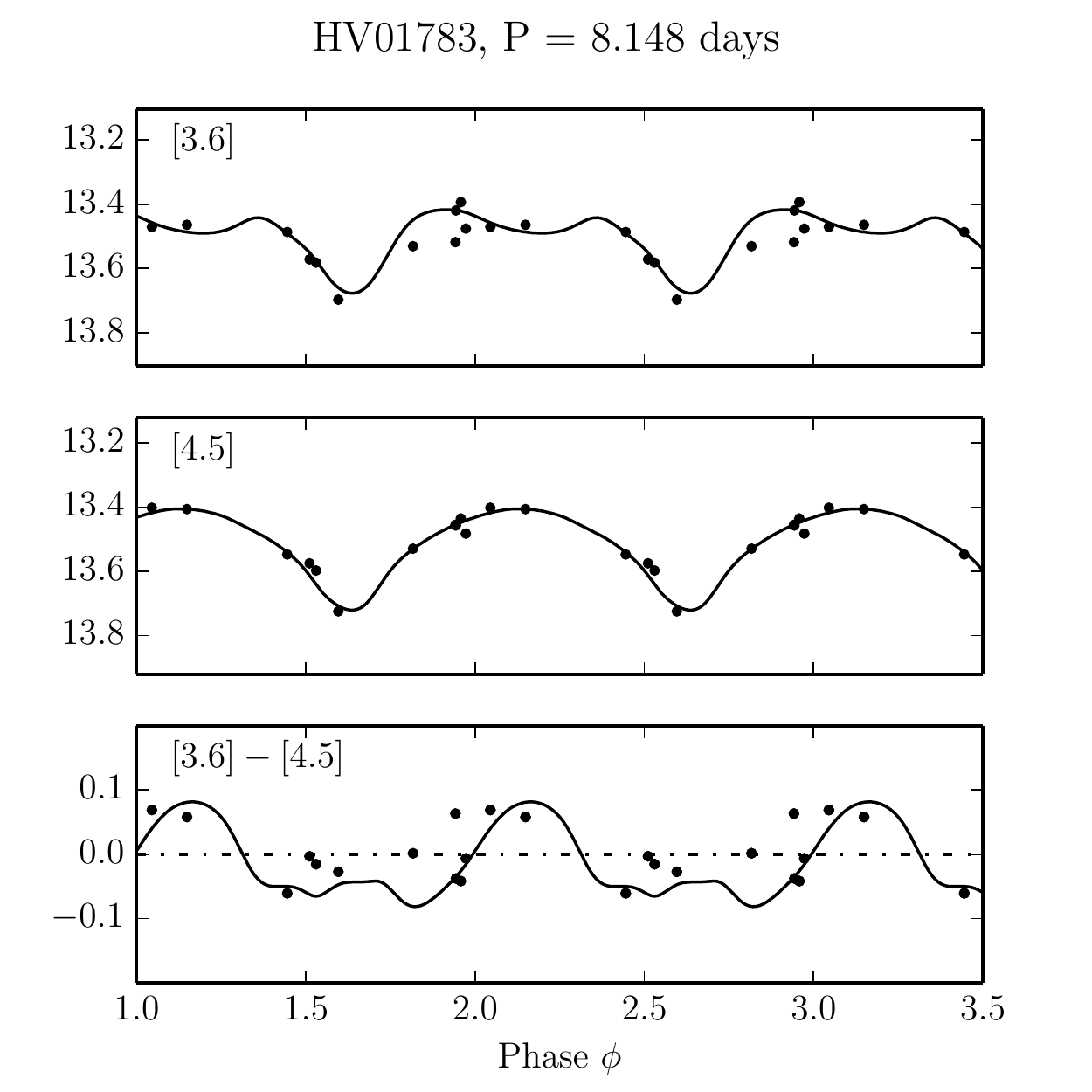}&
\includegraphics[width=50mm]{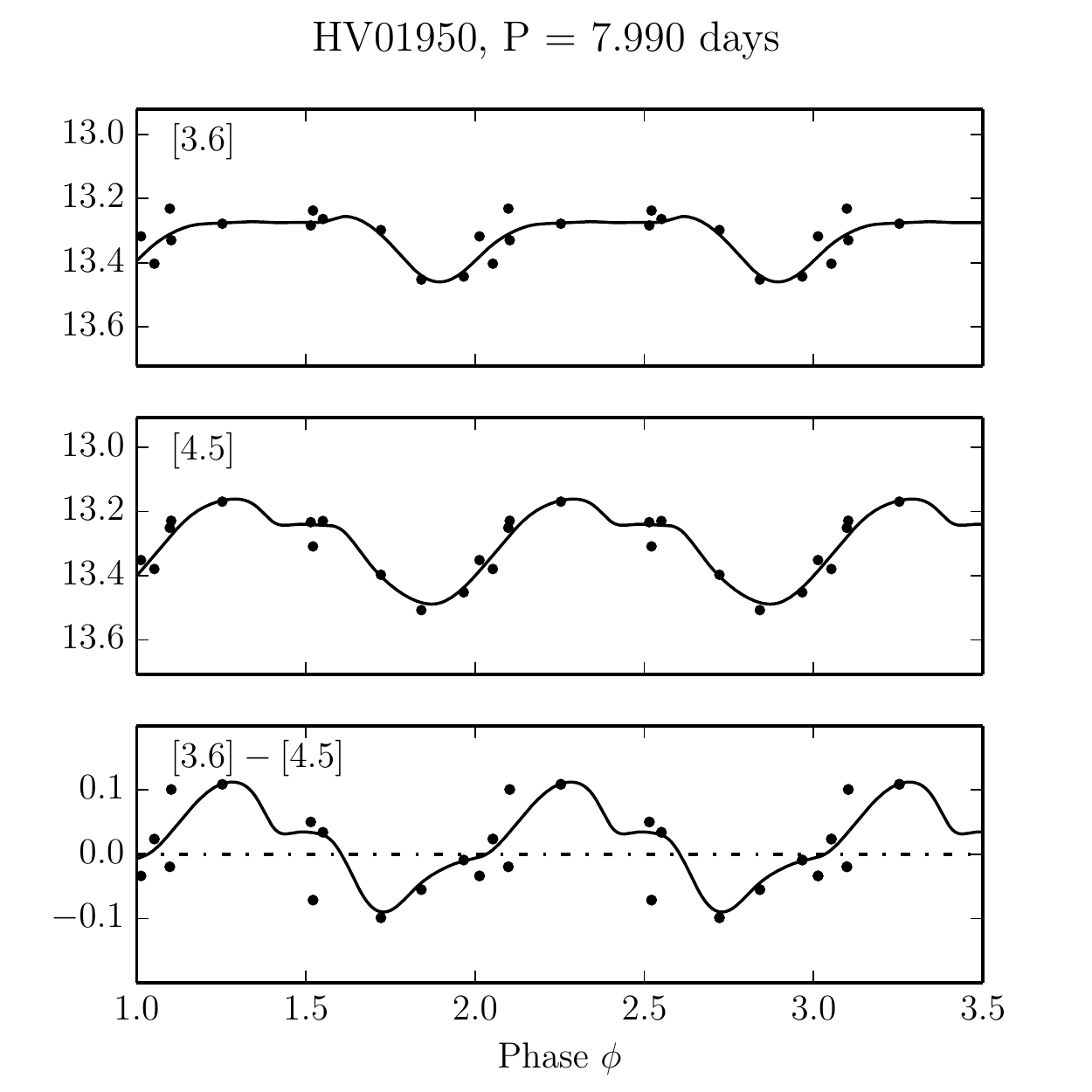} &
\includegraphics[width=50mm]{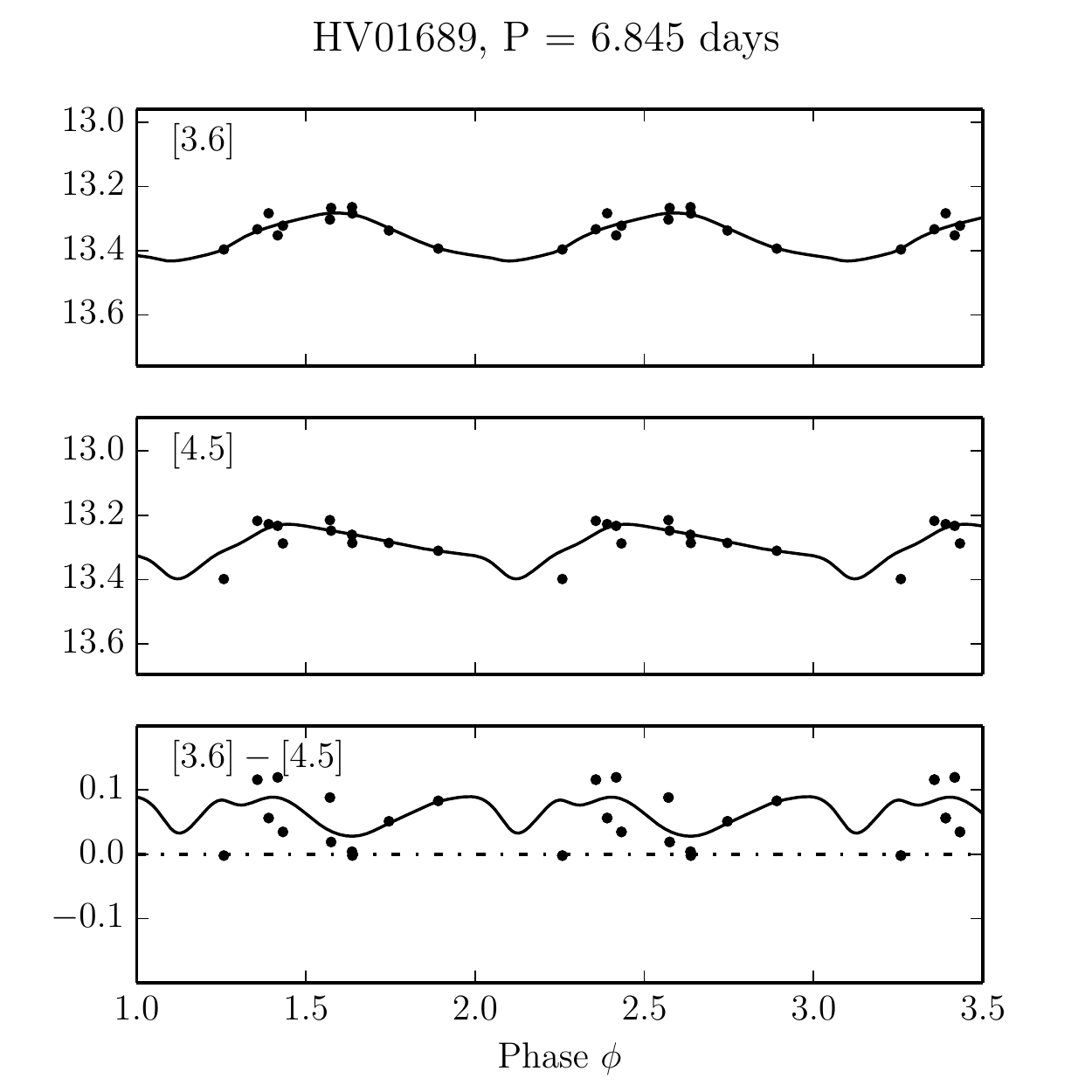} \\ 
\includegraphics[width=50mm]{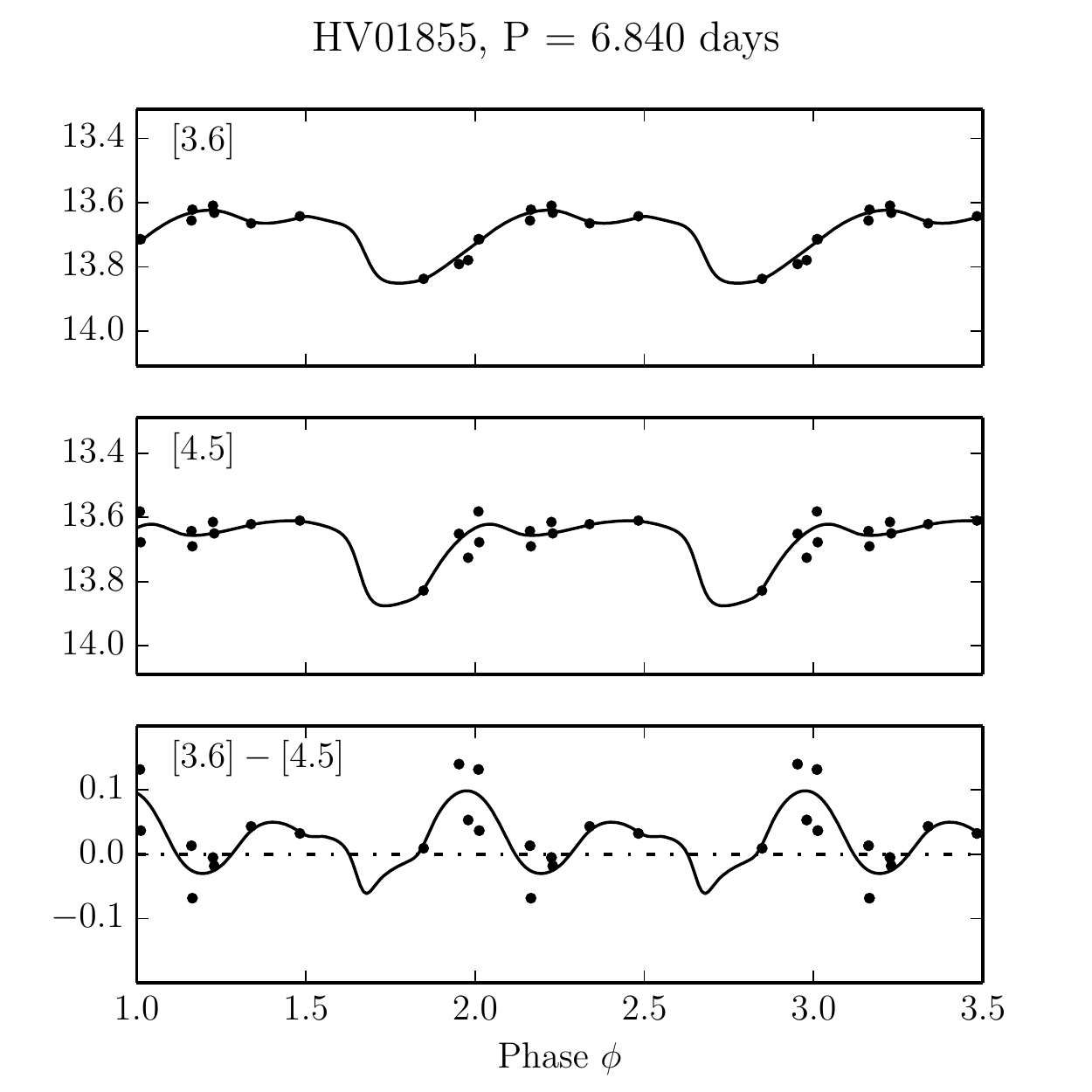} & 
\includegraphics[width=50mm]{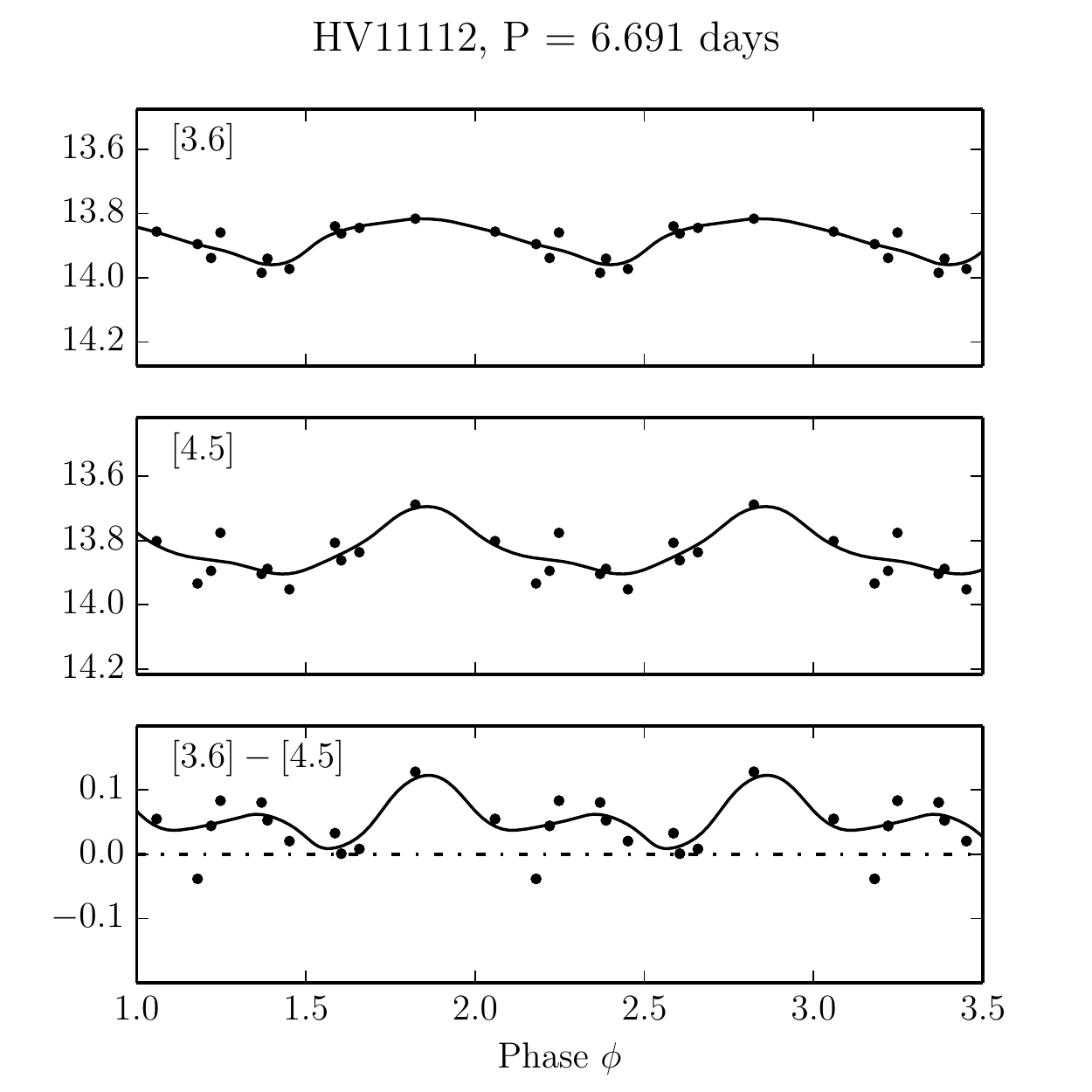} &
\includegraphics[width=50mm]{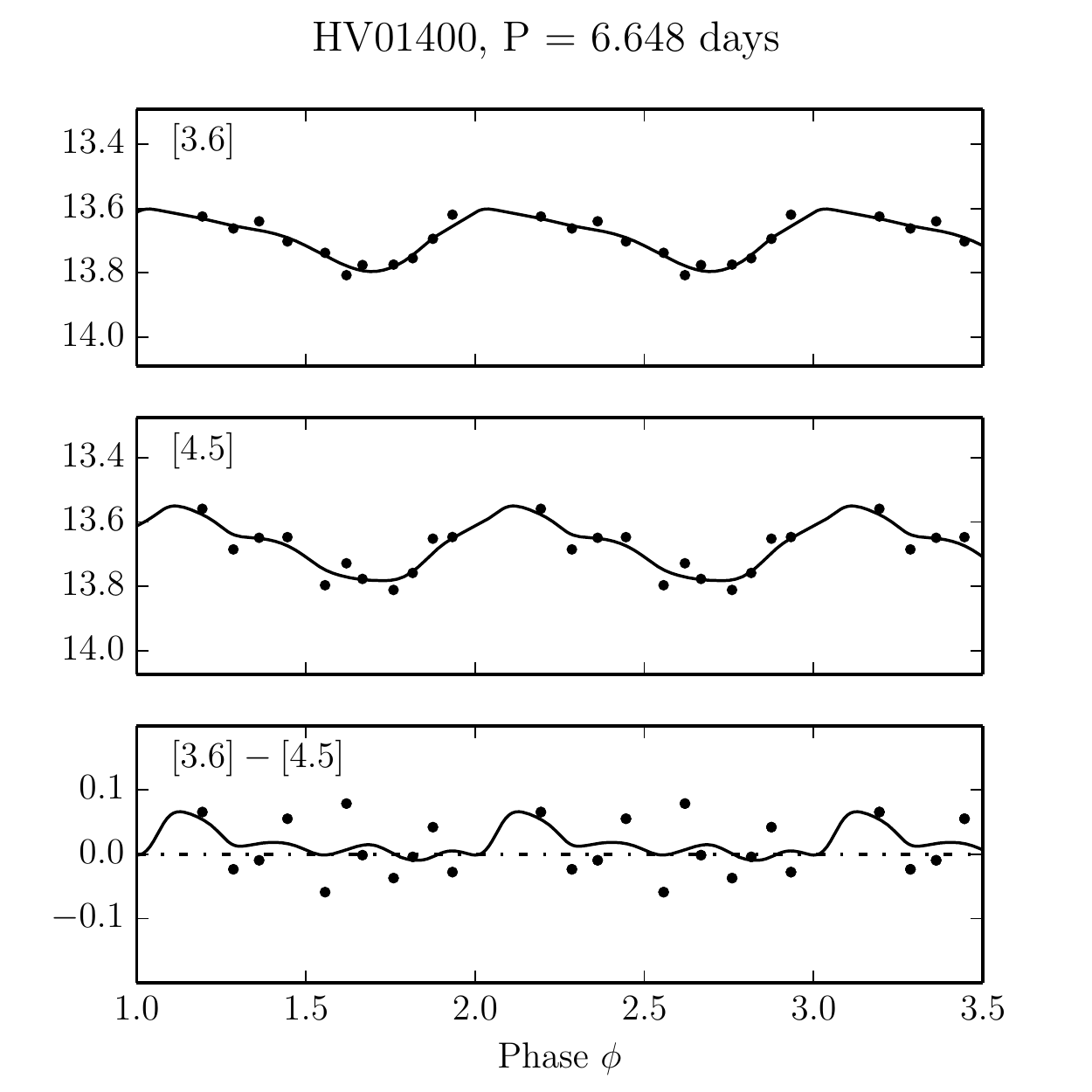} \\
\end{array}$ 
\caption{IRAC light curves of the SMC Cepheid sample, in order of decreasing period. Point sizes are comparable to the uncertainties in the $[3.6]$ (top) and $[4.5]$ (middle) panels. The bottom panels show the variation of the IRAC  $[3.6]-[4.5]$  color with phase. More negative in color corresponds to greater absorption by CO.  The light curves were fit using GLOESS, a Gaussian local estimation algorithm which has been used throughout the CHP (e.g. see S11, M12).  To be included as an Appendix.}
\label{fig:light_curves_appendix}
 \end{center}
\end{figure}

\end{document}